\documentclass{tran-l}

\usepackage[inactive]{srcltx} 
\usepackage{graphicx}

\usepackage{amsmath,amssymb,mathtools,mathrsfs,bm,dsfont,latexsym}

\usepackage{array} 
\usepackage{yhmath} 

\usepackage{multicol}

\usepackage{url}

\newcommand{\cl}[2]{\ensuremath{\mathit{Cl}_{#1,#2}}}
\newcolumntype{L}{>{$}l<{$}}

\newcommand{\bbR}{\ensuremath{\mathbb{R}}}
\newcommand{\bbI}{\ensuremath{\mathbb{I}}}
\newcommand{\bbC}{\ensuremath{\mathbb{C}}}
\newcommand{\bbH}{\ensuremath{\mathbb{H}}}

\newcommand{\reverse}[1]{\widetilde{#1}}
\newcommand{\gradeinverse}[1]{\wideparen{#1}}

\newcommand{\ii}{\ensuremath{\mathrm{i}}}
\newcommand{\jj}{\ensuremath{\mathrm{j}}}
\newcommand{\kk}{\ensuremath{\mathrm{k}}}

\def\A{\mathsf{A}}
\def\B{\mathsf{B}}
\def\C{\mathsf{C}}
\def\X{\mathsf{X}}
\def\m#1{\mathsf{#1}}

\def\e#1{\mathbf{e}_{#1}} 
\newcommand{\dd}{\mathrm{d}}
\newcommand{\ee}{\mathrm{e}}

\newcommand{\ba}{\ensuremath{\mathbf{a}}}

\newcommand{\bi}{\ensuremath{\mathbf{i}}}
\newcommand{\bj}{\ensuremath{\mathbf{j}}}
\newcommand{\bk}{\ensuremath{\mathbf{k}}}
\newcommand{\bv}{\ensuremath{\mathbf{v}}}

\newcommand{\cA}{\ensuremath{\mathcal{A}}}
\newcommand{\cB}{\ensuremath{\mathcal{B}}}

\newcommand{\gs}[1]{\ensuremath{\langle #1\rangle_0}}
\newcommand{\gv}[1]{\ensuremath{\langle #1\rangle_1}}
\newcommand{\gb}[1]{\ensuremath{\langle #1\rangle_2}}
\newcommand{\gt}[1]{\ensuremath{\langle #1\rangle_3}}
\newcommand{\gq}[1]{\ensuremath{\langle #1\rangle_4}}

\newcommand{\magn}[1]{\lvert #1\rvert}

\usepackage{xcolor}

\begin{document}

\title[Spectral square roots of MV]
 {Spectral square roots of the  multivector}


\begin{center}
\author{Adolfas Dargys$^{*}$, Art{\=u}ras Acus$^{**}$}

\address{$^{*}$Center for Physical Sciences and Technology, Semiconductor
Physics Institute \\ Saul{\.e}tekio 3, LT-10257
Vilnius, Lithuania\\
adolfas.dargys@ftmc.lt\newline\indent
$^{**}$Institute of Theoretical Physics and Astronomy, Vilnius
University\\ Saul{\.e}tekio 3, LT-10257 Vilnius, Lithuania\\
arturas.acus@tfai.vu.lt}

\end{center}

\begin{abstract} The problem of  multivector (MV) multiple square roots  in real
geometric  Clifford  algebras $\cl{p}{q}$  with symbolic coefficients is considered. The
method to find  multiple MV square roots that is  based on
R.~Bott's periodicity  table and matrix eigensystem in $\cl{p}{q}$ is proposed.
The method can be applied to MV having both
numerical and symbolic coefficients.  In addition, method
allows to determine  the domain of the existence of thus obtained spectral square roots.
A number of examples is presented for multivectors
in low, $p+q\le 3$, and higher dimensional Clifford algebras, including 4D
(anti)-Euclidean space and relativistic   \cl{1}{3} and \cl{3}{1} algebras.
Tables of the required basis vectors  for conversion of MV to Bott's matrix
representation  have been  found from respective algebra
idempotents using ideal theory and  presented for real Clifford algebras in Appendix.
\end{abstract}


\maketitle


\tableofcontents

\section{Introduction}\label{sec:1}

Prof. A.~Cayley  was the first to carry over the concept of  square root
of a number to matrix in 1872~\cite{Cayley1872}. Apart from non-commutativity
property, in a sharp contrast to real and complex numbers where the
root has two plus/minus values, the
matrix  root  may be multiple, i.e. have more  than two
roots, or the roots may not exist at all. Due to importance in
applications~\cite{Lancaster95,Abou2003,Hitzer2022}, in the recent
monograph~\cite{Higham08} two sections were devoted to numerical analysis of
square and $p$-th roots of matrices. Matrix numerical methods are more common
due to  difficulties with general symbolic matrix theory.

However, in Clifford algebra (CA), also called geometric algebra (GA),
it appears that  the matrix representations (reps) of multivectors
are relatively simple and, thus, general symbolic methods, as shown in the present paper
may be applied directly to find multiple square roots of a symbolic matrix.
The respective matrix representations (reps) of CAs  have been  found by R.~Bott~\cite{Bott1959}
in 1959 and summarized in a Bott table.  In one  or another shape,
the Bott table may be found in nearly all books on Clifford
algebras (see, for example, the popular one~\cite{Lounesto97}).
In a non-commutative CA, the multivectors (MVs),
 may be represented by $2n\times 2n$
real, complex, or Hamilton quaternionic matrices. An
important property of real, complex and quaternionic matrices
which represent  MVs is that such matrices have rather simple structure and  possess high
symmetry, therefore as we shall see, such CA matrix reps may be
beneficial in calculation of the square roots of general MVs.

The most akin to the present work is the  square root of $-1$ that plays a
key role in the Clifford-Fourier transforms and CA based wavelet
theory \cite{Hitzer2011, Hitzer2013,Hitzer2022}. In contrast to
complex number Fourier transform (we shall remind that complex
number algebra is isomorphic to Clifford algebra \cl{0}{1}),  the
existence of more than two  square roots of $-1$ in (non-commutative)
CAs allows to create new geometric kernels for Fourier transforms
and wavelets as well as new transforms such as left-right and
double-sided Fourier transforms~\cite{Hitzer2022}. Also, there
appears a possibility - beyond the analysis in a complex plane - to include multiple
roots into monogenic and holomorphic CA function analysis in
larger vectorial  spaces~\cite{Brackx2001,Gurlebeck2008,Tutschke2007,Tutschke2015},
including relativistic ones so important in the modern physics and cosmology.

Up till now, a main attempt to compute MV square root symbolically
has been  focused on $n=2$ and $n=3$ CAs,
 where $n=p+q$ is $\cl{p}{q}$ algebra dimension, as well as on
quaternions~\cite{Niven1942,Janovska2007,Opfer2017} or their
derivatives such as coquaternions (also called split quaternions)
or nectarines~\cite{Ozdemir2009,Falcao2018,Opfer2017}. The square
root of biquaternion (complex quaternion) was considered in~\cite{Sangwine2006}.
Larger real  algebras \cl{p}{q}, when $n\ge4$ and the
signature is $s=p-q=3(\textrm{mod}\,4)$, were considered
in~\cite{Hitzer2013}. In particular, to find square roots of $-1$ the authors~~\cite{Hitzer2013}
used the CA-to-matrix isomorphisms in order to be able to
characterize algebraically the continuous manifolds of square
roots. In the analysis, the topologically connected conjugacy
classes of square roots are used for this purpose. Such  $-1$ roots may
be useful in construction of the Clifford-Fourier transforms.

In papers \cite{Acus2024v2,Dargys-Acus2020},  we have demonstrated that
in 3-dimensional CAs the  square root  of a  general MV can be extracted in radicals.
New  method that allow to get all  numerical square roots are discussed  and demonstrated as well.
Also, we have shown~\cite{Acus2024v2} that there may be isolated square roots as well as
continuum of roots (infinitely many roots)  on hypersurfaces in 3-dimensional algebras.
In this paper we take use of isomorphism between real Clifford
algebra MVs and their matrix representation (rep) that follows
from R.~Bott's periodicity table. The developed spectral method  allows to calculate isolated roots
of general symbolic  MV and conditions for roots to exist. To find the roots, the
matrix reps are converted to diagonal form and then all possible
sign combinations of real or complex roots on a diagonal spectral matrix are selected. After conversion back to
Bott's representation one gets required MV multiple  square roots. For
use in practice, we have also provided matrix reps of
basis vectors for all algebras when $n<7$. Examples of symbolic
square roots are given for some important CAs.

In Sec.~\ref{notation},  notation is introduced. In
Sec.~\ref{method}, the principle of the spectral diagonalization method to
find MV roots is described. Also, equivalence between MV and its
matrix representation is touched with the emphasis upon features
related to roots. Concrete MV root formulas for low dimensional as
well as 3D and 4D GA's are presented, respectively, in
Sec.~\ref{examples1-2D}, Sec.~\ref{examples3D} and
Sec.~\ref{examples4D}. Adaptation of spectral diagonalization method to MVs
in a numerical form, especially in high dimensional GA's,  is
demonstrated and discussed in Sec.~\ref{numerical}. Finally, as an
illustration of application of  square roots we solve  quadratic
MV equations in~Sec.~\ref{Riccati}. Conclusions are drawn is
Sec.~\ref{conclusions}. In Appendix (Sec.~\ref{basisTables}) the
tables of matrix reps for basis vectors that are required in the
method are calculated and presented.

\section{Multivector  matrix reps}\label{notation}

The Clifford algebra \cl{p}{q} is an associative vectorial and noncommutative
 algebra having $2^n$ basis elements, where
$n=p+q$ is the dimension of the vector space. Geometrically it is
defined by orthogonal unit basis vectors $\e{i}$, where
$i=1,\dots,n$.  $s=p-q$ is the signature of
algebra, where squares of $p$ and $q$  vectors are, respectively, $\e{i}^2=1$ and
 $\e{i}^2=-1$. If $q=0$ the vector space is Euclidean
and if $p=0$ it is anti-Euclidean. Otherwise the space has a mixed
signature. Apart of vectors the space accommodates oriented planes
called bivectors $\e{ij}=\e{i}\e{j}$, oriented volumes called
trivectors $\e{ijk}=\e{i}\e{j}\e{k}$, and oriented supervolumes called
quadvectors, pentavector etc.  The number of subscripts indicates
the grade of basis element, so that the scalar is a grade-0
element, the vector is a grade-1 element, etc. The largest elementary
(super)volume belongs to blade  $\e{ijk\dotsm n}$, which  commonly is called
the pseudoscalar and indicated by $I$. Thus, for example in 3D algebras
(\cl{3}{0},\cl{2}{1},\cl{1}{2} and \cl{0}{3}) the MV consists of
the following elements (basis blades)
$\{1,\e{1},\e{2},\e{3},\e{12},\e{13},\e{23},\e{123}\equiv I\}$.
The first element represents the scalar.

The  MV $\A$ is a sum of different blades multiplied by real
coefficients $a_J$. In 3D algebras the general MV expanded in
coordinates (basis elements) reads,
\begin{equation}\begin{split}\label{mvA}
\A=&a_0+a_1\e{1}+a_2\e{2}+a_3\e{3}+a_{12}\e{12}+a_{23}\e{23}+a_{31}\e{31}+a_{123}I=\\
&a_0+\ba+\cB+a_{123}I,
\end{split}\end{equation} where $a_0=\gs{\A}$ is the scalar,
$\ba\equiv\gv{\A}=a_1\e{1}+a_2\e{2}+a_3\e{3}$ is the vector and
$\cB\equiv\gb{\A}=a_{12}\e{12}+a_{23}\e{23}+a_{31}\e{31}$ represents bivector
part. Thus,   to single out
a sum of scalar and pseudoscalar from MV given by Eq.~\eqref{mvA}, it is enough to
apply grade-0 and grade-3 selectors,
$\gs{\A}+\gt{\A}=a_0+a_{123}I$. Similar rule exists for matrix reps as well
(see Table~\ref{MV-rep} and Eq.~\eqref{matrixToMVA}).
In general, $\langle\m{A}\rangle_k$ selects $k$-th grade element from the
MV~$\m{A}$, where $k=0,1,2,3...$ designates scalar, vector, bivector,
trivector etc. More about  MV properties can be found, for
example, in books~\cite{Lounesto97,Doran03}. Table~\ref{MVmatReps}
lists some useful relations between MVs and their matrix
counterparts, or reps, where  the important property  is included too, i.e., the geometric  product of two MVs
is equivalent to  matrix which represent
MVs product.

\begin{center}
 \begin{table}\label{MV-rep}
\begin{tabular}{l|l|l}
 & MV  & Matrix rep\\\hline
 & & \\[-5pt]
 Multivector (MV) & $\m{A},\ \m{B}$ & $\hat{\m{A}},\ \hat{\m{B}}$\\
 MV geometric product & $\m{A}\m{B}$ & $\hat{\m{A}}\hat{\m{B}}$   \\
Inverse MV $\m{A}^{-1}$ & $\m{A}\m{A}^{-1}=1$& $\hat{\m{A}}\hat{\m{A}}^{-1}=\hat{1}$\\
Reversion & $\reverse{\m{AB}}$ &$\hat{\m{B}}\hat{\m{A}}$\\
 Scalar part of MV& $a_0=\langle\m{A}\rangle_0$& $a_0=m^{-1}\text{Tr}(\hat{\m{A}})$,\ $m=\dim\hat{\m{A}}$\\
Basis vector, bivector,\dots& $\e{i}$, $\e{ij}$,\dots & $\hat{\e{}}_j$, $\hat{\e{}}_{ij}=\hat{\e{}}_i\hat{\e{}}_j$,\dots\\
 & &also see Eq.~\eqref{MVToMatrix}\\
Graded MV &$\m{A}=\sum_{i=0}^n\langle\m{A}\rangle_i$ & $\hat{\m{A}}=\sum_{i=0}^n\hat{\m{A}}_i$\\
Inverse basis vector, \dots &$\e{i}^{-1}$, $\e{ij}^{-1}$,\dots& $(\overline{\hat{\e{}}_i})^{*}$, $(\overline{\hat{\e{}}_{ij}})^{*}$, \dots\\
Rep $a_J\hat{\e{}}_J$ extracted from $\hat{\m{A}}$,  &$a_J{\e{}}_J$ &$a_J\hat{\e{}}_J=m^{-1}\text{Tr}(\hat{\m{A}}\hat{\e{}}_J^{-1})\hat{\e{}}_J$\\
also refer toEq.~\eqref{matrixToMVA}& & \\
& &$\overline{\hat{\m{A}}}$ is a transposed matrix \\
 & & $(\hat{\m{A}})^{*}$ is a conjugate matrix\\
\end{tabular}
\caption{\label{MVmatReps} The equivalence between multivectors
(MVs) and their matrix representations (reps).}
\end{table}
\end{center}

In the Appendix (Sec.~\ref{basisTables}) tables, the 4-th line gives a set of
matrix reps $\hat{\e{}}_i$ for all basis vectors $\e{i}$ that represent  real
CAs.  Reps for  higher grade elements can be obtained by multiplying
respective $\hat{\e{}}_i$ matrices, similarly as it is done for higher grade MVs.
The reader should notice that
in the tables the first basis rep $\hat{\e{}}_1$ is  a
diagonal matrix. Exceptions are anti-Euclidean algebras $Cl_{0,q}$, where all basis vectors square to $-1$ and, therefore, can't be used as an idempotent generators.  In physics, the direction of  associated  $\e{1}$-vector
is connected with the so-called spin quantization axis.
This attribute, i.e., the existence of the quantization
axis,\footnote{In
physics, the quantization axis usually is related to $\e{3}$,
i.e. represents $z$-axis direction rather than  direction of
$\e{1}$  vector.}  was called ``the polarization of space'' or ``fundamental
polar form'' by \'{E}.~Cartan~\cite{Cartan1966}. The fundamental
polar form is an invariant under the group of rotations, or in
physical terms determines all possible eigenstates of an investigated object (the spin).
In the experiment, the latter (quantization axis) is controlled and depends on
apparatus orientation with respect to quantum system.

In the following we shall manipulate between MVs and their matrix
reps. In matrix representation, the MV $\m{A}=a_0+\sum_{i=1}^n
a_i\e{i}+\sum_{i<j}a_{ij}\e{ij}+\dotsc$  has similar structure:
\begin{equation}\label{MVToMatrix}
\hat{\m{A}}=a_0\hat{1}+\sum_{i=1}^n
a_i\hat{\e{}}_i+\sum_{i<j}a_{ij}\hat{\e{}}_{ij}+\dotsc,
\end{equation}
where $\hat{1}$ is a  unit matrix of a considered  algebra and $\hat{\e{}}_{ij}=\hat{\e{}}_{i}\hat{\e{}}_{j}$ is the matrix product of
basis vector reps introduced in  Sec.~\ref{basisTables}.
Thus, as can be seen from  Eq.~\eqref{MVToMatrix}, the knowledge of elementary  reps allows one
to write down any MV in a matrix form immediately.
Conversion of  $m\times m$ matrix rep $\hat{\m{B}}$ back to initial MV form $\m{B}$
may be acquired by rule,
\begin{equation}\label{matrixToMVA}
\m{B}=\sum_J a_J\e{J}=\frac{1}{m}\sum_J\Big(\text{Tr}(\hat{\m{B}}\hat{\e{}}_J^{-1})\Big)\e{J},
\end{equation}
where $J=\{0,i,ij,ijk,\dotsc,ijk\dotsb n\}$ is a multiindex of the
$n$-dimensional MV and Tr~is the matrix trace. The term
$\text{Tr}(\hat{\m{B}}\hat{\e{}}_J^{-1})$  selects $J$-th   coefficient. Since basis matrix satisfies, $\hat{\e{}}_J^{2}=\pm\hat{1}$,
the inverse is either the same matrix or has opposite sign, $\hat{\e{}}_J^{-1}=\pm\hat{\e{}}_J$, therefore, it
is enough to select a correct sign before  $\hat{\mathbf{e}}_J$. Also, one should remember
that the matrix representation of MV in a concrete geometric
algebra consists of narrow class of matrices, for example in 3D algebras
there are $2^{n}=2^3=8$  different $\hat{\e{}}_J$ matrices.

\section{Diagonalization method}\label{method}
\subsection{Square root algorithm}
The algorithm is based on the Bott's table reps and the
equivalence between MV geometric product and matrix product of
reps. Also refer to Table~\ref{MVmatReps}. The
diagonalization method consists of the following steps.
\begin{itemize}
 \item[1.] Select the  MV $\m{A}$ that belongs to a particular \cl{p}{q}.
 \item[2.] Convert the MV $\m{A}$ to
matrix rep $\hat{\m{A}}$ with the help of corresponding table  in
Sec.~\ref{basisTables}. Elementary reps that belong to
bivectors, trivectors etc  are the  products of respective
vector reps from  the selected table. Multiply the
basis elementary reps by respective real coefficients that belong to MV.
 \item[3.] Find eigenvalues and eigenvectors, i.e. eigensystem, of the constructed
rep matrix. Since  the coefficients of $\m{A}$ are  real numbers,
any  program to find eigensystem is suitable for this purpose. If the initial
MV has symbolic coefficients then respective symbolic program, for example, in
\textit{Mathematica} or \textit{Maple} package, should be used.
 \item[4.] Using the eigenvalues, construct a diagonal spectral matrix, and then
extract square roots from diagonal entries. Select one of the plus/minus sign combinations before the roots.
 \item[5.] Then, construct the transformation matrix from the eigenvectors
that will allow you to perform transformation of  the diagonal  matrix back to the
initial  representation $\hat{\m{B}}$, i.e. to a new matrix $\hat{\m{B}}$.
 In the matrix $\hat{\m{B}}$, there is hidden the MV square root
for which sign combinations you have selected on the diagonal matrix are already taken into account.
 \item[6.]With  formula \eqref{matrixToMVA}, convert $\hat{\m{B}}$ to square root MV
$\m{B}$ and check that  all  coefficients are  real. Only pairs of complex conjugate coefficients, which reduce to real numbers,
are allowed as explained by Fig.~\ref{fig:1}.  The square of
the roots must yield the initial MV, $\B^2=\A$.
  \item[7.]  Repeat items $1-6$    with all possible
plus/minus sign combinations for remaining roots.
  \item[8.] The described procedure also
may be used to find square roots of $-1$ that is needed for Clifford-Fourier transform~\cite{Hitzer2022}. Then, in addition, the
following conditions must be satisfied, where $n=p+q$,
\begin{equation*}\begin{split}
&\gs{\A^2}=-1,\qquad\qquad\quad\ \text{root condition}, \\
&\langle\A^2\rangle_k=0,\ 1\le k\le n,\quad\text{constraints.}\\
\end{split}\end{equation*}
\end{itemize}
The described procedure may be applied to find roots of MVs
with symbolic as well as  numerical coefficients.

\subsection{Some practical issues}\label{comments}
Here  some points relevant  to square root method  are discussed shortly.

1. \textit{Maximal number of roots}. The maximal number of spectral
roots is equal to  all possible plus/minus sign combinations on
the diagonal of eigenvalue matrix. This property follows  from Bott's table
(also, refer to the tables in Sec.~\ref{basisTables}, where Bott's
matrix symbol is written next to  algebra name). Bott's matrices may be real, complex
or quaternionic. The maximal number of square roots is $n_t=2^t$.\,
 where $t$ is the dimension of the Bott's matrix rep.
For example, $t=1$ for algebras \cl{0}{1} (complex numbers) and
\cl{0}{2} (quaternions);  $t=2$ for algebras \cl{2}{0}, \cl{3}{0} and
\cl{1}{3} (quaternions);  $t=4$ for algebras \cl{3}{1}, \cl{2}{2} and \cl{2}{1} with reps $\bbR(4)$ and $^2\bbR(2)$,
respectively (Sec.~\ref{basisTables}).
Appearance of repeated roots, complex coefficients or no-root solutions may
diminish the  number of  possible square roots.

2. \textit{Quaternionic reps}. The problems may arise with
quaternionic reps. The simplest way to circumvent the
problem is to replace the quaternionic entries by $2\times 2$
complex matrix blocks that represent Hamilton imaginaries $\mathbf{i},\mathbf{j},\mathbf{k}$.
Since maximal number of the roots must
remain the same as it is in quaternionic rep, therefore the number of
plus/minus sign combinations must be smaller then that given by formula~$n_t=2^t$. For
example, if initial quaternion matrix is $1\times 1$, we have only
two plus/minus roots. If quaternions are replaced by complex
$2\times 2$ matrices then the root signs on diagonals must remain
the same, i.e.  $\{++\}$ and $\{--\}$. Thus, in this case the combinations $\{+-\}$
and $\{-+\}$ are forbidden, otherwise spurious roots will appear
in the solution. For concrete examples refer to
Subsec.~\ref{examples2D} and Subsec.~\ref{subsecCL03}.

3. \textit{Pairs of  plus/minus roots}. In general, the spectral square root
algorithm gives pairs of roots
having opposite plus/minus signs, simply because in the diagonal
root matrix there always appear combinations with opposite signs,
for example $(+,-,+,-)$ and $(-,+,-,+)$  in the case of $4\times
4$ matrix rep. Thus,  the MV roots always consist of root pairs
having plus/minus signs, and it is enough to restrict the calculations
to $n_t/2$ sign combinations.

\begin{figure}[t]
\centering
a)\includegraphics[height=5cm]{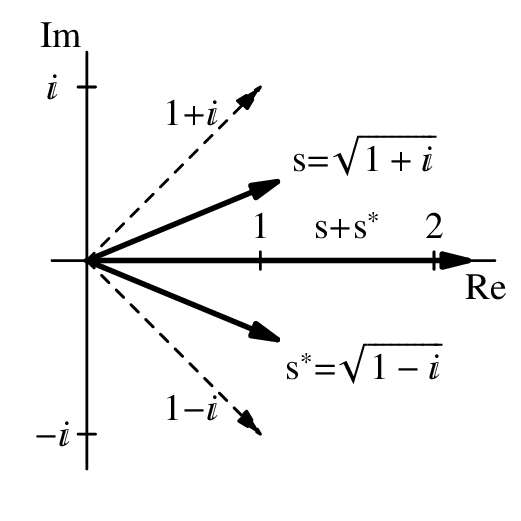}
b)\includegraphics[height=5cm]{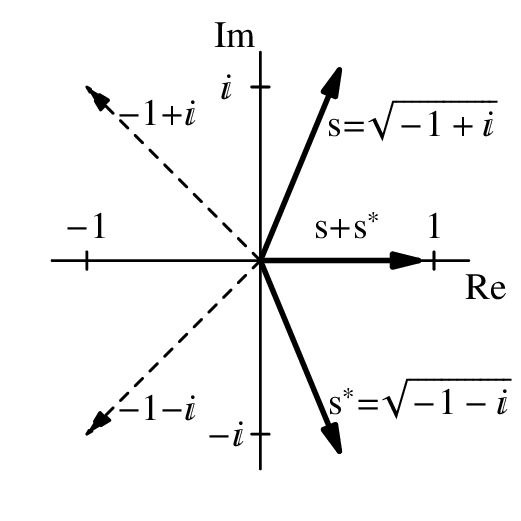}
\caption{Complex plane illustration of sum of square roots $s$ and $s^*$ with
complex conjugate coefficients inside the roots that corresponds to real multivector
$(s+s^{*})A_r$, where $A_r$ may be, for example, an elementary blade. a)
$s+s^{*}=\sqrt{1+\ii}+\sqrt{1-\ii}=2^{5/4}\cos(\pi/8)$. b)
$s+s^{*}=\sqrt{-1+\ii}+\sqrt{-1-\ii}=2^{5/4}\cos(3\pi/8)$.
\label{fig:1}}
\end{figure}

4. \textit{Roots with complex conjugate pairs}.
In real CAs, the coefficients of a square root must be real.
Frequently, after conversion back to initial  MV form the roots appear in pairs with
complex conjugate coefficients as Fig.~\ref{fig:1} illustrates. Such
pairs, in fact,  represent real coefficients and in symbolic calculations they
must be considered  as real expressions.  For example, if $c=c_r+\ii c_i$,  $c^{*}=c_r-\ii c_i$ and
$|c|=\sqrt{c_r^2+c_i^2}$, then one can write~\cite{Korn1961}
\begin{equation}\begin{split}\label{cmplConj}
&\sqrt{c}+\sqrt{c^{*}}=2|c|^{1/2}\cos\big(\tfrac{1}{2}\arctan(c_r,c_i)+k\pi\big),\qquad\  \text{real} \\
&\ii(\sqrt{c}-\sqrt{c^{*}})=-2|c|^{1/2}\sin\big(\tfrac{1}{2}\arctan(c_r,c_i)+k\pi\big).\quad \text{real}\\
\end{split}\end{equation}

Thus, the expressions on left side do represent real numbers.  Two arguments in arc tangent function take into
account a proper quadrant on the complex plane.\footnote{In case
of hyperbolic plane (see Fig.~\ref{fig:2}b), there are four
quadrants, therefore, a two-argument hyperbolic area tangent may be introduced too.
\textit{Mathematica} understands the  two-argument command for
$\bf{ArcTan(x,y)}$  only. }
From Eq.~\eqref{cmplConj} follows that
\begin{equation}\begin{split}\label{cmplRoot}
\sqrt{c}&=|c|^{1/2}\Big(\cos\big(\tfrac{1}{2}\arctan(c_r,c_i)+k\pi\big)+\ii \sin\big(\tfrac{1}{2}\arctan(c_r,c_i)+k\pi\big)\Big)\\
&=|c|^{1/2}\exp\Big(\ii\big(\tfrac{1}{2}\arctan(c_r,c_i)+k\pi\big)\Big),\qquad \text{where\ } k\in\mathbb{N}.
\end{split}\end{equation}

5. \textit{Principal value}.  If $k=0$ the  in~\eqref{cmplRoot}
one has the principal root.  At  $k\ne 0$ the Eq.~\eqref{cmplRoot}  also includes multiple roots
of the complex number.
Problems may arise with the
principal  root value when a double complex  root, i.e. root-in-root
is encountered. Then, one may get either real or imaginary answer
depending on program commands used to extract the root.
Similarly as in \textit{Mathematica},
we will take into account only the principal  square root.


\subsection{\label{eigensystem}Eigensystem and related problems}

Apart from Bott's table, the spectral  root algorithm also
requires to know the diagonal eigenvalue matrix $\A_d$ and respective eigenvector matrix $\hat{\m{T}}$
constructed from  MV rep $\hat{\A}$. They allow
to get all combinations of roots from spectrum (the eigenvalues) and  to
transform scalar roots with different sign combinations  on the diagonal back to   initial  matrix representation,
i.e. to get square root matrix $\hat{\B}$.
After conversion of   $\hat{\B}$ to MV  $\B$ one gets one of the possible MV square roots.
The transformation matrix $\hat{\m{T}}$ that is made up of eigenvectors
can be constructed from matrix  $\hat{\A}$ eigenvectors.
In  \textit{Mathematica}, the eigenvalue list and eigenvector matrix
are generated automatically by command \textbf{Eigensystem[ ]}.
If rep $\hat{\A}$ that represents  MV is real and symmetric then
all  eigenvalues  are real too.
If the matrix is Hermitian then all eigenvalues are real
but the transformation matrix $\hat{\m{T}}$ is complex. For large MVs $\A$, usually both the
diagonal eigenvalue matrix and  $\hat{\m{T}}$ are complex.
As mentioned, the eigenvectors allow to construct the
transformation matrix to bring the diagonal signed  root matrix  back to the initial representation, i.e.
to matrix $\hat{\B}$.  The transformation exists if the matrix $\hat{\A}$  has distinct eigenvalues.
This happens if the  characteristic polynomial of $m\times m$ matrix $\hat{\A}$ has $m$ different roots.

Let's consider a concrete example, when the matrix  $\hat{\m{A}}$ is of  dimension  $4\times 4$.
We assume that $\hat{\m{A}}$ has four eigenvalues $\{a_1,a_2,a_3,a_4\}$ which
may be real, imaginary or complex. First,
construct a diagonal $4\times 4$ matrix $\hat{\A}_d$ with  eigenvalues $a_i$ on the diagonal.
Then, all possible square roots of  $\hat{\A}_d$ are given by all possible combinations of plus/minus signs of roots on the diagonal,
\begin{equation}\label{exampleai}
\hat{\A}_d^{1/2}=
\begin{bmatrix}
\pm\sqrt{a_1}&0&0&0\\
0&\pm\sqrt{a_2}&0&0\\
0&0&\pm\sqrt{a_3}&0\\
0&0&0&\pm\sqrt{a_4}\\
\end{bmatrix},
\end{equation}
where $a_i$ are some  functions of MV coefficients. All in all, there are $16$
combinations of  plus/minus signs in matrix~\eqref{exampleai}
and, respectively, we expect 16 different root matrices. For example, for sign combinations,
$\{+\sqrt{a_1},+\sqrt{a_2},+\sqrt{a_3},+\sqrt{a_4}\}$ and
$\{-\sqrt{a_1},-\sqrt{a_2},-\sqrt{a_3},-\sqrt{a_4}\}$,  we have the
first pair of plus/minus roots. The next two combinations,
$\{-\sqrt{a_1},\sqrt{a_2},\sqrt{a_3},\sqrt{a_4}\}$ and
$\{\sqrt{a_1},-\sqrt{a_2},-\sqrt{a_3},-\sqrt{a_4}\}$, gives the second pair, etc. Thus, in general there are eight plus/minus root
pairs. If one of the eigenvalues is  zero then we shall have
smaller number of roots. For example, if the third root on the diagonal
then the signed second pair is, $\pm\{\sqrt{a_1},\sqrt{a_2},0,\sqrt{a_4}\}$,
$\pm\{-\sqrt{a_1},\sqrt{a_2},0,\sqrt{a_4}\}$.   In case,  all eigenvalues
are zeroes, we assume that spectral  roots are absent.
Thus, in the example we expect 8 eight plus/minus pairs having different
combinations and accordingly a double number of MV roots.

According to Bott's 8-periodicity table~\cite{Lounesto97}, some of CAs are represented by
quaternion matrices, i.e.,  the entries of reps are   quaternions.
The standard  computer programs usually are suitable for work with real and
complex matrices only, hence problems may arise with quaternionic eigens. In
such a case it is convenient  to replace the elementary Hamilton
quaternions $\bi$, $\bj$ and $\bk$,  which satisfy
$\bi^2=\bj^2=\bk^2=-1$, by following $2\times 2$ complex matrices,
respectively,
\begin{equation}\label{quatMatRep}
\hat{\e{}}_1=
\Bigl[\begin{matrix}
\ii&0\\0&\ii
\end{matrix}\Bigr],\quad
\hat{\e{}}_2=
\Bigl[\begin{matrix}
0&1\\-1&0
\end{matrix}\Bigl],\quad
\hat{\e{}}_3=
\Bigl[\begin{matrix}
0&\ii\\ \ii&0
\end{matrix}\Bigl],
\end{equation}
and  perform all computations by  complex and doubled matrix reps.
For example, if MV has been represented by $2\times 2$
quaternionic matrix  then   $t=2$ and
 one gets $n_t=2^2=4$, i.e., one expects not more than 4 different
quaternionic MV roots in order  to be an agreement with
Bott's table. Thus, to be in agreement with quaternionic representation
and get only true roots, the diagonal complex $4\times 4$ matrix
should be restricted, in particular, they  must have not more than four sign
combinations: $\pm\{\sqrt{a_1},\sqrt{a_2},\sqrt{a_3},\sqrt{a_4}\}$,
$\pm\{\sqrt{a_1},\sqrt{a_2},-\sqrt{a_3},-\sqrt{a_4}\}$, which now are
in agreement with quaternionic  sign pairs, $\pm\{Q_1,Q_2\}$ and
$\pm\{Q_1,-Q_2\}$, where $Q_i$ is a general quaternion. Thus, in
this case the $4\times 4$ matrix will give no more than four roots.

Finally, to have the roots in a MV form, we must perform inverse
transformation, $\hat{\m{T}}\hat{\B}\hat{\m{T}}^{-1}$, back  to
primary matrix representation and convert the obtained matrix to
MV form using the formula~\eqref{matrixToMVA}. After
transformation and conversion,  the number of MV square roots
may remain the  same or become  smaller. The latter case may happen
when the transformation matrix  $\hat{\m{T}}$ is  complex
and may bring in complex term(s) into the final  root MV $\B$.

\subsubsection{The square root of  idempotent} The MV is
idempotent if its the square is equal to MV itself~\cite{Lounesto97}.
When $\e{1}^2=1$,  in low dimensional algebras a typical
idempotent is   $\tfrac12(1+\e{1})$. In the first item
of tables in Sec.~\ref{basisTables}, the reader will find the sets of
idempotents of real CAs employed to construct the matrix reps
of basis vectors. The rep of the
idempotent usually is a diagonal matrix  with 1's and 0's on the diagonal, for example,
for  MV in \cl{2}{0}, \cl{1}{1} and \cl{3}{0} algebras one of the idempotents is
$[\begin{smallmatrix}1&0\\0&0\end{smallmatrix}]$. In
\cl{3}{1}, an example  of  idempotent may be
$\tfrac14(1\pm\e{1})(1\pm\e{24})$ that  represents  diagonal
$4\times 4$ matrix with a single $1$ on the diagonal
and the remaining $0$'s. The proposed
diagonalization method fails if it is applied to such type of
matrices. However, there is no need to calculate the square root
of the idempotent since, due to above mentioned property (the square of the
idempotent is equal to initial MV), it should be evident that the
square root of the idempotent must be equal to plus/minus
idempotent itself.

\section{Spectral MV roots in low dimensional CAs}\label{examples1-2D}

\subsection{MV roots in 1D algebras}\label{examples1D}
In 1-dimensional vector space we have two commutative  algebras, \cl{0}{1} and \cl{1}{0}, respectively,
called the complex  and hyperbolic number algebra. According to the
first table in the Appendix (Sec.~\ref{basisTables}) the matrix
representation of \cl{0}{1} is given by complex numbers $\bbC(1)$
and that of \cl{1}{0}  is  given by  diagonal 2-dimensional real
matrices  ${^2\bbR(1)}$.

\subsubsection{\cl{0}{1} algebra, $\bbC(1)$\label{examplesCL01} (complex
numbers)}
In this algebra the MV is $\m{A}=a_0+a_1\e{1}$ where $\e{1}^2=-1$
and $a_0,a_1\in\bbR$. It is isomorphic to complex number, $z=x+\ii
y$,  algebra. Thus, in this case  the  square root of MV coincides with
that of the complex number~\cite{Korn1961},
\begin{equation}\label{rootCL01A}
\m{B}=\pm\sqrt{\m{A}}=(a_0^2+a_1^2)^{1/4}\,\Big(\cos\frac{\varphi+2k\pi}{2}+\e{1}\sin\frac{\varphi+2k\pi}{2}\Big),
\quad \varphi=\arctan(a_0,a_1),
\end{equation}
where $\varphi$ is the argument
expressed through arc tangent that takes into account the
quadrant of the point $(x,y)=(a_0,a_1)$ in complex plane. In
practice, usually root principal value, $k=0$, is used. From
Eq.~\eqref{rootCL01A} follows that  square root of basis vector is
$\sqrt{\pm\e{1}}=\tfrac{1}{\sqrt{2}}(1\pm\e1)$. The limit from
right gives $\lim_{a_0{\to 0^{+}}}\arctan(a_0,a_1)=\pi/2$.
For limit from left, $\lim_{a_0{\to
0^{-}}}\arctan(a_0,a_1)=-\pi/2$, one finds
$\sqrt{-\e{1}}=\tfrac{1}{\sqrt{2}}(1-\e1)$. In case of principal value, $k=0$,
Eq.~\eqref{rootCL01A} can be transformed to $\m{B}=b_0+
\e{1} b_1=\sqrt{\m{A}}$, where
\begin{equation}\label{rootCL01B}
b_0^2=\tfrac{1}{2}\Bigl(a_0+(a_0^2+a_1^2)^{1/2}\Bigr),\quad
b_1^2=\tfrac{1}{2} \Bigl(-a_0+(a_0^2+a_1^2)^{1/2}\Bigr).
\end{equation}
 The latter formula has no singularities.

\begin{figure}[t]
\centering
a)\includegraphics[height=5cm]{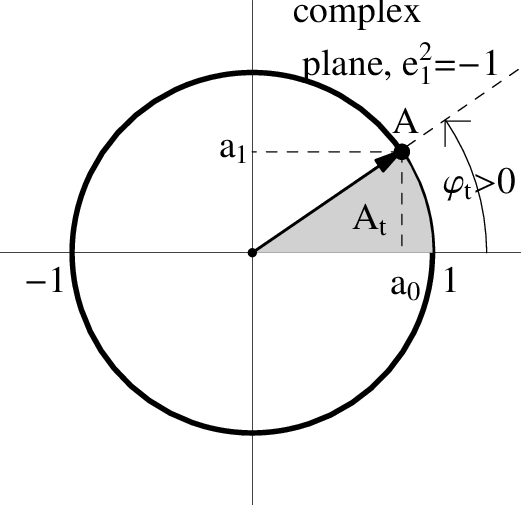}
b)\includegraphics[height=5cm]{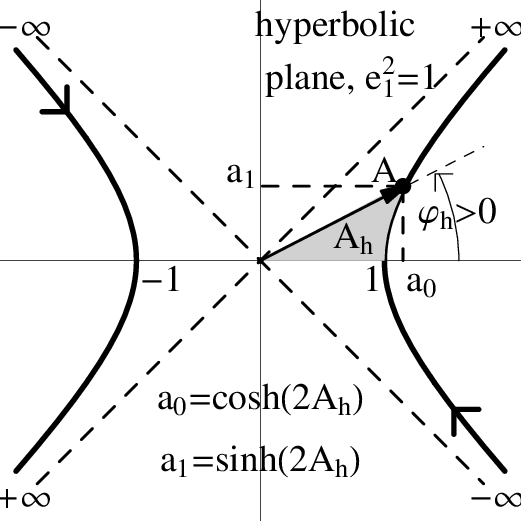}%
\caption{Analogy between unit circle ($x^2+y^2=r^2$, where $r$ is
a circle radius) and rectangular hyperbola ($x^2-y^2=r^2$, where
$r=\text{const}$ is a hyperbolic radius, which is a distance on
the horizontal axis  between coordinate center and intersection of
hyperbola with the horizontal axes). For $r=1$: a) rotation of a
MV $\A$ in a complex plane and b) boost of a MV $\A$ in a
hyperbolic plane. The end of radius-vector $\A=a_0+a_1\e{1}$ moves
along circle or rectangular hyperbola and represents unit MV:
$\magn{\m{A}}=\sqrt{a_0^2+a_1^2}=1$  for circle when $\e{1}^2=-1$,
and $\magn{\m{A}}=\sqrt{a_0^2-a_1^2}=1$ for hyperbola when
$\e{1}^2=1$. The coordinates of $\m{A}$ are $a_0=\cos\varphi_t$and $a_1=\sin\varphi_t$ for circle, and $a_0=\cosh\varphi_h$ and
$a_1=\sinh\varphi_h$ for hyperbola.\label{fig:2}}
\end{figure}

\subsubsection{\cl{1}{0} algebra, ${^2\bbR(1)}$\label{examplesCL1} (hyperbolic numbers)}
In \cl{1}{0} there is a single basis vector $\e{1}$ the square of which  $\e{1}^2=1$.
According to first table in Sec.~\ref{basisTables}, $\e{1}$  can be represented by diagonal  matrix:
\begin{math}[\begin{smallmatrix}1&0\\0&-1\end{smallmatrix}]\end{math}.  The generic  MV $\m{A}=a_0+a_1\e{1}$
is diagonal in matrix representation ${^2\bbR(1)}$,
\begin{equation}
\hat{\m{A}}=\label{a0a1}
  a_0\left[\begin{vmatrix}1&0\\0&1\end{vmatrix}\right]+
 a_1\left[\begin{vmatrix}1&0\\0&-1\end{vmatrix}\right]=
 \left[\begin{vmatrix}a_0+a_1&0\\0&a_0-a_1\end{vmatrix}\right].
\end{equation}
Thus, one can  write the root outright, $\hat{\m{B}}=\sqrt{\hat{\m{A}}}=
\bigl[\begin{smallmatrix}\pm\sqrt{a_0+a_1}&0\\0&\pm\sqrt{a_0-a_1}\end{smallmatrix}\bigr]$,
from which follows that in contrast to complex numbers the hyperbolic
numbers may have up to four sign combinations: $\{++\}$, $\{--\}$, $\{+-\}$,
$\{-+\}$ and, respectively,  four
roots.  Since  $\hat{\m{A}}$ is diagonal the transformation matrix constructed from
eigenvectors is very simple
$\hat{\m{T}}=\hat{\m{T}}^{-1}=\big[\begin{smallmatrix}0&1\\1&0\end{smallmatrix}\big]$.
Transformation of $(++)$  and $(+-)$ root matrices  back to initial representation gives
matrices
\begin{equation}
  \left[\begin{vmatrix}
\sqrt{a_0+a_1} &0\\
 0&\sqrt{a_0-a_1}
\end{vmatrix}\right],\quad
\left[\begin{vmatrix}
-\sqrt{a_0-a_1} &0\\
 0&\sqrt{a_0+a_1}
\end{vmatrix}\right].
\end{equation}
For combinations  $(--)$  and $(-+)$, the signs of the matrices are opposite.
After the conversion to MV form we find two pairs of real roots,
\begin{equation}\label{roothyper}
\B=\sqrt{\m{A}}=\begin{cases}
 \pm\tfrac{1}{2}\bigl((\sqrt{a_0-a_1}+\sqrt{a_0+a_1})-\e1(\sqrt{a_0-a_1}-\sqrt{a_0+a_1}\,)\bigr),\\
 \pm\tfrac{1}{2}\bigl((\sqrt{a_0-a_1}-\sqrt{a_0+a_1})-\e1(\sqrt{a_0-a_1}+\sqrt{a_0+a_1}\,)\bigr).
\end{cases}
\end{equation}
If  $a_0>a_1$ there are four roots.
If $a_0=a_1=1$, the  first terms in the coefficients vanish and  the
Eq.~\eqref{roothyper} reduces to a single pair of roots
$\B=\pm\sqrt{a_0}(1+\e{1})/\sqrt{2}$.  If $a_0<a_1$ the root coefficients are complex
 and such roots must be rejected. Thus, in the latter case the hyperbolic MV has no square roots.

 When $a_0>a_1$ the roots~\eqref{roothyper} can be
expressed by hyperbolic functions, in a form similar to trigonometric ones in
Eq.~\eqref{rootCL01A},
\begin{equation}
\sqrt{\m{A}}=\pm(a_0^2-a_1^2)^{1/4}\lparen\cosh\frac{\varphi_h}{2}+\e{1}\sinh\frac{\varphi_h}{2}\rparen,
\quad\varphi_h=\textrm{Artanh}(a_0,a_1),
\end{equation}
where $\textrm{Artanh}(a_0,a_1)$ is the two-argument area hyperbolic  tangent.
If we draw horizontal  and vertical  axes in plane, then  the
angle $\varphi_h$ will lie between the horizontal axis
and vector that connects coordinate center and point $ (a_0,a_1)$ on the
hyperbola (refer to Fig.~\ref{fig:2}). If $\varphi_h=0$ then
 $a_1=0$, and the MV root becomes  $\sqrt{\m{A}}=\pm\sqrt{a_0}$.
The hyperbolic angle is defined  in a range
$\varphi_h=(-\infty,+\infty)$ and is limited by
asymptotes (dashed lines). If $d_1$ and $d_2$ are distances from a point
on the hyperbola to left and right focus on the horizontal axis then the difference
$|d_1-d_2|$ is independent of $\varphi_h$ and plays the role of hyperbola radius.

\subsection{MV roots in 2D algebras}\label{examples2D}

\subsubsection{\cl{0}{2} algebra, $\bbH(1)$. (Hamilton quaternions)\label{examplesCL02}} Matrix
representation of this algebra is given by quaternions (also,
consult table in Sec.~\ref{basisTables}), i.e., by $1\times 1$
quaternionic matrices.
From this follows that the quaternion has two plus/minus roots
only. To make the problem easier\footnote{Calculations were done by
\textit{Mathematica} package that knows Hamilton's quaternion
algebra. However,  in the package the full quaternion number
calculus is absent as yet.}, we replace the basis
quaternions $\bi$, $\bj$ and $\bk$ by complex matrices,
\begin{equation}
\bi\to\hat{\e{}}_1=
 \bigl[\begin{smallmatrix}
\ii&0\\
0&-\ii
\end{smallmatrix}\bigr],\
\bj\to\hat{\e{}}_2= \bigl[\begin{smallmatrix}
0&1\\
-1&0
\end{smallmatrix}\bigr],\
\bk=\bi\bj\to \hat{\e{}}_{12}=
 \bigl[\begin{smallmatrix}
0&\ii\\
\ii&0
\end{smallmatrix}\bigr],\
\mathbf{ijk}\to \bigl[\begin{smallmatrix}
-1&0\\
0&-1
\end{smallmatrix}\bigr].
\end{equation}
In the matrix rep the quaternion MV,
$\m{A}=a_0+a_1\e{1}+a_2\e{2}+a_{12}\e{12}$, and its determinant
are, respectively,
\begin{equation}
\hat{\m{A}}=\left[\begin{matrix}
 a_0+\ii a_1&a_2+\ii a_{12}\\
 -a_2+\ii a_{12}& a_0-\ii a_1\\
\end{matrix}\right],\quad \det\hat{\m{A}}=a_0^2+a_1^2+a_2^2+a_{12}^2.
\end{equation}
The matrix $\hat{\m{A}}$ has two eigenvalues
$\epsilon_{1,2}=a_0\pm m$, where
$m=(-a_1^2-a_2^2-a_{12}^2)^{1/2}$. Denoting by $\hat{\m{T}}$ the
transformation matrix constructed from eigenvectors of
$\hat{\m{A}}$, one can get the square root from a diagonal
eigenvalue matrix:
\begin{equation}\label{sqrtQuat}
\sqrt{\hat{\m{A}}}=\hat{\m{T}}
  \left[\begin{matrix}\pm\sqrt{a_0-m}&0\\0&\pm\sqrt{a_0+m}\end{matrix}\right]
\hat{\m{T}}^{-1},\quad \hat{\m{T}}=\left[\begin{matrix}\frac{a_1+\ii m}{a_{12}+\ii a_2} &
\frac{a_1-\ii m}{a_{12}+\ii a_2}  \\ 1& 1 \\
\end{matrix}\right],
\end{equation}
where either $\{++\}$ or $\{--\}$ signs must be used only as
pointed out in Subsec.~\ref{eigensystem}. After
conversion of $\sqrt{\hat{\m{A}}}$ back to MV one finds only a
single pair of quaternionic roots:
\begin{equation}\label{rootCL02A}
(\sqrt{\m{A}}\,)_{1,2}=\pm\tfrac{1}{2}\Bigl(\sqrt{a_0+m}+\sqrt{a_0-m}+\tfrac{\gv{\m{A}}+\gb{\m{A}}}{\sqrt{m^2}}
\bigl(\sqrt{a_0+m}-\sqrt{a_0-m}\bigr)\Bigr).
\end{equation}
Since $m$  is imaginary, $m\in\bbI$, the sum and difference of
conjugate complex roots in Eq.~\eqref{rootCL02A} ensure that all
square root coefficients are real numbers as Fig.~\ref{fig:1}
demonstrates. Note, the vector $\gv{\m{A}}$ brings in quaternionic
imaginaries $\bi\equiv\e{1}$ and $\bj\equiv\e{2}$ and the bivector
$\gb{\m{A}}$ brings in the product $\bi\bj=\bk\equiv\e{12}$. The
compensation of imaginary unit in the difference of conjugate
terms comes from $\sqrt{m^2}$. In the limit $m\to 0$, one has the
scalar root $\pm\sqrt{a_0}$.

If angle $\theta=\textrm{arctan}(a_0,|m|)$ is introduced, where
$|m|=\sqrt{a_1^2+a_2^2+a_{12}^2}$,  the roots can be rewritten in
a trigonometric form,
\begin{equation}\label{rootCL02B}
\big(\sqrt{\m{A}}\big)_{1,2}=\pm\bigl(a_0^2+|m|^2\bigr)^{1/4}
\Bigr(\cos\frac{\theta}{2}+\frac{\gv{\m{A}}+\gb{\m{A}}}{|m|}\sin\frac{\theta}{2}\Bigr).
\end{equation}
Thus, we have found that the quaternionic numbers  have two (not
four) square roots. As mentioned, this happens because the complex
eigenvalue root matrix $\sqrt{\hat{\m{A}}}$ in
Eq.~\eqref{sqrtQuat}, in fact, embodies a single indivisible
object, namely, $1\times 1$ quaternionic matrix, as a result the
mixed sign combinations $\{+-\}$ or $\{-+\}$ are forbidden. In
contrast to Eq.~\eqref{rootCL02B}, in the first form,
Eq.~\eqref{rootCL02A}, at $a_0=0$ a singularity appears that may
be  eliminated by inserting into the former the angle value
$\theta=\pi/2$ what gives
\begin{equation}
\big(\sqrt{\m{A}}\big)_{1,2}=\pm\big(|m|+\gv{\m{A}}+\gb{\m{A}}\big)/(2|m|)^{1/2}\,.
\end{equation}

\subsubsection{\cl{1}{1} algebra, $\bbR(2)$} This algebra may have
either four, or two roots, or no roots at all. The irreducible
matrix reps of \cl{1}{1} belong to real $2\times 2$ real matrix
$\bbR(2)$. According to  table in
Sec.~\ref{basisTables} the basis vectors  are represented by
matrices \begin{math}\hat{\e{}}_1=[\begin{smallmatrix}1&0\\0&-1
\end{smallmatrix}]\end{math}, $\hat{\e{}}_2=
[\begin{smallmatrix}0&-1\\1&0
\end{smallmatrix}]$, then $\hat{\e{}}_{12}=
[\begin{smallmatrix}0&-1\\-1&0
\end{smallmatrix}]$. The matrix representation of MV then~is
\begin{equation}
\hat{\m{A}}= \begin{vmatrix}
 a_0+a_1&-a_2-a_{12}\\
 a_2-a_{12}& a_0-a_1
\end{vmatrix},\quad \det\hat{\m{A}}=a_0^2-a_1^2+a_2^2-a_{12}^2.
\end{equation}
The calculation proceeds exactly in the same way as for
\cl{0}{2}, except that now in addition one must include mixed
signs, $\{+-\}$ and $\{-+\}$, in the eigenvalue related matrices.
 If quantity $m=\sqrt{a_1^2-a_2^2+a_{12}^2}\,$, $a_1^2+a_{12}^2>a_2^2$,
 is introduced then
four square roots are found for generic MV, the squares of which give the initial MV
$\m{A}=a_0+a_1\e{1}+a_2\e{2}+a_{12}\e{12}$,
\begin{equation}\begin{split}
&(\sqrt{\m{A}}\,)_{1,2}=\pm\tfrac{1}{2}\Bigl(\bigl(\sqrt{a_0+m}+\sqrt{a_0-m}\bigr)+
\frac{\gv{\m{A}}+\gb{\m{A}}}{m}\bigl(\sqrt{a_0+m}-\sqrt{a_0-m}\bigr)\Bigr), \\
&(\sqrt{\m{A}}\,)_{3,4}=\pm\tfrac{1}{2}\Bigl(\bigl(-\sqrt{a_0+m}+\sqrt{a_0-m}\bigr)-
\frac{\gv{\m{A}}+\gb{\m{A}}}{m}\bigl(\sqrt{a_0+m}+\sqrt{a_0-m}\bigr)\Bigr). \\
\end{split}\end{equation}
The number of roots is regulated by quantity~$m$: if
$a_0>m$  there are four roots, if $a_0=m$ the are two roots and if
$a_0<m$ the spectral roots do not exist.

The structure of the square roots shows that  sums and  differences in the root coefficients  may be complex conjugate numbers.
In such cases the roots are real. For example, for MV $\m{A}=2 \e{2}-1$ one has $m=\ii 2$, as a result the first two roots reduce  to
\[
(\sqrt{\m{A}})_{1,2}=\pm \Bigg( \sqrt{\frac{2}{\sqrt{5}-1}} \e{2}+\sqrt{\frac{1}{2} \Big(\sqrt{5}-1\Big)}\Bigg).
\] The remaining two roots give complex coefficients and thus must be rejected.

\subsubsection{\cl{2}{0} algebra, $\bbR(2)$}The irreducible MV representation of this
algebra also belongs to $\bbR(2)$, and there may be up to four
roots. According to  table in Sec.~\ref{basisTables} the basis
matrices are $\hat{\e{}}_1= [\begin{smallmatrix}1&0\\0&-1
\end{smallmatrix}]$ and $\hat{\e{}}_2=
[\begin{smallmatrix}0&1\\1&0
\end{smallmatrix}]$, the product of which yields bivector matrix,
$\hat{\e{}}_{12}=[\begin{smallmatrix}0&1\\-1&0
\end{smallmatrix}]$. Similar  calculations as for \cl{1}{1} give the same
root formulas, except that now $a_1^2+a_2^2>a_{12}^2$
and $m$  should be replaced by
 \begin{math}m=\sqrt{a_1^2+a_2^2-a_{12}^2}\,\end{math}, what results in a different
domain  for existence of either four or two MV roots having real
coefficients. The four roots exist if
 $a_0>m$.  If $a_0=m$ then pairs of roots are degenerate, i.e.,
 we have two different roots only. If  $a_0<m$,  there are no roots.

\section{Spectral MV roots in 3D algebras}\label{examples3D} The generic
MV in 3-dimensional algebras has the following form,
\begin{equation}\label{mv3D}
\m{A}=a_0+a_1\e{1}+a_2\e{2}+a_3\e{3}+a_{12}\e{12}+a_{13}\e{13}+a_{23}\e{23}+a_{123}\e{123}.
\end{equation}
We shall remind that in the present paper, instead of more natural
index ordering $\e{31}$, respectively  for larger grade basis elements in
higher dimensional algebras, we use $\e{13}$ instead,
\footnote{\label{noteA}The calculations were
performed by \textit{Mathematica} symbolic algebra package that
uses the  inverse degree lexicographic ordering. }

\subsection{\cl{3}{0} algebra, $\bbC(2)$}
The basis vectors (generators) of \cl{3}{0} are represented by
Pauli matrices:
 $\hat{\e{}}_1=[\begin{smallmatrix}1&0\\0&-1
\end{smallmatrix}]$,
 $\hat{\e{}}_2=
[\begin{smallmatrix}0&1\\1&0
\end{smallmatrix}]$,
 $\hat{\e{}}_3=
[\begin{smallmatrix}0&-\ii\\\ii&0
\end{smallmatrix}]$.
Matrix rep of a general MV, Eq.~\eqref{mv3D},~is
\begin{equation} \hat{\m{A}}=
  \left[\begin{vmatrix}
(a_0+a_1)+\ii(a_{23}+a_{123}) & (a_2+a_{12})-\ii(a_3+a_{13})\\
 (a_2-a_{12})-\ii(a_3-a_{13}) &(a_0-a_1)-\ii(a_{23}-a_{123})\\
  \end{vmatrix} \right].
\end{equation}
The  trace is $\textrm{Tr}\hat{\m{A}}=2(a_0+\ii a_{123})$ and the
determinant is
\begin{equation}\begin{split}
\textrm{Det}\hat{\m{A}}&=(a_0+\ii a_{123})^2-(a_1+\ii
a_{23})^2-(a_2-\ii a_{13})^2-(a_3+\ii a_{12})^2\\
&=(a_0+\ii a_{123})^2+(a_{23}-\ii a_{1})^2+(a_{13}+\ii
a_{2})^2+(a_{12}-\ii a_{3})^2.
\end{split}\end{equation}
 The matrix $\hat{\m{A}}$ has two eigenvalues, $\varepsilon_{1}$ and  $\varepsilon_{2}$, and
the following transformation matrix $\hat{\m{T}}$  built from
eigenvectors:
\begin{equation}\begin{split}\label{eigens}
&\varepsilon_{1,2}=a_0+\ii a_{123}\mp\sqrt{\chi},\\
&\chi=(a_1+\ii
a_{23})^2-(a_{13}+\ii a_2)^2-(a_{12}-\ii a_3)^2,\\
&\widetilde{\hat{\m{T}}}=\left[\begin{matrix}
 \frac{-a_1-\ii a_{23}+\sqrt{\chi}}{a_2+a_{12}-\ii{a_3-a_{13}}}&1\\
 \ii\Bigl(\frac{a_1+\ii a_{23}+\sqrt{\chi}}{a_2+a_{12}-\ii{a_3-a_{13}}}\Bigr) &1
\end{matrix}\right],
\end{split}\end{equation}
where the tilde indicates matrix transposition. After extraction
of roots and selection of root signs in diagonal root matrix, and
then after transformation of the root matrix back to initial
representation we get
\begin{equation}\label{transformCL30}
\sqrt{\hat{\m{A}}}=\hat{\m{B}}=\hat{\m{T}}
\left[\begin{matrix}
\pm\sqrt{\epsilon_1} &0\\
0 & \pm\sqrt{\epsilon_2} \\
\end{matrix}\right]
\hat{\m{T}}^{-1}.
\end{equation}
Finally, after conversion back to MV form, we obtain
expressions for individual roots, $\m{B}_i=(\sqrt{\m{A}})_i$.  A
characteristic property of the multivectors $\m{B}_i$ is that its coefficients
consist of sums of complex conjugate pairs, similarly as shown
in Fig.~\ref{fig:1}. This warrants that all coefficients
are real numbers.

It is convenient to introduce complex quantities that are related with
respective algebra spectrum. Thus, in case of \cl{3}{0} the following
\textit{einsatz} is used,
\begin{equation}\label{coeffCL30}\begin{split}
&\alpha\pm\ii\beta=\sqrt{a_0\pm\ii a_{123}-\sqrt{\ba^2+ \cA^2\mp\ii 2\ba\wedge \cA \e{123}}}
=\sqrt{a_0\pm\ii a_{123}-\Delta_{\pm}}\,, \\
&\gamma\pm\ii\delta=\sqrt{a_0\pm\ii a_{123}+\sqrt{\ba^2+
\cA^2\mp\ii 2(\ba\wedge \cA) \e{123}}}=\sqrt{ a_0\pm\ii a_{123}+\Delta_{\pm}}\,,\\
&\varepsilon\pm\ii\varphi \equiv\Delta_{\pm}=\sqrt{\ba^2+
\cA^2\mp\ii 2\ba\wedge \cA \e{123}}=\\
 &\qquad\qquad\sqrt{(a_1\pm\ii
a_{23})^2+(a_{2}\mp\ii a_{13})^2+(a_{3}\pm\ii a_{12})^2},
\end{split}\end{equation}
where under double roots the parts of the spectrum $\eqref{eigens}$  can be discerned.
Also we have introduced the grades: $a_0=\gs{\m{A}}$,
$\ba=\gv{\m{A}}=a_1\e{1}+a_2\e{2}+a_3\e{3}$,
$\cA=\gb{\m{A}}=a_{12}\e{12}+a_{13}\e{13}+a_{23}\e{23}$ and
$\gt{\m{A}}=a_{123}\e{123}$. The letters in Greek
 $\{\alpha,\beta,\gamma,\delta,\varepsilon,\varphi\}$ represent either real or purely imaginary quantities,
thus,  conjugate pairs~in\eqref{coeffCL30} may be converted, for example, to
real coefficients, or to trigonometric functions as  illustrated by
Eq.~\eqref{cmplConj} and Fig.~\ref{fig:1}. From~\eqref{coeffCL30},
one can  explicitly write the coefficients in terms of the grades:
\begin{equation}\begin{split}\label{coeffCL30B}
&\alpha=\tfrac12\bigl(\sqrt{a_0+\ii a_{123}-\Delta_{+}}+\sqrt{a_0-\ii a_{123}-\Delta_{-}}\bigr),\\
&\beta=-\ii\tfrac12\bigl(\sqrt{a_0+\ii a_{123}-\Delta_{+}}-\sqrt{a_0-\ii a_{123}-\Delta_{-}}\bigr),\\
&\gamma=\tfrac12\bigl(\sqrt{a_0+\ii a_{123}+\Delta_{+}}+\sqrt{a_0-\ii a_{123}+\Delta_{-}}\bigr),\\
&\delta=-\ii\tfrac12\bigl(\sqrt{a_0+\ii a_{123}+\Delta_{+}}-\sqrt{a_0-\ii a_{123}+\Delta_{-}}\bigr),\\
&\varepsilon=\tfrac12(\Delta_{-}+\Delta_{+}), \qquad
\varphi=-\ii\tfrac12(\Delta_{-}-\Delta_{+}).
\end{split}\end{equation}
From the last two formula follows that expression
$2(\varepsilon^2+\varphi^2)$, which appears in denominators of
the coefficients below, Eqs~\eqref{CL30B12} and \eqref{CL30B34},
 is real and equal to $4(\ba^2+\cB^2)$.

In \cl{3}{0}, in general, there are four square roots, or two pairs of plus/minus roots,
\begin{equation}
(\sqrt{\m{A}})_{i,j}=\m{B}_{i,j}=\pm(b_0+b_1\e{1}+b_2\e{2}+b_3\e{3}+b_{12}\e{12}+
b_{13}\e{13}+b_{23}\e{23}+b_{123}\e{123}).
\end{equation}
The coefficients of the first pair of plus/minus roots in
$\m{B}_{1,2}$  and  the second plus/minus roots in $\m{B}_{3,4}$
may be expressed in terms of $\alpha,\beta,\dotsm$ as written down
below in a shorthand form, where real vector and bivector
coefficients, $b_1,b_2,b_2$ and $b_{12},b_{13},b_{23}$, are given
in a condensed form, $b_{1,2,3}$ and  $b_{12,13,23}$.

\begin{equation}\label{CL30B12}\begin{split}
&\indent\cl{3}{0}. \textit{The first pair of roots\ }\m{B}=\sqrt{\m{A}}:\\
&b_0=\tfrac12(\alpha+\gamma),\quad b_{123}=\tfrac12(\beta+\delta),\\
 &b_{1,2,3}=\frac{(-1)^{1,2,1}a_{23,13,12}\bigl((\delta-\beta)\varepsilon+
(\alpha-\gamma)\varphi\bigr)+a_{1,2,3}\bigl((\gamma-\alpha)\varepsilon+(\delta-\beta)\varphi\bigr)}
{2(\varepsilon^2+\varphi^2)},\\
&b_{12,13,23}=\frac{(-1)^{1,2,1}\
a_{3,2,1}\bigl((\beta-\delta)\varepsilon+
(\gamma-\alpha)\varphi\bigr)+a_{12,13,23}\bigl((\gamma-\alpha)\varepsilon+(\delta-\beta)\varphi\bigr)}
{2(\varepsilon^2+\varphi^2)}.\\
\end{split}\end{equation}
\begin{equation}\label{CL30B34}\begin{split}
&\indent\cl{3}{0}. \textit{The second pair of roots\ }\m{B}=\sqrt{\m{A}}:\\
& b_0=\tfrac12(\alpha-\gamma),\quad b_{123}=\tfrac12(\beta-\delta),\\
&b_{1,2,3}=\frac{-(-1)^{1,2,3}a_{23,13,12}\bigl((\beta+\delta)\varepsilon-
(\alpha+\gamma)\varphi\bigr)-a_{1,2,3}\bigl((\alpha+\gamma)\varepsilon+(\beta+\delta)\varphi\bigr)}
{2(\varepsilon^2+\varphi^2)},\\
&b_{12,13,23}=\frac{(-1)^{1,2,3}a_{3,2,1}\bigl((\beta+\delta)\varepsilon-
(\alpha+\gamma)\varphi\bigr)-a_{12,13,23}\bigl((\alpha+\gamma)\varepsilon+(\beta+\delta)\varphi\bigr)}
{2(\varepsilon^2+\varphi^2)},\\
\end{split}\end{equation}
Three lower and upper indices correspond respectively to three different
formulas for vector and bivector coefficients, thus,
the numbers in  upper and  lower  indices should be matched up in a correct sequence.
For even MVs, when
$a_1=a_2=a_3=a_{123}=0$, formulas in Eqs.~\eqref{CL30B12} and
\eqref{CL30B34} reduce to the well-known  Euclidean space exponential  rotors.
As mentioned, the denominators may be replaced by vector and bivector,
$2(\varepsilon^2+\varphi^2)=4(\ba^2+\cA^2)$.

The coefficients in \eqref{CL30B12} and \eqref{CL30B34} become
singular when the denominators tend towards zero,
$(\varepsilon^2+\varphi^2)\to 0$, or equivalently when
$\bigl((a_1+\ii a_{23})^2+(a_{2}-\ii a_{13})^2+(a_{3}+\ii
a_{12})^2\bigr)^{1/2}\to 0$. This  limit yields MV
$\m{A}=a_0+a_{123}e_{123}$, the roots of which can be found by the
same algorithm. Indeed, corresponding MV matrix now is diagonal,
$\hat{\m{A}}=\bigl[\begin{smallmatrix}a_0+\ii a_{123}&0\\0&a_0+\ii
a_{123}\end{smallmatrix}\bigr]$. The pairs of square roots that
correspond to either opposite, $\{++\}$ and $\{--\}$, or mixed,
$\{+-\}$ and $\{-+\}$, signs  before root symbols in diagonal
matrices then give roots that may be reduced to Euler exponential form,
\begin{equation}\begin{split}
(\sqrt{\m{A}})_{1,2}&=\pm\tfrac12\Bigl((\sqrt{a_0-\ii
a_{123}}+\sqrt{a_0+\ii a_{123}})\e{2}-
\ii(\sqrt{a_0-\ii a_{123}}-\sqrt{a_0+\ii a_{123}})\e{13}\Bigr)\\
 &=\pm(a_0^2+a_{123}^2)^{1/4}(\e{2}\cos\frac{\theta}{2}-\e{13}\sin\frac{\theta}{2})=
\pm\e{2}(a_0^2+a_{123}^2)^{1/4}e^{\e{123}\theta/2}, \\
(\sqrt{\m{A}})_{3,4}&=\pm\ii\tfrac{1}{2}\Bigl((\sqrt{a_0-\ii
a_{123}}-\sqrt{a_0+\ii a_{123}})\e{3}+
\ii(\sqrt{a_0-\ii a_{123}}+\sqrt{a_0+\ii a_{123}})\e{12}\Bigr) \\
 &=\pm(a_0^2+a_{123}^2)^{1/4}(\e{3}\sin\frac{\theta}{2}-\e{12}\cos\frac{\theta}{2})=
\mp\e{12}(a_0^2+a_{123}^2)^{1/4}e^{\e{123}\theta/2},\\
\end{split}\end{equation}
where $\theta=\arctan(a_{123}/a_0)$.  If $a_0\to 0_{+}$ then
$\theta\to \pi/2$, and if $a_0\to 0_{-}$ then $\theta\to -\pi/2$.
In the limit $\theta=\pm\pi/2$, the pairs of square roots  merge
and instead of four one gets two different roots:
\begin{equation}
(\sqrt{\m{A}})_{1,2}=\sqrt{a_{123}}(\e{2}-\e{13})/\sqrt{2}\,,\quad
(\sqrt{\m{A}})_{3,4}=-\sqrt{a_{123}}(\e{3}+\e{12})/\sqrt{2}\,.
\end{equation}
Thus, the both expressions in the limit $\theta=\pm\pi/2$
determine the square root of pseudoscalar in \cl{3}{0}.
Nonetheless, in generic Eqs.~\eqref{CL30B12} and \eqref{CL30B34}
one may construct  combinations of vector and bivector
coefficients that give removable singularity as illustrated in
\textit{Example~2} below. As a check, we have also calculated the
square roots using \textit{real} 4x4 matrix representation of MV in
\cl{3}{0} under restriction on the sign combinations on respective diagonal
pairs in the root entries.

\vspace{2mm} \indent\textit{Example~1. Numerical MV}. In
\cl{3}{0}, the MV $\m{A}=-1+\e{3}-\e{12}+\tfrac{1}{2}\e{123}$ in
matrix form is
$\hat{\m{A}}=\Big[\begin{smallmatrix}-1+\ii/2&-1-\ii\\1+\ii&-1+\ii/2\end{smallmatrix}\Big]$.
The eigenvalues of the latter are complex numbers,
$\{-2+2\ii/3,-\ii/2\}$, and the transformation matrix
$\hat{\m{T}}$ constructed from eigenvectors is
$\Big[\begin{smallmatrix}\ii&-\ii\\1&1\end{smallmatrix}\Big]$.
Transformation back by~\eqref{transformCL30} and conversion of
resulting matrix to MV gives four roots:
$\m{B}_{1,2}=\pm\tfrac{1}{2}(\e{3}+\e{12}-2\e{123})$ and
$\m{B}_{3,4}=\pm\tfrac{1}{2}(-1+2\e{12}-2\e{123})$, the squares of
which give initial MV~$\m{A}$. The same  roots were obtained
from coefficients in Eq.~\eqref{CL30B12} and Eq.~\eqref{CL30B34}
after insertion of MV numerical coefficients. If general form of
MV is used, $\m{A}=a_0+a_3\e{3}+a_{12}\e{12}+a_{123}\e{123}$, then
Eq.~\eqref{CL30B12} and Eq.~\eqref{CL30B34} may be employed to
find the domain of existence of roots: All nonzero root
coefficients must be real. Analysis shows that in the considered
example for all input coefficients $a_0,a_3,a_{12},a_{123}$ the
roots exist, because all root coefficients were found to consist
of pairs of complex conjugate terms (see  Fig.~\ref{fig:1}).

\vspace{2mm}\textit{Example~2}. The MV $\m{A}=\e{1}+\e{12}$ has no
roots, because required transformation matrix,
$\hat{\m{T}}=\big[\begin{smallmatrix}1&-1\\1&-1\end{smallmatrix}\big]$,
is singular, i.e. its determinant $\textrm{Det}\hat{\m{T}}=0$ and
trace $\textrm{Tr}\hat{\m{T}}=0$, and thus matrix
$\hat{\m{T}}^{-1}$ that is needed in the transformation back to
initial representation does not exist. This is in agreement with
the Sullivan's formula~\cite{Sullivan1993} in  Eq.~\eqref{Sullivan}.

Finally, the example below illustrates that spectral  method
may be adapted  to find the roots when the determinant
of the transformation  matrix is zero and how to calculate
functions of  MV argument.

\textit{Example 3.}   In \cl{3}{0},
the transformation matrix $\hat{\m{T}}$ of the MV presented below
is singular (its determinant is zero).
\[\begin{split}
  &\m{A}=-1+2\e{1}+\e{2}+2\e{3}-2\e{12}-2\e{13}+\e{23}-\e{123},\\
  &\hat{\m{A}}=\left[\begin{vmatrix}
1&-1\\
3+\ii 4& -3-\ii 2\\
\end{vmatrix}\right],\qquad\hat{\m{T}}=
\left[\begin{vmatrix}
2-\ii&0\\
5& 0
\end{vmatrix}\right].
\end{split}\]
The eigenvalues of matrix  $\hat{\m{A}}$  are degenerate, $\varepsilon_{1,2}=-(1+\ii)$. However,
the main difficulty is that the transformation matrix $\hat{\m{T}}$
constructed from eigenvectors has zero determinant and therefore cannot be used for
transformations between MV and matrix reps.
The algorithm can be fixed if infinitesimal quantity $\varepsilon$ is added to one of  basis elements.
For example, after replacement  $2\e{2}\to(2+\varepsilon)\e{2}$
the  eigenvalues and  transformation matrix become
  non-degenerate  and  non-singular,
 \[
 [\begin{smallmatrix}\varepsilon_1&0\\0&\varepsilon_2\end{smallmatrix}]
=\Big[\begin{smallmatrix}-1-\ii-\sqrt{\varepsilon(4-4\ii+\varepsilon)}&0\\
0&-1-\ii+\sqrt{\varepsilon(4-4\ii+\varepsilon)}\end{smallmatrix}\Big]
,\
 \hat{\m{T}}=
 \Big[  \begin{smallmatrix}
 \ii\frac{(-2-\ii)+\sqrt{\varepsilon(4-4\ii+\varepsilon)}}{(4-3\ii)+\varepsilon} & -\ii\frac{(2+\ii)+\sqrt{\varepsilon(4-4\ii+\varepsilon)}}{(4-3\ii)+\varepsilon}\\
1 & 1\\
 \end{smallmatrix}\Big].
\]

The MV square root is a multivalued function. However, the  matrix diagonalization method can be extended to simpler cases,
for example, to calculate functions of MV.
Let's compute the exponential of MV argument by spectral method.
To get MV  exponential function $\exp{\m{A}}$,  at first we have
to calculate the following back transformation to initial representation,
 \[
 \hat{\m{T}} [\begin{smallmatrix}\exp{(\varepsilon_1)}&0\\0&\exp{(\varepsilon_2)}
 \end{smallmatrix}]\hat{\m{T}}^{-1}\]
 and then take the limit $\varepsilon\to 0$. After
  conversion back to MV form we get finally,
 \[\begin{split}
 \textrm{e}^{\m{A}}&=\tfrac12\big(\ee^{-1-\ii}+\ee^{-1+\ii}\big)-
  \tfrac{\ii}{2}\big(\ee^{-1-\ii}-\ee^{-1+\ii}\big)\e{123}+\\
 &\tfrac12\big((2+\ii)\ee^{-1-\ii}+(2-\ii)\ee^{-1+\ii}\big)\e{1}+
   \tfrac12\big((1+2\ii)\ee^{-1-\ii}+(1-2\ii)\ee^{-1+\ii}\big)\e{2}+ \\
  & \big((1-\ii)\ee^{-1-\ii}+(1+\ii)\ee^{-1+\ii}\big)\e{3}+
     \big((-1-\ii)\ee^{-1-\ii}-(1-\ii)\ee^{-1+\ii}\big)\e{12}+\\
  &\tfrac12\big((-2+\ii)\ee^{-1-\ii}-(2+\ii)\ee^{-1+\ii}\big)\e{13}+
   \tfrac12\big((1-2\ii)\ee^{-1-\ii}+(1+2\ii)\ee^{-1+\ii}\big)\e{23}.
\end{split} \]
  The result  can be rewritten in trigonometric functions,
\[\begin{split}
\textrm{e}^{\m{A}}&=\textrm{e}^{-1}\big(\cos 1+(2\cos 1+\sin 1)\e{1}+(\cos 1+2\sin 1)\e{2}
+(2\cos 1-\sin 1)\e{3}-\\
&2(\cos 1+\sin 1)\e{12} -(2\cos 1-\sin 1)\e{13} +(\cos 1-2\sin 1)\e{23} -\sin 1\,\e{123}\big).
\end{split}\]
The  exponential can be obtained  by  different (related to MV spectrum) method~\cite{Acus2025a}, where basis-free
representation and roots of minimal polynomial are used.

\subsection{\cl{1}{2} algebra, $\bbC(2)$}
In this algebra general MV has four roots. Since  algebra generators  are
the complex  matrices,
 $\hat{\e{}}_1=\bigl[\begin{smallmatrix}1&0\\0&-1
\end{smallmatrix}\bigr]$,
 $\hat{\e{}}_2=
\bigl[\begin{smallmatrix}0&-1\\1&0
\end{smallmatrix}\bigr]$,
 $\hat{\e{}}_3=
\bigl[\begin{smallmatrix}0&-\ii\\-\ii&0
\end{smallmatrix}\bigr]$,
the procedure to find square roots and properties are similar to
those in \cl{3}{0}, therefore we present only final results.
In \cl{1}{2}, the matrix rep of a generic MV $\m{A}$ is
\begin{equation} \hat{\m{A}}=
\begin{bmatrix}
(a_0+a_1)+\ii(a_{23}+a_{123}) & -(a_2+a_{12})-\ii(a_3+a_{13})\\
 (a_2-a_{12})-\ii(a_3-a_{13}) &(a_0-a_1)-\ii(a_{23}-a_{123})\\
\end{bmatrix},
\end{equation}
and the determinant is
\begin{equation}
\textrm{Det}\hat{\m{A}}=(a_0+\ii a_{123})^2-(a_1+\ii
a_{23})^2+(a_2+\ii a_{13})^2+(a_3-\ii a_{12})^2.
\end{equation}
Now, instead of Eqs~\eqref{coeffCL30} the terms with different signs
appear in double roots for coefficients,
\begin{equation}\begin{split}\label{coeffCL21}
&\alpha\pm\ii\beta=\sqrt{a_0\mp\ii a_{123}-\sqrt{(a_{12}\mp\ii a_{3})^2+(a_{13}\pm\ii a_{2})^2-(a_{23}\pm\ii a_{1})^2}}\,, \\
&\gamma\pm\ii\delta=\sqrt{a_0\pm\ii a_{123}+\sqrt{(a_{12}\pm\ii a_{3})^2+(a_{13}\mp\ii a_{2})^2-(a_{23}\mp\ii a_{1})^2}}\,, \\
&\epsilon\pm\ii\varphi=\sqrt{\ba^2+B^2\mp(\ba\wedge
B)\e{123}}=\sqrt{(a_{12}\pm\ii a_{3})^2+(a_{13}\mp\ii
a_{2})^2-(a_{23}\mp\ii a_{1})^2}\,.
\end{split}\end{equation}
where $\{\alpha,\beta,\gamma,\delta,\epsilon,\varphi\}\in\textrm{either }\bbR$ or
$\bbI$. $B$ is the bivector. The square root vector  $\{b_1,b_2,b_2\}$ and bivector
$\{b_{12},b_{13},b_{23}\}$ coefficients in
$\sqrt{\m{A}}=\m{B}=b_0+b_1\e{1}+b_2\e{2}+b_3\e{3}+b_{12}\e{12}+b_{13}\e{13}+b_{23}\e{23}+b_{123}\e{123}$
are (the notation is similar to that explained for \cl{3}{0}).
\begin{equation}\label{CL12coefA}\begin{split}
&\indent\cl{1}{2}. \textit{The first pair of roots\ }\m{B}=\pm\sqrt{\m{A}}\,.\\
& b_0=\tfrac12(\alpha+\gamma),\quad b_{123}=\tfrac12(-\beta+\delta),\\
 &b_{1,2,3}=\frac{(-1)^{1,1,2}a_{23,13,12}\bigl((\beta+\delta)\varepsilon+
(\alpha-\gamma)\varphi\bigr)+a_{1,2,3}\bigl((\gamma-\alpha)\varepsilon+(\beta+\delta)\varphi\bigr)}
{2(\varepsilon^2+\varphi^2)},\\
&b_{12,13,23}=\frac{(-1)^{1,2,2}\
a_{3,2,1}\bigl((\beta+\delta)\varepsilon+
(\alpha-\gamma)\varphi\bigr)+a_{12,13,23}\bigl((\gamma-\alpha)\varepsilon+(\beta+\delta)\varphi\bigr)}
{2(\varepsilon^2+\varphi^2)}.\\
\end{split}\end{equation}
\begin{equation}\label{CL12coefB}\begin{split}
&\indent\cl{1}{2}. \textit{The second pair of roots\ }\m{B}=\pm\sqrt{\m{A}}\,.\\
&b_0=\tfrac12(\alpha-\gamma),\quad b_{123}=-\tfrac12(\beta+\delta),\\
&b_{1,2,3}=\frac{(-1)^{1,1,2}a_{23,13,12}\bigl((\beta-\delta)\varepsilon+
(\alpha+\gamma)\varphi\bigr)+(-1)^{2,2,1}a_{1,2,3}\bigl((\alpha+\gamma)\varepsilon+(-\beta+\delta)\varphi\bigr)}
{2(\varepsilon^2+\varphi^2)},\\
&b_{12,13,23}=\frac{(-1)^{1,2,2}a_{3,2,1}\bigl((\beta-\delta)\varepsilon+
(\alpha+\gamma)\varphi\bigr)+(-1)^{2,1,1}a_{12,13,23}\bigl((\alpha+\gamma)\varepsilon+(-\beta+\delta)\varphi\bigr)}
{2(\varepsilon^2+\varphi^2)}.\\
\end{split}\end{equation}
Concrete expressions  for
$\{\alpha,\beta,\gamma,\delta,\varepsilon,\varphi\}$ that appear
in Eqs~\eqref{CL12coefA} and \eqref{CL12coefB} can be deduced from
formulae in~\eqref{coeffCL21} using summations and substraction.

It may be useful  to compare  the  roots in $2\times 2$ matrix form
with D.~Sullivan's simpler $2\times 2$ matrix root formula which
should be valid for the both algebras,  \cl{3}{0}  and \cl{1}{2}.  D.~Sullivan's matrix root
formula is~\cite{Sullivan1993}
\begin{equation}\label{Sullivan}
\sqrt{\hat{\m{A}}}=\epsilon_2\frac{\hat{\m{A}}+\epsilon_1\sqrt{\textrm{Det}\hat{\m{A}}}\,\hat{1}}
{\sqrt{\textrm{Tr}\hat{\m{A}}+2\epsilon_1
\sqrt{\textrm{Det}\hat{\m{A}}}}},
\end{equation}
where $\hat{1}$ is the $2\times 2$ unit matrix and $\textrm{Tr}$
is the trace of $\hat{\m{A}}$. $\epsilon_{1}$ and
$\epsilon_{2}$ are equal to $\pm 1$ combinations what gives
four matrix roots.  We have found that after conversion of
formula~\eqref{Sullivan} to MV form,  all roots do coincide with ours,
although root formulas have somewhat different shapes due to persistent
appearance of complex conjugate pairs in our spectral method.
However, it must be remembered that
D.~Sullivan's formula is limited to real and complex $2\times 2$ matrices only.

\vspace{2mm}\textit{Example~1}.
 In \cl{1}{2}, the roots of MV
 $\m{A}=\e{1}-2\e{23}$ are~\cite{Acus2024v2}
\[\begin{split}
&\m{B}_{1,2}=\pm\tfrac12\big(c_2(-\e{1}+\e{123})-c_1(1+\e{23})\big),\\
&\m{B}_{3,4}=\pm\tfrac12\big(-c_1(\e{1}+\e{123})+c_2(-1+\e{23})\big), \\
\end{split}\] where $c_1=\sqrt{-2+\sqrt{5}}$ and
$c_2=\sqrt{2+\sqrt{5}}$. The coefficients, Eq.~\eqref{CL12coefA}
and \eqref{CL12coefB}, allow to find the roots of symbolic  MV,
$\m{A}=a_1\e{1}+a_{23}\e{23}$, as well. Let's  introduce
$p_{\pm}=\sqrt{(a_1\pm\ii a_{23})^2}$ and
$m_{\pm}=-\sqrt{(a_1\pm\ii a_{23})^2}$, then the first pair,
Eq.~\eqref{CL12coefA}, will give the following coefficients of
roots expressed through $a_1$ and $a_{23}$,
\begin{equation*}\begin{split}
b_0=&\,\tfrac14\big(\sqrt{p_{+}}+\sqrt{p_{-}}+\sqrt{m_{+}}+\sqrt{m_{-}}\,\big), \\
b_1=&\,\big((p_{-}\sqrt{p_{+}}+p_{+}\sqrt{p_{-}}-p_{+}\sqrt{m_{-}}-p_{-}\sqrt{m_{+}})a_1+\\
 &\quad\ii(p_{-}\sqrt{p_{+}}-p_{+}\sqrt{p_{-}}+p_{+}\sqrt{m_{-}}-p_{-}\sqrt{m_{+}})a_{23}\big)/(4p_{+}p_{-}), \\
b_{23}=&\,-\ii\big((p_{-}\sqrt{p_{+}}-p_{+}\sqrt{p_{-}}+p_{+}\sqrt{m_{-}}+m_{-}\sqrt{m_{+}})a_1+\\
 &\qquad(p_{-}\sqrt{p_{+}}+p_{+}\sqrt{p_{-}}-p_{+}\sqrt{m_{-}}+m_{-}\sqrt{m_{+}})a_{23}\big)/(4p_{+}p_{-}), \\
b_{123}=&\,\ii\tfrac14\big(\sqrt{p_{+}}-\sqrt{p_{-}}+\sqrt{m_{+}}-\sqrt{m_{-}}\,\big). \\
\end{split}\end{equation*}
The roots consist  of complex-conjugate pairs, Fig.~\ref{fig:1}.
Remaining coefficients are $b_2=b_3=b_{12}=b_{13}=0$. From this
follows that all coefficients are real and the domain of the first
pair of roots of $\pm\sqrt{a_1\e{1}+a_{23}\e{23}}$ is $-\infty
>a_1, a_{12}>\infty$.

\subsection{\cl{2}{1} algebra, $^2\bbR(2)$}
As shown in~\cite{Acus2024v2},  in this algebra there may be up to
16 individual square roots. The  diagonalization method confirms
once more that the roots come out of all possible combinations of
plus/minus signs on a diagonal of spectral matrix. Basis vectors
in \cl{2}{1} algebra have  block-diagonal real matrix
representation (see respective table in Sec.~\ref{basisTables}),
\begin{equation}
 \hat{\e{}}_1=\Biggl[\begin{smallmatrix}1&0&0&0\\0&-1&0&0\\0&0&-1&0\\0&0&0&1\\
\end{smallmatrix}\Biggr],
\quad
 \hat{\e{}}_2=
\Biggl[\begin{smallmatrix}0&1&0&0\\1&0&0&0\\0&0&0&-1\\0&0&-1&0\\
\end{smallmatrix}\Biggr],
\quad
 \hat{\e{}}_3=
\Biggl[\begin{smallmatrix}0&-1&0&0\\1&0&0&0\\0&0&0&1\\0&0&-1&0\\
\end{smallmatrix}\Biggr].
\end{equation}
The matrix representation of a general MV is
\begin{equation}\label{matrixCL21}
\hat{\m{A}}=\Biggl[
\begin{smallmatrix}
 a_0+a_1+a_{23}+a_{123}&a_2-a_3+a_{12}-a_{13} & 0 & 0\\
 a_2+a_3-a_{12}-a_{13}& a_0-a_1-a_{23}+a_{123} & 0 & 0\\
 0 & 0&a_0-a_1+a_{23}-a_{123}&-a_2+a_3+a_{12}-a_{13} \\
 0 & 0& -a_2-a_3-a_{12}-a_{13}& a_0+a_1-a_{23}-a_{123}\\
\end{smallmatrix}\Biggr].
\end{equation}
The determinant consists of multipliers that are related to
matrix two blocks in Eq.~\eqref{matrixCL21},
\begin{equation}\begin{split}
\textrm{Det}\hat{\m{A}}=&\big((a_0+a_{123})^2+(a_1-a_{23})^2+(a_2+a_{13})^2+(a_3-a_{12})^2\big)\times\\
&\big((a_0-a_{123})^2+(a_1+a_{23})^2+(a_2-a_{13})^2+(a_3-a_{12})^2\big).
\end{split}\end{equation}

The coefficients needed for square root MVs can be
expressed by a set of six parameters
$\{\alpha,\beta,\gamma,\delta,\varphi_1,\varphi_2\}\in$ either $\bbR$ or
$\bbI$. The following  relations (einsatz) among parameters and coefficients
of MV root
$\pm(b_0+b_1\e{1}+b_2\e{2}+b_3\e{3}
+b_{12}\e{12}+b_{13}\e{13}+b_{23}\e{23}+b_{213}\e{123})$
was found from analysis of eigensystem that is related to MV
matrix representation, Eq.~\eqref{matrixCL21},
\begin{equation}\begin{split}\label{paramCL21}
&\ii\varphi_1=\sqrt{(a_1+a_{23})^2+(a_2-a_{13})^2-(a_3-a_{12})^2}\,,\\
&\ii\varphi_2=\sqrt{(a_1-a_{23})^2+(a_2+a_{13})^2-(a_3+a_{12})^2}\,,\\
&\alpha\pm\ii\beta=\sqrt{a_0+a_{123}\pm\ii\varphi_1}\,,\\
&\gamma\pm\ii\delta=\sqrt{a_0-a_{123}\pm\ii\varphi_2}\,.\\
\end{split}\end{equation}
The real coefficients that appear in the square root MV
are given below.  The upper/lower signs on right-hand expressions \eqref{CL21B1}- \eqref{CL21B8}
conform with the left/right double  indices in the square root
coefficients.
\begin{equation}\label{CL21B1}\begin{split}
&\indent\cl{2}{1}. \textit{Coefficients of the first root pairs,\ }\m{B}=\pm\sqrt{\m{A}}\,.\\
& b_0=\tfrac12(\alpha+\gamma),\quad b_{123}=\tfrac12(\alpha-\gamma),\\
&b_{1,23}=\Bigl(\pm(a_1-a_{23})\delta\varphi_1+(a_1+a_{23})\beta\varphi_2 \Bigr)/(2\varphi_1\varphi_2), \\
&b_{2,13}=\Bigl((a_2+a_{13})\delta\varphi_1\pm(a_2-a_{13})\beta\varphi_2 \Bigr)/(2\varphi_1\varphi_2), \\
&b_{3,12}=\Bigl((a_3+a_{12})\delta\varphi_1\pm(a_3-a_{12})\beta\varphi_2 \Bigr)/(2\varphi_1\varphi_2). \\
\end{split}\end{equation}
\begin{equation}\label{CL21B2}\begin{split}
&\indent\cl{2}{1}. \textit{Coefficients of the second  root pairs,\ }\m{B}=\pm\sqrt{\m{A}}\,.\\
& b_0=\tfrac{1}{2}(\alpha+\ii\delta),\quad b_{123}=\tfrac{1}{2}(\alpha-\ii\delta),\\
&b_{1,23}=\Bigl(\mp\ii(a_1-a_{23})\gamma\varphi_1+(a_1+a_{23})\beta\varphi_2\Bigr)/(2\varphi_1\varphi_2), \\
&b_{2,13}=-\Bigl(\ii(a_2+a_{13})\gamma\varphi_1\mp(a_2-a_{13})\beta\varphi_2)\Bigr)/(2\varphi_1\varphi_2, \\
&b_{3,12}=-\Bigl(\ii(a_3+a_{12})\gamma\varphi_1\mp(a_3-a_{12})\beta\varphi_2\Bigr)/(2\varphi_1\varphi_2), \\
\end{split}\end{equation}
\begin{equation}\label{CL21B3}\begin{split}
&\indent\cl{2}{1}. \textit{Coefficients of the third pair of roots,\ }\m{B}=\pm\sqrt{\m{A}}\,.\\
& b_0=\tfrac{1}{2}(\alpha-\ii\delta),\quad b_{123}=\tfrac{1}{2}(\alpha+\ii\delta),\\
&b_{1,23}=\Bigl(\pm\ii(a_1-a_{23})\gamma\varphi_1+(a_1+a_{23})\beta\varphi_2\Bigr)/(2\varphi_1\varphi_2), \\
&b_{2,13}=\Bigl(\ii(a_2+a_{13})\gamma\varphi_1\pm(a_2-a_{13})\beta\varphi_2\Bigr)/(2\varphi_1\varphi_2), \\
&b_{3,12}=\Bigl(\ii(a_3+a_{12})\gamma\varphi_1\pm(a_3-a_{12})\beta\varphi_2\Bigr)/(2\varphi_1\varphi_2). \\
\end{split}\end{equation}
\begin{equation}\label{CL21B4}\begin{split}
&\indent\cl{2}{1}. \textit{Coefficients of the fourth pair of roots,\ }\m{B}=\pm\sqrt{\m{A}}\,.\\
& b_0=\tfrac{1}{2}(\ii\beta+\gamma),\quad b_{123}=\tfrac{1}{2}(\ii\beta-\gamma),\\
&b_{1,23}=\Bigl(\pm(a_1-a_{23})\delta\varphi_1-\ii(a_1+a_{23})\alpha\varphi_2\Bigr)/(2\varphi_1\varphi_2), \\
&b_{2,13}=\Bigl((a_2+a_{13})\delta\varphi_1\pm\ii(-a_2+a_{13})\alpha\varphi_2\Bigr)/(2\varphi_1\varphi_2), \\
&b_{3,12}=\Bigl((a_3+a_{12})\delta\varphi_1\pm\ii(-a_3+a_{12})\alpha\varphi_2\Bigr)/(2\varphi_1\varphi_2). \\
\end{split}\end{equation}
\begin{equation}\label{CL21B5}\begin{split}
&\indent\cl{2}{1}. \textit{Coefficients of the fifth pair of roots,\ }\m{B}=\pm\sqrt{\m{A}}\,.\\
& b_0=-\tfrac{1}{2}(\ii\beta-\gamma),\quad b_{123}=-\tfrac{1}{2}(\ii\beta+\gamma),\\
&b_{1,23}=\Bigl(\pm(a_1-a_{23})\delta\varphi_1+\ii(a_1+a_{23})\alpha\varphi_2\Bigr)/(2\varphi_1\varphi_2), \\
&b_{2,13}=\Bigl((a_2+a_{13})\delta\varphi_1\pm\ii(a_2-a_{13})\alpha\varphi_2\Bigr)/(2\varphi_1\varphi_2), \\
&b_{3,12}=\Bigl((a_3+a_{12})\delta\varphi_1\pm\ii(a_3-a_{12})\alpha\varphi_2\Bigr)/(2\varphi_1\varphi_2). \\
\end{split}\end{equation}
\begin{equation}\label{CL21B6}\begin{split}
&\indent\cl{2}{1}. \textit{Coefficients of the sixth pair of roots,\ }\m{B}=\pm\sqrt{\m{A}}\,.\\
& b_0=\tfrac{1}{2}(\alpha-\gamma),\quad b_{123}=\tfrac{1}{2}(\alpha+\gamma),\\
&b_{1,23}=\Bigl(\pm(-a_1+a_{23})\delta\varphi_1+(a_1+a_{23})\beta\varphi_2\Bigr)/(2\varphi_1\varphi_2), \\
&b_{2,13}=-\Bigl((a_2+a_{13})\delta\varphi_1\pm(-a_2+a_{13})\beta\varphi_2\Bigr)/(2\varphi_1\varphi_2), \\
&b_{3,12}=-\Bigl((a_3+a_{12})\delta\varphi_1\pm(-a_3+a_{12})\beta\varphi_2\Bigr)/(2\varphi_1\varphi_2). \\
\end{split}\end{equation}
\begin{equation}\label{CL21B7}\begin{split}
&\indent\cl{2}{1}. \textit{Coefficients of the seventh pair of roots,\ }\m{B}=\pm\sqrt{\m{A}}\,.\\
& b_0=\ii\tfrac{1}{2}(\beta+\delta),\quad b_{123}=\ii\tfrac{1}{2}(\beta-\delta),\\
&b_{1,23}=-\ii\Bigl(\pm(a_1-a_{23})\gamma\varphi_1+(a_1+a_{23})\alpha\varphi_2\Bigr)/(2\varphi_1\varphi_2), \\
&b_{2,13}=-\ii\Bigl((a_2+a_{13})\gamma\varphi_1\pm(a_2-a_{13})\alpha\varphi_2\Bigr)/(2\varphi_1\varphi_2), \\
&b_{3,12}=-\ii\Bigl((a_3+a_{12})\gamma\varphi_1\pm(a_3-a_{12})\alpha\varphi_2\Bigr)/(2\varphi_1\varphi_2). \\
\end{split}\end{equation}
\begin{equation}\label{CL21B8}\begin{split}
&\indent\cl{2}{1}. \textit{Coefficients of the eighth pair of roots,\ }\m{B}=\pm\sqrt{\m{A}}\,.\\
& b_0=-\ii\tfrac{1}{2}(\beta-\delta),\quad b_{123}=-\ii\tfrac{1}{2}(\beta+\delta),\\
&b_{1,23}=\ii\Bigl(\pm(-a_1+a_{23})\gamma\varphi_1+(a_1+a_{23})\alpha\varphi_2\Bigr)/(2\varphi_1\varphi_2), \\
&b_{2,13}=-\ii\Bigl((a_2+a_{13})\gamma\varphi_1\pm(-a_2+a_{13})\alpha\varphi_2\Bigr)/(2\varphi_1\varphi_2), \\
&b_{3,12}=-\ii\Bigl((a_3+a_{12})\gamma\varphi_1\pm(-a_3+a_{12})\alpha\varphi_2\Bigr)/(2\varphi_1\varphi_2). \\
\end{split}\end{equation}

\vspace{3mm} From analysis of root coefficients, one can state
that  depending on concrete values of the coefficients in the
initial  MV $\m{A}$, in \cl{2}{1} algebra there may be eight pairs
or smaller number of real square roots. The number of roots is
controlled by values of parameters in Eqs~\eqref{paramCL21}. They must give real
coefficients in $\m{B}=\sqrt{\m{A}}$ for roots that exist. All coefficients
of $\m{B}$ represent real numbers if  conditions are satisfied
$\{\alpha,\gamma\}\in\bbR$ and $\{\beta,\delta\}\in\bbI$, and
pairs of products of parameters satisfy additional conditions: $\{\delta\varphi_1,\beta\varphi_2\}\in\bbR$;
$\{\gamma\varphi_1,\beta\varphi_2\}\in\bbI$;  $\{\delta\varphi_1,\alpha\varphi_2\}\in\bbI$;
$\{\gamma\varphi_1,\alpha\varphi_2\}\in\bbI$.
In fact, the presented conditions determine the domain of the existence of multivector roots.
Below a number of illustrative examples
that represent various kinds of roots $\B$ in \cl{2}{1} are given.

\vspace{2mm} \indent\textit{Example 1}. In \cl{2}{1} a simple MV,
$\m{A}=2+\e{1}+\e{13}$, has eight pairs of plus/minus roots of
$\m{A}$~\cite{Acus2024v2}.  Below, as usual, the sign combinations in brackets
indicate concrete plus/minus root signs used for diagonal root
matrix entries. The roots denoted by $\m{B}_{i,j}$ coincide with those
found  earlier in~\cite{Acus2024v2}, where distinct method to find
the roots has been proposed and which is based on the condition
for MV coefficients under the square root to be real numbers. In
the formulae below,  $c_1=\sqrt{2+\sqrt{2}}$, $c_2=\sqrt{2-\sqrt{2}}$\,,
\begin{equation*}\begin{split}
&1.\ (++++),\ \m{B}_{7,8}=\pm\tfrac{1}{2\sqrt{2}}\Bigl(\sqrt{2}(c_1+c_2)+(c_1-c_2)\e{1}+(c_1-c_2)\e{13}\Bigr),\\
&2.\ (-+++),\ \m{B}_{1,2}=\pm\tfrac{1}{2\sqrt{2}}\Bigl((c_1+c_2)\e{2}-(c_1+c_2)\e{23}-\sqrt{2}(c_1-c_2)\e{123})\Bigr),\\
&3.\ (+-++),\ \m{B}_{9,10}=\pm\tfrac{1}{2\sqrt{2}}\Bigl(\sqrt{2}c_2-c_2\e{1}-c_1\e{2}-c_2\e{13}+c_1\e{23}+\sqrt{2}c_1\e{123}\Bigr),\\
&4.\ (++-+),\ \m{B}_{11,12}=\pm\tfrac{1}{2\sqrt{2}}\Bigl(\sqrt{2}c_1+c_1\e{1}-c_2\e{2}+c_1\e{13}+c_2\e{23}-\sqrt{2}c_2\e{123}\Bigr),\\
&5.\ (+++-),\ \m{B}_{15,16}=\pm\tfrac{1}{2\sqrt{2}}\Bigl(\sqrt{2}c_2-c_2\e{1}+c_1\e{2}-c_2\e{13}-c_1\e{23}-\sqrt{2}c_1\e{123}\Bigr),\\
&6.\ (--++),\ \m{B}_{3,4}=\pm\tfrac{1}{2\sqrt{2}}\Bigl((-c_1+c_2)\e{2}+(c_1-c_2)\e{23}+\sqrt{2}(c_1+c_2)\e{123}\Bigr),\\
&7.\ (-+-+),\ \m{B}_{5,6}=\pm\tfrac{1}{2\sqrt{2}}\Bigl(\sqrt{2}(c_1-c_2)+(c_1+c_2)\e{1}+(c_1+c_2)\e{13}\Bigr),\\
&8.\ (-++-),\ \m{B}_{13,14}=\pm\tfrac{1}{2\sqrt{2}}\Bigl(\sqrt{2}c_1+c_1\e{1}+c_2\e{2}+c_1\e{13}-c_2\e{23}+\sqrt{2}c_2\e{123}\Bigr).\\
\end{split}\end{equation*}
In the brackets the adverse four-sign combinations, for example
replacement of $(-+++)$ by $(+---)$, give complementary roots that
have opposite overall sign. We have checked that the coefficients
of all roots are real numbers indeed.  In particular, the parameters
in Eqs~\eqref{CL21B1}-\eqref{CL21B8} have the values:
$\alpha=\gamma=\tfrac12(c_1+c_2)$;  $\beta=\delta=-\ii\tfrac12(c_1-c_2)$;
$\ii\{\varphi_1,\varphi_2\}=\{\sqrt{2},\sqrt{2}\}$;
$\{\delta\varphi_1,\beta\varphi_2\}=\{-c_1+c_2,-c_1+c_2\}/\sqrt{2}$;
$\{\ii\gamma\varphi_1,\beta\varphi_2\}=\{c_1+c_2,-c_1+c_2\}/\sqrt{2}$;
$\{\delta\varphi_1,\ii\alpha\varphi_2\}=\{-c_1+c_2,c_1+c_2\}/\sqrt{2}$;
$\{\delta\varphi_1,\beta\varphi_2\}=\{-c_1+c_2,-c_1+c_2\}/\sqrt{2}$;
$\ii\{\gamma\varphi_1,\alpha\varphi_2\}=\{c_1+c_2,c_1+c_2\}/\sqrt{2}$.

In Eqs~\eqref{paramCL21} the parameters  expressed by coefficients of MV  $\m{A}$ may
be employed to determine the domain of existence of the roots. In the
considered example,  $\A$ has three nonzero  coefficients, $a_0=2$,
$a_1=1$ and $a_{13}=1$, which  (after the remaining coefficients have been nullified),
yield the following equations, 
\begin{equation*}\begin{split}
&\varphi_1=\varphi_2=-\ii\sqrt{a_1^2+a_{13}^2},\quad\textrm{and\ } 2\varphi_1\varphi_2=-2(a_1^2+a_{13}^2)\in\bbR\,,\\
&\alpha=\gamma=\tfrac12\Bigl(\sqrt{a_0+\sqrt{a_1^2+a_{13}^2}}+\sqrt{a_0-\sqrt{a_1^2+a_{13}^2}}\Bigr),\\
&\beta=\delta=-\ii\tfrac12\Bigl(\sqrt{a_0+\sqrt{a_1^2+a_{13}^2}}+\sqrt{a_0-\sqrt{a_1^2+a_{13}^2}}\Bigr),\\
\end{split}\end{equation*}
where $\{\varphi_1,\varphi_2\}\in\bbI$, $2\varphi_1\varphi_2\in\bbR$, and
$\{\alpha,\gamma\}\in\bbR$ and $\{\beta,\delta\}\in\bbI$. These
obtained conditions are determined by $a_0$ and $\sqrt{a_1^2+a_{13}^2}$.
In particular, if  the condition $a_0> \sqrt{a_1^2+a_{13}^2}$ is  satisfied then all 16 roots exist.
Under such condition, the imaginary numbers  in all final square root
coefficients in  $\B_{i,j}$  vanish.   If $a_0\le\sqrt{a_1^2+a_{13}^2}$, the roots do not exist.

\vspace{2mm}
 \indent\textit{Example~2}. In \cl{2}{1}, $\m{A}=2+\e{1}+2\e{3}+\e{13}$
has four real roots, i.e., two real root pairs that are related to signs
$\{\pm,\pm,\pm,\pm\}$ and $\{\pm,\pm,\mp,\mp\}$, respectively,
 \begin{equation*}\begin{split}
&\m{B}_{1,2}=\pm\tfrac{1}{2\sqrt{2}}\Bigl(\sqrt{2}(g_1+g_2)-\ii(g_1-g_2)\e{1}-
2\ii(g_1-g_2)\e{3}-\ii(g_1-g_2)\e{123})\Bigr), \\
&\m{B}_{3,4}=\pm\tfrac{1}{2\sqrt{2}}\Bigl(\ii(g_1-g_2)\e{2}+2\ii(g_1-g_2)\e{12}-
\ii(g_1-g_2)\e{23})+\sqrt{2}(g_1+g_2)\e{123}\Bigr),\\
\end{split}\end{equation*} where $g_1=\sqrt{2+\ii\sqrt{2}}$ and $g_2=\sqrt{2-\ii\sqrt{2}}$\,.
The remaining roots are complex, thus, they must be rejected.

\subsection{\cl{0}{3} algebra, $^2\bbH(1)$}\label{subsecCL03}
From matrix rep of this algebra  follows that we  may have
either four roots, two roots, or no roots at all. The
table in Sec.~\ref{basisTables} shows that reps of  basis
vectors are $2\times 2$ diagonal quaternion matrices,
 $\hat{\e{}}_1=\bigl[\begin{smallmatrix}-\mathbf{i}&0\\0&\mathbf{i}
\end{smallmatrix}\bigr]$,
 $\hat{\e{}}_2=\bigl[\begin{smallmatrix}-\mathbf{j}&0\\0&\mathbf{j}
\end{smallmatrix}\bigr]$,
 $\hat{\e{}}_3=
\bigl[\begin{smallmatrix}-\mathbf{k}&0\\ 0&\mathbf{k}
\end{smallmatrix}\bigr]$.  After the replacement of Hamilton quaternionic imaginaries by matrices,
 \begin{math} \mathbf{i}\to
 \bigl[\begin{smallmatrix}
\ii&0\\0&-\ii
\end{smallmatrix}\bigr],\quad
\mathbf{j}\to \bigl[\begin{smallmatrix}
0&1\\-1&0
\end{smallmatrix}\bigr],\quad
\mathbf{k}=\to
 \bigl[\begin{smallmatrix}
0&\ii\\ \ii&0
\end{smallmatrix}\bigr],
\end{math}
we have  $4\times 4$ complex matrix reps:
\begin{equation}\label{CL30rep}
\hat{\e{}}_1=\Biggl[\begin{smallmatrix}-\ii&0&0&0\\0&\ii&0&0\\0&0&\ii&0\\0&0&0&-\ii\\
\end{smallmatrix}\Biggr],
\quad
 \hat{\e{}}_2=
\Biggl[\begin{smallmatrix}0&-1&0&0\\1&0&0&0\\0&0&0&1\\0&0&-1&0\\
\end{smallmatrix}\Biggr],
\quad
 \hat{\e{}}_3=
\Biggl[\begin{smallmatrix}0&-\ii&0&0\\-\ii&0&0&0\\0&0&0&\ii\\0&0&\ii&0\\
\end{smallmatrix}\Biggr],
\end{equation}  squares of which are minus unit matrices.
Bivectors and pseudoscalar reps, as usual, may be found
just by multiplying  matrices,  $\hat{\e{}}_1$, $\hat{\e{}}_2$, and
$\hat{\e{}}_3$. The initial MV $\m{A}$ in a
matrix form  now is
\begin{equation}\label{matrixCL03}
\hat{\m{A}}=\Biggl[
\begin{smallmatrix}
 (a_0+a_{123})-\ii(a_{1}-a_{23})&-(a_2+a_{13})-\ii(a_{3}-a_{12}) & 0 & 0\\
 (a_2+a_{13})-\ii(a_{3}-a_{12})& (a_0+a_{123})+\ii(a_{1}-a_{23}) & 0 & 0\\
 0 & 0&(a_0-a_{123})+\ii(a_{1}+a_{23})&(a_2-a_{13})+\ii(a_{3}+a_{12}) \\
 0 & 0& -(a_2-a_{13})+\ii(a_{3}+a_{12})& (a_0-a_{123})-\ii(a_{1}+a_{23}\\
\end{smallmatrix}\Biggr],\end{equation}
which  has four eigenvalues:
\begin{equation}\begin{split}
\epsilon_{1,2}&=a_0-a_{123}\mp\ii\sqrt{(a_1+a_{23})^2+(a_2-a_{13})^2+(a_3+a_{12})^2}, \\
\epsilon_{3,4}&=a_0+a_{123}\mp\ii\sqrt{(a_1+a_{23})^2+(a_2-a_{13})^2+(a_3+a_{12})^2}.
\end{split}\end{equation}
The pairs of diagonal square  root matrices therefore must have the
following sign combinations:
$\pm\{\sqrt{\epsilon_1},\sqrt{\epsilon_2},\sqrt{\epsilon_3},\sqrt{\epsilon_4}\}$
and
$\pm\{\sqrt{\epsilon_1},\sqrt{\epsilon_2},-\sqrt{\epsilon_3},-\sqrt{\epsilon_4}\}$.

As in previous instances, the einsatz
\begin{equation}\begin{split}\label{coeffCL03}
&\alpha\pm\ii\beta=\sqrt{a_0+a_{123}\pm\ii\varphi_2}\,, \\
&\gamma\pm\ii\delta=\sqrt{a_0-a_{123}\mp\ii\varphi_1}\,, \\
\end{split}\end{equation}
where $\alpha,\beta,\gamma,\delta\in$ either $\bbR$ or $\bbI$, will help us
to find the coefficients in the root MVs. In~\eqref{coeffCL03}, the functions $\varphi_1$
and $\varphi_2$ are real-valued that can be expressed in either
coordinates or in a coordinate-free form:
 \begin{equation}\begin{split}\label{CL03phi12}
  &\varphi_1=-\ii\sqrt{-(a_{13}-a_{2})^2-(a_{23}+
a_{1})^2-(a_{12}+a_{3})^2}=-\ii\sqrt{\ba^2+B^2-2(\ba\wedge B)\e{123}}\,,\\
  &\varphi_2=-\ii\sqrt{-(a_{13}+a_{2})^2-(a_{23}-
a_{1})^2-(a_{12}-a_{3})^2}=-\ii\sqrt{\ba^2+B^2+2(\ba\wedge
B)\e{123}}\,,
\end{split}\end{equation}
where $B$ is the bivector. Because Bott's  rep is a 2D-matrix,  the largest number of different  roots
should be four. The coefficients are computed in
exactly the same way as described  earlier. The real coefficients of
  the root MV
$\pm\m{B}=\pm(b_0+b_1\e{1}+b_2\e{2}+b_3\e{3}
+b_{12}\e{12}+b_{13}\e{13}+b_{23}\e{23}+b_{123}\e{123})$ are:
\begin{equation}\label{CL03B12}\begin{split}
&\indent\cl{0}{3}. \textit{The coefficients of the first pair of
roots\
}\m{B}=\pm\sqrt{\m{A}}\,.\\
& b_0=\tfrac12(\alpha+\gamma),\quad b_{123}=\tfrac12(\alpha-\gamma),\\
&b_{1,23}=\Bigl(\pm(a_1-a_{23})\beta\varphi_1+(a_1+a_{23})\delta\varphi_2 \Bigr)/(2\varphi_1\varphi_2), \\
&b_{2,13}=\Bigl((a_2+a_{13})\beta\varphi_1\pm(a_2-a_{13})\delta\varphi_2 \Bigr)/(2\varphi_1\varphi_2), \\
&b_{3,12}=\Bigl(\pm(a_3-a_{12})\beta\varphi_1+(a_3+a_{12})\delta\varphi_2 \Bigr)/(2\varphi_1\varphi_2)\,. \\
\end{split}\end{equation}
Below,  in the second  root pair, the vector and bivector coefficients have opposite signs.
\begin{equation}\label{CL03B34}\begin{split}
&\indent\cl{0}{3}. \textit{The coefficients of the second pair of
roots
}\m{B}^{\prime}=\pm\sqrt{\m{A}}=\pm(b_0^\prime+\\
&\indent b_1^\prime\e{1}+ b_2^\prime\e{2}+b_3^\prime\e{3}+b_{12}^\prime\e{12}+
b_{13}^\prime\e{13}+b_{23}^\prime\e{23}+b_{123}^\prime\e{123}):\\
 &b_0^\prime=\tfrac12(-\alpha+\gamma),\quad b_{123}^\prime=-\tfrac12(\alpha+\gamma),\\
 &b_1^\prime=-b_1,\ b_2^\prime=-b_2,\ b_3^\prime=-b_3,\\
 &b_{12}^\prime=-b_{12},\ b_{13}^{\prime}=-b_{13},\ b_{23}^\prime=-b_{23}.
\end{split}\end{equation}
Since the  singularities in vector and bivector coefficients  appear in the
denominators via product $\varphi_1\varphi_2$, the roots
of a general MV in the generic expressions \eqref{CL03B12} and
\eqref{CL03B34} may be singular if the coefficients in
\eqref{CL03phi12}   satisfy $a_i=a_{jk}=0$, $i\ne j\ne k$. Calculations show that
the considered singularities are removable.
However, in this case it is simpler to find the roots  from the very beginning,
i.e. to apply the spectral method directly to a simpler MV rather than to resort to the
generic formulas, Eq.~\eqref{CL03B12} and \eqref{CL03B34}.

\vspace{3mm}
 \textit{Example~1.} In \cl{0}{3}, the first and
second pairs of roots of $\m{A}=a_0+a_{123}\e{123}$  as found from
generic expressions,  Eq.~\eqref{CL03B12} and \eqref{CL03B34}, and
from direct calculations with $\m{A}=a_0+a_{123}\e{123}$ by the
present method, are found to be
\begin{equation}\begin{split}
&\B_{12}=\pm\tfrac12\Big(\sqrt{a_0+a_{123}}+\sqrt{a_0-a_{123}}+(\sqrt{a_0+a_{123}}-\sqrt{a_0-a_{123}})\e{123}\Big),\\
&\B_{34}=\mp\tfrac12\big(\sqrt{a_0+a_{123}}-\sqrt{a_0-a_{123}}+(\sqrt{a_0+a_{123}}+\sqrt{a_0-a_{123}})\e{123}\big).\\
\end{split}\end{equation}
The formulas show that the spectral roots exist if $a_0>0$ and
$|a_{123}|\le a_0$. If $a_0=a_{123}$, the  root formulas reduce to a
single formula that is valid if $a_0>0$.

 \textit{Example~2.} \cl{0}{3},  matrix representation   MV
$\m{A}=a_3\e{3}+a_{12}\e{12}$ is
\[\hat{\m{A}}=\ii\Bigg[\begin{smallmatrix}
0& a_{12}-a_3&0&0 \\
 a_{12}-a_3&0&0&0\\
0&0&0& a_{12}+a_3\\
0&0& a_{12}+a_3&0\\
\end{smallmatrix} \Bigg].
\]
The determinant is $\det\hat{\m{A}}=(a_3^2-a_{12}^2)^2$,
which is zero if $a_3=-a_{12}$. From this follows that the MV
$\e{3}\pm\e{12}$ has no roots.
In the latter case the
coefficients, Eq.~\eqref{CL03B12} and \eqref{CL03B34}, have
essential (unremovable) singularity.

\textit{Example~3}. In \cl{0}{3} the MV
$\m{A}=-\e{3}+\e{12}+4\e{123}$ has no spectral roots for required
sign combinations  $\{\pm\pm\pm\pm\}$ and $\{\pm\pm\mp\mp\}$
on  the diagonal spectral matrix. Because of that, it is useful to consider
more general symbolic MV
$\m{A}=a_{3}\e{3}+a_{12}\e{12}+a_{123}\e{123}$ that has a block-diagonal  rep,
\begin{equation}\label{matrixEx3}
\hat{\m{A}}=\Biggl[
\begin{smallmatrix}
 a_{123}&-\ii(a_{3}-a_{12}) & 0 & 0\\
-\ii(a_{3}-a_{12})& a_{123} & 0 & 0\\
 0 & 0&-a_{123}&\ii(a_{3}+a_{12}) \\
 0 & 0& \ii(a_{3}+a_{12})&- a_{123}\\
\end{smallmatrix}\Biggr].
\end{equation}
The trace of $\hat{\m{A}}$ is  zero, however,   the determinant is not zero. Thus,
the present spectral  method may be applied to find symbolic
 roots $\m{B}=\sqrt{\m{A}}$.
It was found that   the roots of~\eqref{matrixEx3} have the
following structure  $\m{B}=b_0+b_{3}\e{3}+b_{12}\e{12}+b_{123}\e{123}$.  For sign combination  $\{++--\}$
 the coefficients  $b_{0}$ and $b_{123}$ are
\begin{equation}\label{b0123}
b_{0,123}=\tfrac14\Big((\sqrt{a_{123}-h_1}+\sqrt{a_{123}+
h_1})\mp(\sqrt{-a_{123}-h_2}+\sqrt{-a_{123}+h_2})\Big).
\end{equation}
The remaining coefficients  $b_{3}$ and $b_{12}$ are
\begin{equation}\label{b312}
b_{3,12}=\frac14\Big(\mp\frac{h_1(\sqrt{a_{123}-h_1}-\sqrt{a_{123}+h_1})}{a_{12}-a_3}
-\frac{h_2(\sqrt{-a_{123}-h_2}-\sqrt{-a_{123}+h_2})}{a_{12}+a_3}\Big),
\end{equation}
where $h_1=\sqrt{-(a_{12}-a_3)^2}=\ii|a_{12}-a_3|$ and  $h_2=\sqrt{-(a_{12}+a_3)^2}=\ii|a_{12}+a_3|$
are imaginary quantities. At $a_{2}=\pm a_3$ there is singularity that can be removed.
Then, after insertion of $a_3=-1$, $a_{12}=1$ and  $a_{123}=4$  we get
$b_{0,123}=\frac14(\mp4\ii+\sqrt{4-2\ii}+\sqrt{4+2\ii})$ and
$b_{3,12}=\mp\ii\frac14(-\sqrt{4-2\ii}+\sqrt{4+2\ii})$. The conjugate square roots
give real quantities,  however $\mp4\ii$ remains uncompensated. In conclusion, this root does not exist.
Similar situation is with  \{++++\} root.
To check the results, also the roots have been calculated from general root coefficients, Eqs  \eqref{CL03B12}
and \eqref{CL03B34}, after replacing the symbols by numbers. It must stressed that the
example demonstrates that only the roots  that are related to MV discrete spectrum do not exist.
If different methods \cite{Acus2024v2,Dargys-Acus2020} were applied  it may appear that
the roots that are not related to root spectrum may exist. Indeed, we have found~\cite{Acus2024v2}
 that the considered MV, $\m{A}=-\e{3}+\e{12}+4\e{123}$, has a continuum of roots
 rather than isolated roots.

\section{Spectral MV roots in 4D algebras}\label{examples4D}
4-dimensional algebras are represented either by $\bbR(4)$ or
$\bbH(2)$ matrices. The generic MV in these algebras has
the following  form,
\begin{equation}\begin{split}\label{mvCL40}
\m{A}=&\,a_0+a_1\e{1}+a_2\e{2}+a_3\e{3}+a_4\e{4}+a_{12}\e{12}+a_{13}\e{13}+a_{23}\e{23}+\\
 &a_{14}\e{14}+a_{24}\e{24}+a_{34}\e{34}+a_{1234}\e{1234}.
\end{split}\end{equation}
There are six bivector planes.  Geometrically, the pairs of planes with
different indices, for example $\e{12}$ and $\e{34}$, commute and, as a result,
 represent  nonintersecting oriented planes in 4D
vector space.

\subsection{Roots of MV in \cl{4}{0}, $\bbH(2)$}\label{examples4DEuclidean}
From tables in Sec.~\ref{basisTables} we have the following
quaternion matrix reps for basis vectors,
\begin{equation}\label{ijkReps}
\hat{\e{}}_1=\Big[\begin{matrix} 1&0\\0&-1  \end{matrix}\Big],\quad
\hat{\e{}}_2=\Big[\begin{matrix} 0&1\\1&0  \end{matrix}\Big],\quad
\hat{\e{}}_3=\Big[\begin{matrix} 0&-\mathbf{i}\\\mathbf{i}&0\end{matrix}\Big],\quad
\hat{\e{}}_4=\Big[\begin{matrix} 0&-\mathbf{j}\\\mathbf{j}&0\end{matrix}\Big],
\end{equation}
where $\bi,\bj$ and $\bk\equiv\bi\bj$ are the Hamilton imaginary units that
satisfy $\bi^2=\bj^2=\bk^2=-1$ and $\bi\bj\bk=-1$. In terms of
Hamilton  units the matrix rep of Eq.~\eqref{mvCL40} is
\begin{equation}\begin{split}\label{quatrMatrix}
 &\hat{\m{A}}=\\
 &\bigl[\begin{smallmatrix}
 (a_0+a_1)+(a_{23}+a_{123})\bi+(a_{24}+a_{124})\bj-(a_{34}+a_{134})\bk& (a_2+a_{12})-(a_3+a_{13})\bi-(a_4+a_{14})\bj-(a_{234}+a_{1234})\bk \\
 (a_2-a_{12})+(a_3-a_{13})\bi+(a_4-a_{14})\bj+(-a_{234}+a_{1234})\bk& (a_0-a_1)-(a_{23}-a_{123})\bi-(a_{24}-a_{124})\bj-(a_{34}-a_{134})\bk\\
\end{smallmatrix}\bigr]
\end{split}\end{equation}
 As in Subsec.~\ref{subsecCL03} we shall go over to complex matrix reps.
 After replacement of Hamilton imaginaries in \eqref{ijkReps}  by respective matrices,
 \begin{math} \mathbf{i}\to
 \bigl[\begin{smallmatrix}
\ii&0\\
0&\ii
\end{smallmatrix}\bigr],\
\mathbf{j}\to \bigl[\begin{smallmatrix}
0&1\\
-1&0
\end{smallmatrix}\bigr],\
\mathbf{ij}=\mathbf{k}\to
 \bigl[\begin{smallmatrix}
0&\ii\\
\ii&0
\end{smallmatrix}\bigr],
\end{math}
the basis vector  reps become
\begin{equation}\label{4x4CL40reps}
 \hat{\e{}}_1=\Bigg[\begin{smallmatrix}1&0&0&0\\0&1&0&0\\0&0&-1&0\\0&0&0&-1\end{smallmatrix}\Bigg],\
 \hat{\e{}}_2=\Bigg[\begin{smallmatrix} 0&0&1&0\\0&0&0&1\\1&0&0&0\\0&1&0&0 \end{smallmatrix}\Bigg],\
 \hat{\e{}}_3=\Bigg[\begin{smallmatrix} 0&0&-\ii&0\\0&0&0&\ii\\ \ii&0&0&0\\0&-\ii&0&0 \end{smallmatrix}\Bigg],\
 \hat{\e{}}_4=\Bigg[\begin{smallmatrix} 0&0&0&-1\\0&0&1&0\\0&1&0&0\\-1&0&0&0
\end{smallmatrix}\Bigg].
\end{equation}
Representations of higher grades may be found by multiplying
matrices in~\eqref{4x4CL40reps}. For example, the
bivector matrices are given by six  product,
$\hat{\e{}}_{ij}=\hat{\e{}}_i\hat{\e{}}_j$, $i\ne j$.

In \cl{4}{0}, the  generic  MV rep is the following complex-valued
matrix,
\begin{equation}\begin{split}
&\qquad\qquad \hat{\m{A}}=\\
 &\Bigg[\begin{smallmatrix}
 (a_0+a_1)+\ii(a_{23}+a_{123})          & (a_{24}+a_{124})-\ii (a_{34}+a_{134}) & (a_2+a_{12})-\ii (a_3+ a_{13}) &
   -(a_4+a_{14})-\ii(a_{234}+a_{1234}) \\
 -(a_{24}+a_{124})-\ii(a_{34}+ a_{134}) & (a_0+a_1)-\ii (a_{23}+ a_{123})   & (a_4+a_{14})-\ii(a_{234}-
   a_{1234}) & (a_2+a_{12})+\ii (a_3+a_{13}) \\
 (a_2-a_{12})+\ii(a_3+a_{13})           & (a_4-a_{14})-\ii (a_{234}-a_{1234}) & (a_0-a_1)-\ii (a_{23}-a_{123}) &
   -(a_{24}-a_{124})-\ii(a_{34}-a_{134}) \\
  (-a_4+a_{14})-\ii(a_{234}-a_{1234})  & (a_2-a_{12})-\ii (a_3-a_{13})   & (a_{24}-a_{124})-\ii(a_{34}-
   a_{134}) & (a_0-a_1)+\ii(a_{23}-a_{123})
\end{smallmatrix}\Bigg]\end{split}\end{equation}
Note that  individual matrices of the vector and
pseudoscalar are Hermitian  while those of bivector and trivector
are anti-Hermitian.

\subsubsection{Simple MV in \cl{4}{0}}

The even MVs are connected with rotations and consequently with spinor
group~\cite{Lounesto97}. Since the square of even MV remains  even,  it follows
the square root should be an even MV too.  Let the MV has  following simple form,
\begin{equation}\label{mvCL40A}
\m{A}=a_0+a_{14}\e{14}+a_{24}\e{24}+a_{34}\e{34}+a_{1234}\e{1234}=a_0+\cA+a_{1234}\e{1234},
\end{equation} where $\cA\equiv\gb{\m{A}}=a_{14}\e{14}+a_{24}\e{24}+a_{34}\e{34}$
is the bivector part, the magnitude of which is
$\magn{\cA}=\sqrt{-\cA^2}=\sqrt{a_{14}^2+a_{24}^2+a_{34}^2}$.
The matrix~$\hat{\m{A}}$ has four eigenvalues,
$\{a_0,a_{24}-a_{34},0,-a_{14},-\ii a_{1234}\}$, and respective eigenvectors. From the latter
follows  the transformation matrix,
\begin{equation}
\hat{\m{T}}=\frac{1}{a_{24}-\ii a_{34}}
\begin{bmatrix}
\ii(a_{24}-\ii a_{34})&-\ii(a_{24}-\ii a_{34})&\ii(a_{24}-\ii a_{34})&-\ii(a_{24}-\ii a_{34})\\
a_{14}+\magn{\cA}&a_{14}-\magn{\cA}&a_{14}-\magn{\cA}&a_{14}+\magn{\cA}\\
-\ii(a_{14}+\magn{\cA})&\ii(a_{14}-\magn{\cA})&-\ii(a_{14}-\magn{\cA})&\ii(a_{14}+\magn{\cA})\\
a_{24}-\ii a_{34}&a_{24}-\ii a_{34}&a_{24}-\ii a_{34}&a_{24}-\ii
a_{34}
\end{bmatrix}.
\end{equation}
From computer analysis follows that instead of expected two pairs of roots
there remains a single  (plus/minus) pair with real coefficients, $\sqrt{\m{A}}=\pm\m{B}$, where
\begin{equation}\label{coeffB}
\m{B}=b_0+b_{12}\e{12}+b_{13}\e{13}+b_{23}\e{23}+b_{14}\e{14}+b_{24}\e{24}+b_{34}\e{34}+b_{1234}\e{1234}.
\end{equation}
If complex conjugate pairs,
$\{\varepsilon_{1+},\varepsilon_{1-}\}$ and
  $\{\varepsilon_{2+},\varepsilon_{2-}\}$ are introduced, where
\begin{equation}\begin{split}\label{eigenvalCL40}
&\varepsilon_{1+}=\sqrt{a_0+ a_{1234}+\ii\magn{\cA}},\quad
\varepsilon_{1-}=\sqrt{a_0+ a_{1234}-\ii\magn{\cA}},\\
&\varepsilon_{2+}=\sqrt{a_0-a_{1234}+\ii\magn{\cA}},\quad
\varepsilon_{2-}=\sqrt{a_0-a_{1234}-\ii\magn{\cA}},\\
\end{split}\end{equation}
then  in \eqref{coeffB}  the real    coefficients  of $\m{B}$
for root  signs $\{\pm\pm\pm\pm\}$ are
\begin{equation}\begin{split}
&b_0=\tfrac14\big((\epsilon_{1+}+\epsilon_{1-})+(\epsilon_{2-}+\epsilon_{2+})\big),\\
&b_{12}=\ii a_{34}(\varepsilon_{1+}-\varepsilon_{1-}-\varepsilon_{2+}+\varepsilon_{2-})(4\magn{\cA}),\\
&b_{13}=-\ii a_{24}(\varepsilon_{1+}-\varepsilon_{1-}-\varepsilon_{2+}+\varepsilon_{2-})/(4\magn{\cA}),\\
&b_{23}=\ii a_{14}(\varepsilon_{1+}-\varepsilon_{1-}-\varepsilon_{2+}+\varepsilon_{2-})/(4\magn{\cA}),\\
&b_{14}=\ii a_{14}(-\varepsilon_{1+}+\varepsilon_{1-}-\varepsilon_{2+}+\varepsilon_{2-})/(4\magn{\cA}),\\
&b_{24}=\ii a_{24}(-\varepsilon_{1+}+\varepsilon_{1-}-\varepsilon_{2+}+\varepsilon_{2-})/(4\magn{\cA}),\\
&b_{34}=\ii a_{34}(-\varepsilon_{1+}+\varepsilon_{1-}-\varepsilon_{2+}+\varepsilon_{2-})/(4\magn{\cA}),\\
&b_{1234}= \tfrac14\big((\epsilon_{1+}+\epsilon_{1-})-(\epsilon_{2+}+\epsilon_{2-})\big)/(4\magn{\cA}),\\
\end{split}\end{equation}
where $\magn{\cA}=\sqrt{a_{14}^2+a_{24}^2+a_{34}^2}$ is the
bivector magnitude. The coefficients of remaining roots with
$\{\pm\pm\mp\mp\}$  signs on the diagonal are equal to zero.

\vspace{3mm}
 Now, instead of \eqref{mvCL40A} we shall assume that an
even MV contains the  bivectors of different kind, namely,
\begin{equation}\label{mvCL40B}
\m{A}=a_0+a_{12}\e{12}+a_{13}\e{13}+a_{23}\e{23}+a_{1234}\e{1234}=a_0+\cA+a_{1234}\e{1234},
\end{equation}
where $\cA\equiv\gb{\m{A}}=a_{12}\e{12}+a_{13}\e{13}+a_{23}\e{23}$
is the bivector, the magnitude of which is
$\magn{\cA}=\sqrt{-\cA^2}=\sqrt{a_{12}^2+a_{13}^2+a_{23}^2}\,$.
Note that in both cases the squares of bivectors are equal  to $-1$.
Similar calculations show that in this case there is a single pair
of roots, $\sqrt{\m{A}}=\pm\m{B}$,  with sign combinations $\{\pm\pm\pm\pm\}$, namely,
\begin{equation}
\m{B}=b_0+b_{12}\e{12}+b_{13}\e{13}+b_{23}\e{23}+b_{14}\e{14}
+b_{24}\e{24}+b_{34}\e{34}+b_{1234}\e{1234},
\end{equation} where all coefficients are real:
\begin{align*}
b_0&=(\alpha+\gamma)/2, & b_{1234}&=(\alpha-\gamma)/2,\\
b_{12}&=a_{12}(\beta+\delta)/(2\magn{\cA}), & b_{13}&=a_{13}(\beta+\delta)/(2\magn{\cA}), \\
b_{23}&=a_{23}(\beta+\delta)/(2\magn{\cA}), & b_{14}&=a_{23}(-\beta+\delta)/(2\magn{\cA}),\\
b_{24}&=a_{13}(-\beta+\delta)/(2\magn{\cA}), & b_{34}&=a_{12}(-\beta+\delta)/(2\magn{\cA}),\\
\end{align*}
where
\begin{equation}\begin{split}
&\alpha=\tfrac12\big(\sqrt{a_0+a_{1234}+\ii\magn{\cA}}+\sqrt{a_0+a_{1234}-\ii\magn{\cA}}\big),\\
&\beta=-\tfrac{\ii}{2}\big(\sqrt{a_0+a_{1234}+\ii\magn{\cA}}-\sqrt{a_0+a_{1234}-\ii\magn{\cA}}\big),\\
&\gamma=\tfrac12\big(\sqrt{a_0-a_{1234}+\ii\magn{\cA}}+\sqrt{a_0-a_{1234}-\ii\magn{\cA}}\big),\\
&\delta=-\tfrac{\ii}{2}\big(\sqrt{a_0-a_{1234}+\ii\magn{\cA}}-\sqrt{a_0-a_{1234}-\ii\magn{\cA}}\big).\\
\end{split}\end{equation}
The  coefficients, $\{\alpha,\beta,\gamma,\delta\}\in\bbR$,
similarly as in Eq.~\eqref{coeffCL30}, may be rewritten in a more
compact form:
$\alpha\pm\ii\beta=\sqrt{a_0+a_{1234}\pm\ii\magn{\cA}}$ and
$\gamma\pm\ii\delta=\sqrt{a_0-a_{1234}\pm\ii\magn{\cA}}$. When
$\magn{\cB}=0$, the root has a singularity which  can be eliminated
by finding the  limit $\magn{\cA}\to 0$,
\begin{equation*}
\lim_{\magn{\cA}\to
0}\sqrt{\m{A}}=\tfrac12\big((\sqrt{a_0+a_{1234}}+\sqrt{a_0-a_{1234}})+\e{1234}(\sqrt{a_0+a_{1234}}+\sqrt{a_0-a_{1234}})\big).
\end{equation*}
 Thus, in the latter case the square roots of $\m{A}=a_0+a_{1234}\e{1234}$ exist
 if   under roots the  expression $a_0\pm a_{1234}$ is positive.

\subsubsection{Even  MV in \cl{4}{0}}\label{evenMVCL40} 
Experiments  with computer program show that calculations  proceed smoothly if bivectors,
$\gb{\m{A}}^{\prime}=a_{12}\e{12}+a_{13}\e{13}+a_{23}\e{23}$
and $\gb{\m{A}}^{\prime\prime}=a_{14}\e{14}+a_{24}\e{24}+a_{34}\e{34}$, in a process of calculation remain in the following shortcut form:
\begin{equation}\begin{split}
&\varepsilon_{\pm}^2=(a_{12}\pm a_{34})^2+(a_{13}\mp a_{24})^2+(a_{14}\pm a_{23})^2\,,\\
&\alpha\pm\ii\beta=\sqrt{a_0+a_{1234}\pm\ii\varepsilon_{-}}\,,\\
&\gamma\pm\ii\delta=\sqrt{\gs{\m{A}}-\gq{\m{A}}\pm\ii\varepsilon_{+}}\,.\\
\end{split}\end{equation}
Note that instead of grade $\gb{\m{A}}$ during computation the symbols
$\varepsilon_{+}$ and $\varepsilon_{-}$ are used, where nonintersecting bivector planes in 4-dimensional
space are grouped together in the pairs. If $\gb{\m{A}}$ is divided into two
parts $\gb{\m{A}}=\gb{\m{A}^\prime}+\gb{\m{A}^{\prime\prime}}$,
then $\varepsilon_{\pm}$ can be represented in a coordinate-free
form as
\begin{equation}
\varepsilon_{\pm}=\sqrt{-\gb{\m{A}^\prime}\cdot\gb{\m{A}^\prime}-\gb{\m{A}^{\prime\prime}}\cdot\gb{\m{A}^{\prime\prime}}\pm
2\gb{\m{A}^\prime}\wedge\gb{\m{A}^{\prime\prime}}}\,.
\end{equation}
After  the above preventive measures have been included in the program, the calculations have lasted an acceptable period of time, and
we  have found two pairs of real square  roots.

The coefficients of the first pair,
$(\sqrt{\m{A}})_{1,2}=\pm\m{B}=\pm(b_0+b_{12}\e{12}+b_{13}\e{13}+b_{23}\e{23}+b_{14}\e{14}+
b_{24}\e{24}+b_{34}\e{34}+a_{1234}\e{1234})$, are
\begin{equation}\begin{split}
&b_0=\tfrac12(\alpha+\gamma),\quad b_{1234}=\tfrac12(\alpha-\gamma)\e{1234}, \\
&b_{12}=\tfrac12\big((a_{12}-a_{34})\beta/\varepsilon_{-}+(a_{12}+a_{34})\delta/\varepsilon_{+}\big),\\
&b_{13}=\tfrac12\big((a_{13}+a_{24})\beta/\varepsilon_{-}+(a_{13}-a_{24})\delta/\varepsilon_{+}\big),\\
&b_{23}=\tfrac12\big((a_{23}-a_{14})\beta/\varepsilon_{-}+(a_{23}+a_{14})\delta/\varepsilon_{+}\big),\\
&b_{14}=\tfrac12\big((a_{14}-a_{23})\beta/\varepsilon_{-}+(a_{14}+a_{23})\delta/\varepsilon_{+}\big),\\
&b_{24}=\tfrac12\big((a_{24}+a_{13})\beta/\varepsilon_{-}+(a_{24}-a_{13})\delta/\varepsilon_{+}\big),\\
&b_{34}=\tfrac12\big((a_{34}-a_{12})\beta/\varepsilon_{-}+(a_{34}+a_{12})\delta/\varepsilon_{+}\big).\\
\end{split}\end{equation}
The coefficients of the second root pair $(\sqrt{\m{A}})_{3,4}$ are
\begin{equation}\begin{split}
&b_0=\tfrac12(-\alpha+\gamma),\quad b_{1234}=-\tfrac12(\alpha+\gamma)\e{1234}, \\
&b_{12}=\tfrac12\big(-(a_{12}-a_{34})\beta/\varepsilon_{-}+(a_{12}+a_{34})\delta/\varepsilon_{+}\big),\\
&b_{13}=\tfrac12\big(-(a_{13}+a_{24})\beta/\varepsilon_{-}+(a_{13}-a_{24})\delta/\varepsilon_{+}\big),\\
&b_{23}=\tfrac12\big(-(a_{23}-a_{14})\beta/\varepsilon_{-}+(a_{23}+a_{14})\delta/\varepsilon_{+}\big),\\
&b_{14}=\tfrac12\big(-(a_{14}-a_{23})\beta/\varepsilon_{-}+(a_{14}+a_{23})\delta/\varepsilon_{+}\big),\\
&b_{24}=\tfrac12\big(-(a_{24}+a_{13})\beta/\varepsilon_{-}+(a_{24}-a_{13})\delta/\varepsilon_{+}\big),\\
&b_{34}=\tfrac12\big(-(a_{34}-a_{12})\beta/\varepsilon_{-}+(a_{34}+a_{12})\delta/\varepsilon_{+}\big).\\
\end{split}\end{equation}

\subsection{Roots of MV in \cl{1}{3}, $\bbH(2)$. Dirac matrices}\label{examples4DRelativity13}
The  data table in Sec.~\ref{basisTables} indicates the following
quaternion matrix reps for basis vectors in \cl{1}{3},
\begin{equation}\label{diracQuaternMat}
 \hat{\e{}}_1=\Big[\begin{matrix} 1&0\\0&-1  \end{matrix}\Big],\quad
 \hat{\e{}}_2=\Big[\begin{matrix} 0&-1\\1&0  \end{matrix}\Big],\quad
 \hat{\e{}}_3=\Big[\begin{matrix} 0&-\mathbf{i}\\-\mathbf{i}&0  \end{matrix}\Big],\quad
 \hat{\e{}}_4=\Big[\begin{matrix} 0&-\mathbf{j}\\-\mathbf{j}&0  \end{matrix}\Big].
\end{equation}
After replacement of Hamilton's imaginary units  by matrices
 \begin{math} \mathbf{i}\to
 \bigl[\begin{smallmatrix}
\ii&0\\
0&\ii
\end{smallmatrix}\bigr],\quad
\mathbf{j}\to \bigl[\begin{smallmatrix}
0&1\\
-1&0
\end{smallmatrix}\bigr],\quad
\mathbf{ij}=\mathbf{k}\to
 \bigl[\begin{smallmatrix}
0&\ii\\
\ii&0
\end{smallmatrix}\bigr]
\end{math}, one obtains   Dirac matrices,\footnote{The matrices in
Eq.~\eqref{diracMat} differ from Dirac matrices used in
physics by unitary transformation.}
\begin{equation}\label{diracMat}
 \hat{\e{}}_1=\Bigg[\begin{smallmatrix} 1&0&0&0\\0&1&0&0\\0&0&-1&0\\0&0&0&-1\end{smallmatrix}\Bigg],\
 \hat{\e{}}_2=\Bigg[\begin{smallmatrix} 0&0&-1&0\\0&0&0&-1\\1&0&0&0\\0&1&0&0 \end{smallmatrix}\Bigg],\
 \hat{\e{}}_3=\Bigg[\begin{smallmatrix} 0&0&-\ii&0\\0&0&0&\ii\\-\ii&0&0&0\\0&\ii&0&0 \end{smallmatrix}\Bigg],\
 \hat{\e{}}_4=\Bigg[\begin{smallmatrix} 0&0&0&-1\\0&0&1&0\\0&-1&0&0\\1&0&0&0
\end{smallmatrix}\Bigg].
\end{equation}
 In relativistic physics, basis vector
$\e{1}$ represents the direction of time and $\{\e{2},\e{3},\e{4}\}$ represent space
components. The bivectors $\{\e{12},\e{13},\e{14}\}$
that comprise the  time vector $\e{1}$ are called timelike, while
remaining bivectors $\{\e{23},\e{24},\e{34}\}$ that represent rotations in the space are called
spacelike. The squares of the former/latter have plus/minus sign,
respectively. The geometric product, as usual, is  provided by
matrix respective  rep products. The order of matrix group  with a
unit matrix included is equal to $16$, which is equal to  basis vector number plus unity in
\cl{1}{3} algebra. The first rotor
$\m{R}\equiv\m{A}=a_0+a_{23}\e{23}+a_{24}\e{24}+a_{34}\e{34}+a_{1234}\e{1234}$
represents all possible  rotations described by  trigonometric
functions in three orthogonal planes, while the second,
$\m{A}=a_0+a_{12}\e{12}+a_{13}\e{13}+a_{14}\e{14}+a_{1234}\e{1234}$,
describes rotors related to hyperbolic planes,
Fig.~\ref{fig:2}b. In physics, the latter are called the
relativistic boosts instead.

\subsubsection{\label{rotationRel}Roots of the spinor related to rotations in physical space.}

Under group of space rotations (Euler rotations) the time basis
vector $\e{1}$ remains unaffected, whereas the vectors $\{\e{2},\e{3},\e{4}\}$
or respective projections (coefficients) are linearly transformed.
The distinctive property of rotational bivectors
$\{\e{23},\e{24},\e{34}\}$ constructed from such  vectors is that
their squares are equal to $-1$ and, thus, the rotations are described by
trigonometric functions. Thus, $\e{23}^2=\e{24}^2=\e{34}^2=-1$ and
$\cA=\gb{\m{A}}=a_{23}\e{23}+a_{24}\e{24}+a_{34}\e{34}$, the
magnitude of which is
$\magn{\gb{\m{A}}}=\sqrt{a_{23}^2+a_{24}^2+a_{34}^2}$. The  matrix
rep of the related rotor is
\begin{equation}\hat{\m{A}}=\begin{bmatrix}
a_0+\ii a_{23} &a_{24}+\ii a_{34} &0 & -\ii a_{1234}\\
-a_{24}+\ii a_{34} &a_0-\ii a_{23} &-\ii a_{1234} &0 \\
0 &-\ii a_{1234} &a_0-\ii a_{23} &-a_{24}+\ii a_{34} \\
-\ii a_{1234} &0 &a_{24}+\ii a_{34} &a_0+\ii a_{23} \\
\end{bmatrix}.\end{equation}
As in previous cases, it is convenient to introduce the following
parameters,
\begin{equation}\begin{split}
&\varepsilon_{\pm}=\pm\ii 2 a_{1234}\magn{\gb{\m{A}}},\text{ where }\magn{\gb{\m{A}}}=\magn{\cA}=\sqrt{a_{23}^2+a_{24}^2+a_{34}^2}\,,\\
&\beta_{\pm}= \sqrt{ -a_{1234}^2+\magn{\gb{\cA}}^2\pm\varepsilon_{\pm}}\,. \\
\end{split}
\end{equation}
Then the spectrum of initial MV may be expressed by
\begin{equation}
\varepsilon_{1,2}=a_0\pm\ii(a_{1234}+\magn{\gb{\m{A}}}),\quad
\varepsilon_{3,4}=a_0\pm\ii(a_{1234}-\magn{\gb{\m{A}}})\,.
\end{equation}

The  complex conjugate pairs  (as illustrated in Fig.~\ref{fig:1}),
the sum of which  give real $\{\pi_{R1},\pi_{R2}\}\in\bbR$ and imaginary
$\{\pi_{I1},\pi_{I2}\}\in\bbI$ square root numbers
related to spectrum, then are introduced,
\begin{equation}\begin{split}\label{piR12}
&\pi_{R1}=\tfrac12\big(\sqrt{a_0+\ii(a_{1234}+\magn{\gb{\m{A}}})}+\sqrt{a_0-\ii(a_{1234}+\magn{\gb{\m{A}}})}\big)=(\sqrt{\varepsilon_1}+\sqrt{\varepsilon_2})/2,\\
&\pi_{R2}=\tfrac12\big(\sqrt{a_0+\ii(a_{1234}-\magn{\gb{\m{A}}})}+\sqrt{a_0-\ii(a_{1234}-\magn{\gb{\m{A}}})}\big)=(\sqrt{\varepsilon_3}+\sqrt{\varepsilon_4})/2,\\
&\pi_{I1}=\tfrac12\big(\sqrt{a_0+\ii(a_{1234}+\magn{\gb{\m{A}}})}-\sqrt{a_0-\ii(a_{1234}+\magn{\gb{\m{A}}})}\big)=(\sqrt{\varepsilon_1}-\sqrt{\varepsilon_2})/2,\\
&\pi_{I2}=\tfrac12\big(\sqrt{a_0+\ii(a_{1234}-\magn{\gb{\m{A}}})}-\sqrt{a_0-\ii(a_{1234}-\magn{\gb{\m{A}}})}\big)=(\sqrt{\varepsilon_3}-\sqrt{\varepsilon_4})/2.\\
\end{split}\end{equation}
In terms of pairs in $\{\pi_{R1},\pi_{R2},\pi_{I1},\pi_{I2}\}$ the MV
square roots are.

  \textit{First pair of roots}.  The real coefficients of roots
$(\sqrt{\m{A}})_{1,2}=\pm\m{B}$ which correspond  to sign combinations  $\{\pm\pm\pm\pm\}$ on the
diagonal root matrix,  are
\begin{alignat}{2}\label{coeffCL40rot}
b_0&=(\pi_{R1}+\pi_{R2})/2, &\quad  b_{1234}&=-\ii(\pi_{I1}+\pi_{I2})/2,\nonumber\\
b_{12}&=a_{34}(\pi_{R1}-\pi_{R2})/(2\magn{\gb{\m{A}}}), &\quad  b_{23}&=\ii a_{23}(-\pi_{I1}+\pi_{I2})/(2\magn{\gb{\m{A}}}),\nonumber\\
b_{13}&=a_{24}(-\pi_{R1}+\pi_{R2})/(2\magn{\gb{\m{A}}}), &\quad  b_{24}&=\ii a_{24}(-\pi_{I1}+\pi_{I2})/(2\magn{\gb{\m{A}}}),\nonumber\\
 b_{14}&=a_{23}(\pi_{R1}-\pi_{R2})/(2\magn{\gb{\m{A}}}),
&\quad b_{34}&=\ii a_{34}(-\pi_{I1}+\pi_{I2})/(2\magn{\gb{\m{A}}}).
\end{alignat}

\textit{Second pair of roots}. Square roots, $(\sqrt{\m{A}})_{3,4}=\pm\m{B}$,
which correspond to  $\{\pm\pm\mp\mp\}$  sign combinations on  the
diagonal, coincide with \eqref{coeffCL40rot}.
Therefore, all in all there are two different roots only.

\vspace{3mm} \textit{Example~1}. Square root of rotor
$\m{A}=\cos\theta+\e{23}\sin\theta$, where $\e{23}$ indicates an
oriented plane and $\theta$ is an angle of rotation in plane $\e{23}$.  Since
$\e{23}^2=-1$, for this simple case one can make use of the Euler
formula to extract the root:
$\sqrt{\m{A}}=(\exp{(\e{23}\theta}))^{1/2}=\cos(\theta/2)+\e{23}\sin(\theta/2)$.
Now, let us apply the general formulas for coefficients,
Eqs~\eqref{coeffCL40rot}. Since
$a_0=\cos\theta$ and $a_{23}=\magn{\gb{\m{A}}}=\sin\theta$, the
$\pi$-coefficients are
$\pi_{R1}=\pi_{R2}=\big(\sqrt{\cos\theta+\ii\sin{\theta}}+\sqrt{\cos\theta-\ii\sin{\theta}}\big)/2$
and
$\pi_{I1}=-\pi_{I2}=\big(\sqrt{\cos\theta+\ii\sin{\theta}}-\sqrt{\cos\theta-\ii\sin{\theta}}\big)/2$,
then we get  the following (real) root  coefficients:
\begin{equation*}\begin{split}
&b_0=\tfrac12(\pi_{R1}+\pi_{R2})=\tfrac12\big(\sqrt{\cos\theta+\ii\sin{\theta}}+\sqrt{\cos\theta-\ii\sin{\theta}}\big),\\
&b_{23}=\ii\tfrac12\sin\theta\big(-\pi_{I1}+\pi_{I2})/\sin\theta=\ii\tfrac12(\sqrt{\cos\theta-\ii\sin{\theta}}-\sqrt{\cos\theta+\ii\sin{\theta}}\big).
\end{split}\end{equation*} The
remaining coefficients are equal to zero. Thus, the square root is
$\sqrt{\m{A}}=\pm(b_0+b_{14}\e{14}+b_{23}\e{23})=\pm\big(\cos{(\theta/2)}+\e{23}\sin{(\theta/2)}\big)$,
square of which gives the initial MV $\m{A}$. The both results were
obtained  for rotors in the  plane $\e{23}$. The coefficients in~\eqref{coeffCL40rot}
allow to construct arbitrary (trigonometric) space rotors in \cl{1}{3}. Thus, the square root of the trigonometric  rotor
halves the angle of rotation in an arbitrary spacelike plane.

\subsubsection{\label{boostRel}Roots of the spinor related to boost}

The remaining rotations, which are  called Minkowski
rotations, include time basis vector $\e1$ and are
performed by combinations of three remaining bivectors $\{\e{12},\e{13},\e{14}\}$.
Since the squares of  bivectors are equal to $+1$, such
rotations are connected with hyperbolic functions and even rotors,
$\m{A}=a_0+a_{12}\e{12}+a_{13}\e{13}+a_{14}\e{14}+a_{1234}\e{1234}\equiv s+\cA+q$.
In matrix form the rotor~is
\begin{equation}
\hat{\m{A}}=\begin{bmatrix} a_0 & 0 &-a_{12}-\ii a_{13}&-x_{14}-\ii a_{1234}\\
0 & a_0&x_{14}-\ii a_{1234}&-a_{12}+\ii a_{13}\\
-a_{12}+\ii a_{13}&x_{14}-\ii a_{1234}&a_0&0\\
-a_{14}-\ii a_{1234}&-x_{12}-\ii a_{13}&0&a_0\\
\end{bmatrix}.
\end{equation}
which has the spectrum,
\begin{equation}\label{boostCL13}
\varepsilon_{1,2}=a_0+\magn{\gb{\m{A}}}\pm\ii a_{1234},\quad
\varepsilon_{3,4}=a_0-\magn{\gb{\m{A}}}\pm\ii a_{1234},
\end{equation} where
$\magn{\gb{\m{A}}}=\magn{\cA}=\sqrt{a_{12}^2+a_{13}^2+a_{14}^2}$\,.
Thus, in general, one expects four roots. Calculations show that,
in fact,  there are two roots only with real coefficients
related to  sign combinations $\{\pm\pm\pm\pm\}$ on the diagonal.  Note,
apart from scalar and pseudoscalar, now the roots include all
six bivectors,
\begin{equation}
\m{B}_{1,2}=\pm(b_0+b_{12}\e{12}+b_{13}\e{13}+b_{23}\e{23}+b_{14}\e{14}
+b_{24}\e{24}+b_{34}\e{34}++b_{1234}\e{1234}).
\end{equation}
We introduce following pairs of complex conjugate shortcuts,
\begin{equation}\begin{split}
&\pi_{R1}=(\sqrt{\epsilon_1}+\sqrt{\epsilon_2})/2,\quad
\pi_{R2}=(\sqrt{\epsilon_3}+\sqrt{\epsilon_4})/2,\\
&\pi_{I1}=(\sqrt{\epsilon_1}-\sqrt{\epsilon_2})/2,\quad
\pi_{I2}=(\sqrt{\epsilon_3}-\sqrt{\epsilon_4})/2.\\
\end{split}\end{equation}
For a square root $\sqrt{\m{A}}=\m{B}$ that correspond to
$\{\pm\pm\pm\pm\}$ signs on the diagonal root matrix we find the
following real coefficient with plus/minus signs:
\begin{alignat*}{2}
b_0&=\big(\pi_{R1}+\pi_{R2}\big)/2,&\quad
 b_{1234}&=-\ii\big(\pi_{I1}+\pi_{I2}\big)/2,\\
b_{12}&=-a_{12}\big(\pi_{R1}-\pi_{R2}\big)/(2\magn{\gb{\m{A}}}),&\quad
 b_{23}&=\ii a_{14}\big(\pi_{I1}-\pi_{I2}\big)/(2\magn{\gb{\m{A}}}),\\
b_{13}&=-a_{13}\big(\pi_{R1}-\pi_{R2}\big)/(2\magn{\gb{\m{A}}}),&\quad
 b_{24}&=-\ii a_{13}\big(\pi_{I1}-\pi_{I2}\big)/(2\magn{\gb{\m{A}}}),\\
b_{14}&=-a_{14}\big(\pi_{R1}-\pi_{R2}\big)/(2\magn{\gb{\m{A}}}),&\quad
 b_{34}&=\ii a_{12}\big(\pi_{I1}-\pi_{I2}\big)/(2\magn{\gb{\m{A}}}).\\
\end{alignat*}
The second pair of roots with diagonal  signs $\{\pm\pm\mp\mp\}$  has all
imaginary coefficients, therefore, the latter roots must be
rejected. In conclusion, in relativistic \cl{1}{3} algebra  there
are only two spectral square roots in the  case of boosts.

\textit{Remark}. It is well known that there are isomorphisms between low dimension GAs and even subalgebra
of high dimension GAs or isomorphisms between Clifford algebras of same dimension. For example, \cl{3}{0} is isomorphism to even subalgebra $\mathit{Cl}_{13}^{+}$.
Therefore, instead of  calculating  the root of even MV $\e{14}+\e{1234}+\e{13}+\e{12}+1$ in \cl{1}{3} he/she may have
a wrong impression that it would be simpler
to find the root from isomorphic MV, $\e{1}+\e{123}+\e{2}+\e{3}+1$, in \cl{3}{0}, or just make use of Sullivan's matrix
formula~\eqref{Sullivan}, and then with help of isomorphism return from \cl{3}{0} back to \cl{1}{3}.
However, such a procedure is illegitimate, since spectral method don't ensure that we will find all square roots. Therefore the application of an isomorphism in general may yield a non spectral root. 

\subsection{Roots of MV in \cl{3}{1}, $\bbR(4)$. Majorana matrices}\label{examples4DRelativity31}

From  table in Sec.~\ref{basisTables}, we have the
following  $4\times 4$ real reps for basis matrix-vectors usually called
Majorana matrices~\cite{Majorana1937},\footnote{This is the most
frequently cited E.~Majorana's article.  According to
B.~Pontecorvo's recollections~\cite{Cifarelli2020}, the famous
Enrico Fermi recommended E.~Majorana to publish the new idea,
however, ``Remembering what happened with the `neutron' discovery,
Fermi wrote the article himself and submitted the work, under
Ettore Majorana's name.... Without Fermi's initiative, we would
know nothing about the Majorana spinors and the Majorana
neutrinos''. Unfortunately, the reader will not find Majorana's
matrices in the mentioned paper.}
\begin{equation}
 \hat{\e{}}_1=\Bigg[\begin{smallmatrix} 1&0&0&0\\0&-1&0&0\\0&0&-1&0\\0&0&0&1\end{smallmatrix}\Bigg],\
 \hat{\e{}}_2=\Bigg[\begin{smallmatrix} 0&1&0&0\\1&0&0&0\\0&0&0&1\\0&0&1&0 \end{smallmatrix}\Bigg],\
 \hat{\e{}}_3=\Bigg[\begin{smallmatrix} 0&0&1&0\\0&0&0&-1\\1&0&0&0\\0&-1&0&0 \end{smallmatrix}\Bigg],\
 \hat{\e{}}_4=\Bigg[\begin{smallmatrix} 0&-1&0&0\\1&0&0&0\\0&0&0&-1\\0&0&1&0
\end{smallmatrix}\Bigg].
\end{equation}
In the matrix rep, the even MV
$\m{A}=a_0+a_{12}\e{12}+a_{13}\e{13}+a_{23}\e{23}+a_{14}\e{14}+a_{24}\e{24}+a_{34}\e{34}
+a_{1234}\e{1234}$ assumes the form
\begin{equation}\hat{\m{A}}=
\begin{bmatrix}
a_0+a_{24}&a_{12}-a_{14} & -a_{1234}+a_{13}&-a_{23}-a_{34} \\
-a_{12}-a_{14} & a_0-a_{24}&a_{23}-a_{34}  &a_{1234}+a_{13} \\
a_{1234}-a_{13} & -a_{23}-a_{34} & a_0+a_{24}& -a_{12}+a_{14}\\
a_{23}-a_{34} &-a_{1234}-a_{13} & a_{12}+a_{14}& a_0-a_{24} \\
\end{bmatrix},
\end{equation}
the eigenvalues of which are
\begin{equation}\begin{split}\label{r1r2}
&\varepsilon_{1,2}=a_0-\ii a_{1234}\mp r_1,\quad
\varepsilon_{3,4}=a_0+\ii a_{1234}\mp r_2,  \\
&r_1=\sqrt{(a_{14}-\ii a_{23})^2-(a_{13}-\ii a_{24})^2-(a_{12}+\ii a_{34})^2}\,, \\
&r_2=\sqrt{(a_{14}+\ii a_{23})^2-(a_{13}+\ii a_{24})^2-(a_{12}-\ii a_{34})^2}\,. \\
\end{split}\end{equation}
Calculations show that there are two pairs of plus/minors roots
for a generic spinor in \cl{3}{1}. The remaining roots have either
imaginary or complex coefficients and therefore  must be rejected.  The square root
$\m{B}=b_0+b_{12}\e{12}+b_{13}\e{13}+b_{23}\e{23}+b_{14}\e{14}+b_{24}\e{24}+b_{34}\e{34}+b_{1234}\e{1234}$
with $\{++++\}$ has the following coefficients,
\begin{equation}\begin{split}\label{rootsCL31A}
 &b_0=\tfrac14(\sqrt{\epsilon_1}+\sqrt{\epsilon_2}+\sqrt{\varepsilon_3}+\sqrt{\varepsilon_4}),
 \quad b_{1234}=\ii\tfrac14(\sqrt{\varepsilon_1}+\sqrt{\varepsilon_2}-\sqrt{\varepsilon_3}-\sqrt{\varepsilon_4}), \\
 &b_{12}=-(a_{12}+\ii a_{34})\chi_1-(a_{12}-\ii a_{34})\chi_2,\\
 &b_{13}=-(a_{13}-\ii a_{24})\chi_1-(a_{13}+\ii a_{24})\chi_2,\\
 &b_{14}=-(a_{14}-\ii a_{23})\chi_1-(a_{14}+\ii a_{23})\chi_2,\\
 &b_{23}=-\ii\big((a_{14}-\ii a_{23})\chi_1-(a_{14}+\ii a_{23})\chi_2\big),\\
 &b_{24}=-\ii\big((a_{13}-\ii a_{24})\chi_1-(a_{13}+\ii a_{24})\chi_2\big),\\
 &b_{34}=\ii\big((a_{12}+\ii a_{34})\chi_1-(a_{12}-\ii a_{34})\chi_2\big),\\
\end{split}\end{equation}
where
$\chi_1=\tfrac14(\sqrt{\varepsilon_1}-\sqrt{\varepsilon_2})\big)/r_1$
and
$\chi_2=\tfrac14(\sqrt{\varepsilon_3}-\sqrt{\varepsilon_4})\big)/r_2$.
The second root with $\{----\}$ has opposite coefficient signs, where $r_1$ and $r_2$ are given~in~\eqref{r1r2}.

The third  root with $\{+-+-\}$, apart from sign combinations, has
similar coefficients,
\begin{equation}\begin{split}\label{rootsCL31B}
 &b_0=-\tfrac14(\sqrt{\varepsilon_1}-\sqrt{\varepsilon_2}+\sqrt{\varepsilon_3}-\sqrt{\varepsilon_4}),
 \quad b_{1234}=\ii\tfrac14(\sqrt{\varepsilon_1}-\sqrt{\varepsilon_2}-\sqrt{\varepsilon_3}+\sqrt{\varepsilon_4}), \\
 &b_{12}=(a_{12}+\ii a_{34})\chi_1+(a_{12}-\ii a_{34})\chi_2,\\
 &b_{13}=(a_{13}-\ii a_{24})\chi_1+(a_{13}+\ii a_{24})\chi_2,\\
 &b_{14}=(a_{14}-\ii
a_{23})\chi_1+(a_{14}+\ii a_{23})\chi_2,\\
 &b_{23}=\ii\big((a_{14}-\ii a_{23})\chi_1-(a_{14}+\ii a_{23})\chi_2\big),\\
 &b_{24}=\ii\big((a_{13}-\ii a_{24})\chi_1-a_{13}+\ii a_{24})\chi_2\big),\\
 &b_{34}=-\ii\big((a_{12}+\ii a_{34})\chi_1-(a_{12}-\ii a_{34})\chi_2\big),\\
\end{split}\end{equation}
where
$\chi_1=\tfrac14(\sqrt{\epsilon_1}+\sqrt{\epsilon_2})\big)/r_1$
and
$\chi_2=\tfrac14(\sqrt{\varepsilon_3}+\sqrt{\varepsilon_4})\big)/r_2$.
For $\{-+-+\}$ the coefficients signs are opposite.

The four roots with sign combinations $\{\pm\pm\mp\mp\}$ and  $\{\pm\mp\mp\mp\}$ have
purely imaginary coefficients  and thus must be rejected.

\vspace{3mm}
 \textit{Example~1}, \cl{3}{1}.
  Square root of scalar-pseudoscalar,
$\m{A}=\eta+\xi\e{1234}$ where $\eta,\xi\in\bbR$, has two
pairs of roots:
\begin{equation*}\begin{split}
&\m{B}_{1,2}=\pm\tfrac12\big(\sqrt{\eta+\ii\xi}+\sqrt{\eta-\ii\xi}
-\ii(\sqrt{\eta+\ii\xi}-\sqrt{\eta-\ii\xi}\,)\e{1234}\big),\\
&\m{B}_{3,4}=\pm\tfrac12\ii\big((\sqrt{\eta-\ii\xi}-\sqrt{\eta+\ii\xi})\e{13}
+\ii(\sqrt{\eta-\ii\xi}+\sqrt{\eta+\ii\xi}\,)\e{24}\big).
\end{split}\end{equation*}
The coefficients of complex conjugate roots  can be transformed to trigonometric functions as
illustrated in  Eq.~\eqref{rootCL01A}.

\vspace{3mm}
 \textit{Example~2},  \cl{3}{1}. Square root of
$\bv=a_1\e{1}+a_2\e{2}+a_3\e{3}+a_4\e{4}$.  The  root existence domain
is  determined by square of vector,
$\bv^2=a_1^2+a_2^2+a_3^2-a_4^2$.
\begin{enumerate}
\item{ $\{\pm\pm\pm\pm\}$}, Roots are absent because the
coefficients are complex numbers for the both real and imaginary
values of $\sqrt{\bv^2}$.
\item{$\{\pm\pm\mp\mp\}$},
$\sqrt{\bv}=\mp\frac12\frac{1-\ii}{(\bv^2)^{1/4}}\sqrt{\bv^2}-\frac{1+\ii}{2(\bv^2)^{1/4}}\bv$. \\
the  roots exist if $\bv^2<0$, i.e. $v_4^2>v_1^2+v_2^2+v_3^2$.
\item{$\{\pm\mp\mp\pm\}$},
$\sqrt{\bv}=\pm\frac12\frac{1+\ii}{(\bv^2)^{1/4}}\big(-a_2\e{12}+\frac{a_2^2-a_4^2}{a_3}\e{13}-
a_4\e{14}+\frac{a_1a_4+\ii
a_2\sqrt{\bv^2}}{a_3}\e{34}+\frac{a_1a_4+\ii
a_2\sqrt{\bv^2}}{a_3}\e{123}+a_1\e{124}-\frac{\bv^2-a_1^2}{a_3}\e{234}\big)+\frac12\frac{1-\ii}{(\bv^2)^{1/4}}\big(\frac{-\ii
a_1a_2+a_4\sqrt{\bv^2}}{a_3}\e{23}+\sqrt{\bv^2}\e{24}+\frac{-\ii
a_1a_2+a_4\sqrt{\bv^2}}{a_3}\e{134}\big)$,\ the  roots exist if\ $\bv^2<0$.
\item{$\{\pm\mp\pm\mp\}$},
$\pm\sqrt{\bv}=\frac12\frac{1+\ii}{(\bv^2)^{1/4}}\big(-\frac{a_4\bv^2+\ii
a_1a_2\sqrt{\bv^2}}{a_3\sqrt{\bv^2}}\e{23}-\frac{\bv^2}{\sqrt{\bv^2}}\e{24}+
\frac{a_2\bv^2+\ii
a_1a_4\sqrt{\bv^2}}{a_3\sqrt{\bv^2}}\e{34}+\frac{a_2\bv^2+\ii
a_1a_4\sqrt{\bv^2}}{a_3\sqrt{\bv^2}}\e{123}-\frac{a_4\bv^2+\ii
a_1a_2\sqrt{\bv^2}}{a_3\sqrt{\bv^2}}\e{134}\big)+\frac12\frac{1-\ii}{(\bv^2)^{1/4}}\big(a_2\e{12}
-\frac{a_2^2-a_4^2}{a_3}\e{13}+a_4\e{14}-a_1\e{124}+\frac{\bv^2-a_1^2}{a_3}\e{234}\big)$,\
\ the roots do not exist (they are imaginary).
\end{enumerate}
Thus, in the coefficient domain the roots exist  if $\bv^2<0$
only. Remaining roots related to signs $\{\pm\pm\pm\mp\}$,
$\{\pm\pm\mp\pm\}$, $\{\pm\mp\pm\pm\}$ and $\{\mp\pm\pm\pm\}$ have
complex coefficients, thus they must be rejected. Therefore, instead of
expected 16 roots in $\sqrt{\bv}$  there remains only 4 roots with real
coefficients. We shall remind once more that
imaginary parts of the coefficients simplify out so that all
coefficients are real.

\section{\label{numerical}Numerical diagonalization. The roots in \cl{4}{1}}
For high dimensional algebras, $n>4$,  calculations in
coordinates may be a laborious task in a  case of general symbolic MV.
Experimentation reveals that the
transformation matrix  $\hat{\m{T}}$  generated by computer program  frequently  has a form of
a rational function (compare simple case represented by
Eq.~\eqref{eigens}) that simplifies substantially if entanglement is avoided, i.e.,
if numerator and denominator are factored in an appropriate form.
However, these problems vanish,  including elimination of removable  singularities, if the MV has
a numerical form.  Thus, the proposed  diagonalization method can be easily extended  to
high dimension GAs when the MV has numerical form. Below, a few  examples
are presented.

Let's calculate discrete roots of $\sqrt{-1}$ in \cl{4}{1}. From
Sec.~\ref{basisTables}, generators of reps are
\begin{equation}\begin{split}\label{basisCL41}
& \hat{\e{}}_1=\Bigg[\begin{smallmatrix} 1&0&0&0\\0&-1&0&0\\0&0&-1&0\\0&0&0&1\end{smallmatrix}\Bigg],\
 \hat{\e{}}_2=\Bigg[\begin{smallmatrix} 0&1&0&0\\1&0&0&0\\0&0&0&1\\0&0&1&0\end{smallmatrix}\Bigg],\
 \hat{\e{}}_3=\Bigg[\begin{smallmatrix} 0&0&1&0\\0&0&0&-1\\1&0&0&0\\0&-1&0&0\end{smallmatrix}\Bigg],\\
& \hat{\e{}}_4=\Bigg[\begin{smallmatrix} 0&0&-\ii&0\\0&0&0&\ii\\\ii&0&0&0\\0&-\ii&0&0\end{smallmatrix}\Bigg],\
 \hat{\e{}}_5=\Bigg[\begin{smallmatrix} 0&-1&0&0\\1&0&0&0\\0&0&0&-1\\0&0&1&0\end{smallmatrix}\Bigg].\
\end{split}\end{equation}
Now, it is a simple matter to apply the spectral diagonalization
method. One finds the following 16 square roots for
different sign combinations:
\begin{equation}\begin{split}
 &\{\mp\pm\pm\pm\},\quad \sqrt{-1}=\pm\tfrac12(\e{34}+\e{134}-\e{2345}+\e{12345}),\\
 &\{\pm\mp\pm\pm\},\quad \sqrt{-1}=\pm\tfrac12(\e{34}-\e{134}+\e{2345}+\e{12345}),\\
 &\{\pm\pm\mp\pm\},\quad \sqrt{-1}=\pm\tfrac12(-\e{34}+\e{134}+\e{2345}+\e{12345}),\\
 &\{\pm\pm\pm\mp\},\quad \sqrt{-1}=\pm\tfrac12(-\e{34}-\e{134}-\e{2345}+\e{12345}),\\
 &\{\pm\pm\pm\pm\},\quad \sqrt{-1}=\pm\e{12345};\qquad\{\pm\pm\mp\mp\},\quad \sqrt{-1}=\pm\e{34},\\
 &\{\pm\mp\mp\pm\},\quad \sqrt{-1}=\pm\e{2345};
 \qquad\ \{\pm\mp\pm\mp\},\quad \sqrt{-1}=\pm\e{134}.\\
 \end{split}\end{equation}
In \cl{4}{1} a  root related to  basis vector
$\sqrt{-1}=\e{5}$ is to be mentioned too.  In the paper~\cite{Hitzer2013}  only five $\sqrt{-1}$ roots
for \cl{4}{1} were found.\footnote{In~\cite{Hitzer2013}, different matrix representations
were used, namely, $(E_{11},-E_{22},-E_{33},E_{44})$,
$(E_{12},E_{21},E_{34},E_{43})$,
$\ii(-E_{12},E_{21},-E_{34},E_{43})$,
$(E_{13},-E_{24},E_{31},-E_{42})$,
$(-E_{13},E_{24},E_{31},-E_{42})$, in the notation of
Sec.~\ref{basisTables}. We have repeated calculation using our
basis, Eq.~\eqref{basisCL41},  and have obtained the same results
as in~\cite{Hitzer2013}\,.}

Now, let's calculate the square roots of $\m{A}=1+\e{1}+2.\e{12}+3.\e{123}+4.\e{1234}+5.\e{12345}$
when  fixed-point notation is used during computation.  In \cl{4}{1},
in  matrix representation the MV $\m{A}$ is
\begin{equation}\hat{\m{A}}=
 \begin{bmatrix}
2.+\ii 5.&2.+\ii 4. & 0 & 3\\
-2.-\ii 4.&\ii 5. & -3. & 0\\
0& 3. & -\ii 5.&-2+\ii 4. \\
3.& 0 &2. -\ii 4.&-2+\ii 5. \\
 \end{bmatrix}.
\end{equation}
Below, only three significant digits
are given out  of the eight  significant digits used in the computation.
The considered  MV $\m{A}$ has eight pairs of roots.
The first  $\{\pm\pm\pm\pm\}$  pair is
\begin{equation}\begin{split}
&\pm\Big(1.769+(0.18\e{1}-0.06\e{3}-0.566\e{5})+(0.306\e{12}+0.03\e{23}+0.02\e{14}-0.08\e{34}+\\
&0.005\e{25}+0.423\e{45})+(0.511\e{123}+0.013\e{134}+0.134\e{125}+0.007\e{235}-0.05\e{145}+\\
&0.269\e{345})+(0.689\e{1234}+0.1\e{1245}-0.144\e{2345})+1.482\e{12345}\Big). \\
 \end{split}\end{equation}
Note that  all coefficients are real numbers. The roots of the second pair, $\{\mp\pm\pm\pm\}$, are real too,
\begin{equation}\begin{split}
&\pm\Big(0.544+(-0.269\e{1}+0.654\e{3}-1.461\e{5})+(-0.4\e{12}-0.327\e{23}+0.051\e{14}+0.872\e{34}+\\
&0.525\e{25}+0.898\e{45})+(-0.73\e{123}-0.034\e{134}+1.744\e{125}+0.788\e{235}-0.654\e{145}-\\
&0.715\e{345})+(-0.99\e{1234}-1.308\e{1245}-0.694\e{2345})+0.758\e{12345}\Big. \\
 \end{split}\end{equation}
The coefficients of remaining  14 roots  were found to be  real numbers as well.

Finally, comparison of numerical roots calculated by spectral method with a more general
numerical  method proposed in ~\cite{Acus2024v2} revealed that the set of discrete roots in the latter is equal or larger
than the number of spectral square roots.  Also, apart from discrete roots,
continuous roots may appear, i.e. the roots containing single or more real free parameter(s) as shown in~\cite{Hitzer2013,Acus2024v2}.


\section{\label{Riccati} Quadratic MV equations}

A simple  quadratic MV equation is
\begin{equation}
\X^2+\A\X+\X\A+\B=0.
\end{equation}
Noting that $\X^2+\A\X+\X\A=(\X+\A)^2-\A^2$ it is
an easy task  to write down  MV  solution,
\begin{equation}
\X=-\A+\sqrt{\A^2-\B}\,.
\end{equation}
Since, as we have seen,  the MV roots require pairs with plus/minus signs, it is enough to
take into account only different pairs that follow from coefficients of MV $(\A^2-\B)$.

The time-dependent differential  matrix Riccati equation
\begin{equation}\label{RiccatiEqDif}
\frac{\dd\m{X}}{\dd
t}=-\m{B(t)}+\m{C(t)}\m{X}+\m{X}\m{D(t)}+\m{X}\m{A(t)}\m{X},
 \end{equation}
plays a prominent role in optimal filter design, control and
system theory~\cite{Abou2003,Lancaster95}. In
Eq.~\eqref{RiccatiEqDif}, $\m{X}$, $\m{A}$, $\m{B}$, $\m{C}$ are
 MVs, where   $\m{X}$ represents unknown. For  stationary case all MVs are
constant and Eq.~\eqref{RiccatiEqDif} reduces to nonlinear Riccati equation
\begin{equation}\label{RiccatiEq}
\m{X}\m{A}\m{X}+\m{C}\m{X}+\m{X}\m{D}=\m{B},
 \end{equation}
If $\m{D}=\m{C}$ and $\m{C}$ is the center of  algebra, i.e.,
$\m{C}$ commutes with all multivectors,  then resulting Clifford-Riccati equation
\begin{equation}\label{XAXCX}
\m{X}\m{A}\m{X}+\m{C}\m{X}+\m{X}\m{C}=\m{B}
\end{equation}
can be solved.  For example, in \cl{3}{0}
and \cl{4}{1} algebras the center is a sum of scalar and
pseudoscalar, $\m{C}=\eta+\xi I$, $\eta,\xi\in\bbR$. Then,
solution of  Eq.~\eqref{XAXCX}~is
\begin{equation}\label{XAXabc}
\m{X}=\Big(-\m{C}\pm\sqrt{\m{BA}+\m{C}^2}\,\Big)\m{A}^{-1}.
\end{equation}
If MVs are the  scalars, $\m{A}=a$, $\m{B}=c$ and $\m{C}=b/2$,
then the  MV equation~\eqref{XAXabc} simplifies to
$x_{1,2}=(-b\pm\sqrt{b^2+4ac})/2a$, which is the solution of the
well-known scalar quadratic equation. The fundamental difference between scalar $x_{1,2}$ and
multivector  $\m{X}$  solutions is that the  former  belongs
to  commutative set while the latter belongs to
non-commutative one and concrete algebra.  Thus, to solve the equation~\eqref{XAXabc} completely, one
must know how to extract the square root from  $\m{BA}+\m{C}^2$.
The algorithm for inverse MV
$\A^{-1}$  can be found  in~\cite{Acus2018}.
Finally, let us illustrate the solution~\eqref{XAXabc} for
\cl{3}{0} when all known MVs in Riccati-Clifford
equation~\eqref{XAXCX} consist of sum of scalar and pseudoscalar
($\e{123}\equiv I$), for example: $\A=1+2I$, $\B=2+3I$, $\C=3+4I$. Then, $I^2=-1$
and $\A^{-1}=(1-2I)/5$. For  plus sign in Eq.~\eqref{XAXabc}, one
finds that the solution is  scalar-pseudoscalar too,
\[\X=\frac{53\sqrt{2}+4\sqrt{541}-22\sqrt{11+\sqrt{1082}}+
\big(-51\sqrt{2}+2\sqrt{541}+4\sqrt{11+\sqrt{1082}}\big)I}{10\sqrt{11+\sqrt{1082}}}.
\]
Insertion of $\X$ back into~\eqref{XAXCX} shows that the solution
satisfies the Riccati-Clifford equation.


\section{\label{conclusions}Conclusions}
In the paper, spectral method to compute square roots of multivector (MV) in real
Clifford algebras \cl{p}{q} is proposed and investigated in detail. The method uses
isomorphism between MVs and matrices, the Bott's periodicity table and matrix eigensystem.
The  method  can be applied to  simple as well as general
MVs  that belong to real Clifford algebra, in symbolic and numerical forms.
A number of examples are presented that illustrate how to use the algoritm in practice
as well as  how  to determine a  domain of the existence of MV square  roots.
The algorithm was also applied to real spinors of \cl{3}{1} and
\cl{1}{3} algebras which are important in relativistic quantum and
spacetime theories~\cite{Doran03}. Usually,  both algebras are considered equivalent
in constructing relativity theory. However, we
have found  that the equivalence breaks down if MV square root
appears. This is because a spinor  in the real \cl{3}{1} algebra
has only two plus/minus roots, while in the
real \cl{1}{3} algebra there may be up to 12 square roots. We believe that presented  in the paper
MV  root equations  may be useful in
analysis of nonlinear MV  algebraic and differential equations. Also, they may replace matrices at present
used in robotics, control and other computer systems.

At present, the most popular and helpful  methods  to find MV square roots are
numerical ones~\cite{Hitzer2011,Hitzer2013,Acus2024v2,Dargys-Acus2020}.
The first endeavour to calculate square roots in symbolic form  can be found in~\cite{Hitzer2013,Dargys-Acus2020}.
Finally, it should be emphasized that the proposed algorithm is based on MV spectrum and,
thus,  the maximal number of roots is limited by dimension of Bott's irreducible
matrix representation of multivector.   As far as we know,
a general algorithm  that predicts all possible multiple roots, including discrete and continuous ones,
was proposed for 3D geometric algebras in~\cite{Acus2024v2}.


\section{Appendix: Tables of basis vector reps}\label{basisTables}
Below tables of basis vector reps in a matrix form for real
Clifford algebras as well as  how to make  use of the tables are
presented. This is a part of full tables for real and complex GAs
that also include expression for general spinor $\Psi$ in the
ideal basis, its matrix representation $\hat{\Psi}$ and
normalization of the spinor~\cite{Acus2024a}. The tables were
calculated by idempotent and ideal theory.

For illustration, we consider   \cl{2}{2} algebra, which
has  four basis vectors (generators) $\{\e{1},\e{2},\e{3},\e{4}\}$. $[\bbR(4)]$
indicates that the Bott's reps of the algebra belong to $4\times
4$ real matrices.   Individual  items in tables below indicate:

 \begin{enumerate}
 \item  \textit{Primitive idempotents}.  There is
a large number of primitive idempotents $P_i$ that can be used to
construct a spinor. We have chosen a pair $\{\e{1},\e{23}\}$ of
commuting base elements, $\e{1}$ and $\e{12}$, which square to $+1$
and construct the following primitive idempotent:
$P_1=\frac{1}{4}(1+\e{1})(1+\e{23})$.
Remaining idempotents, $P_2=\frac{1}{4}(1+\e{1})(1-\e{23})$,
$P_3=\frac{1}{4}(1-\e{1})(1+\e{23})$ and   $P_4=\frac{1}{4}(1-\e{1})(1-\e{23})$,
differ by signs only.

Once the idempotent $P_i$ is
known, we multiply all basis elements, $2^n=2^4=16$, of \cl{2}{2}
by $P_i$ from right to get a left ideal $S(i)=\{\cl{2}{2}\}P_i$. Thus, one gets that the left ideal $S$
consists of four different members,
$S=\{P,\e{2}P,\e{4}P,\e{24}P\}=\{\tfrac{1}{4}(1+\e{1}+\e{23}+\e{123}),\tfrac{1}{4}(\e{2}+\e{3}-\e{12}-\e{13}),
\tfrac{1}{4} (\e{4}-\e{14}+\e{234}-\e{1234}),
\tfrac{1}{4}(\e{24}+\e{34}+\e{124}+\e{134})\}$.

\item \textit{Two-sided ideal (division ring)}. To find a list
of two sided ideal (division ring) $K$,  multiply the obtained
left ideal elements by the same idempotent from  left,
$K(i)=P(i)\{\cl{2}{2}\}P(i)$. The division ring of $\cl{2}{2}$ contains a
single element $K=\{K(1)\}=P1P=\tfrac{1}{4}(1+\e{1}+\e{23}+\e{123})$.

Being an idempotent, $K(1)$ represents a unit element of the real field,
$K(1)^2=K(1)$, which echoes the scalar number property, $1\cdot
     1=1$. Similarly, if $K(2)$-list would  contain two elements,
 then the second element will play the role of an
     imaginary unit. And, if $K(4)$ would contain four different elements
     it will be isomorphic to quaternion ring. To summarize: a single element plays role of a real unit.
A list of two elements is equivalent to complex field. The for element list must be
     interpreted as quaternion units $q_0=1, q_1=\bi, q_2=\bj$ and $q_3=\bk$.

\item \textit{Ideal basis}. In
$\cl{2}{2}$ case the division ring contains only single element,
therefore, the ideal basis coincides with the ideal itself.

The left ideal basis is obtained by
comparing all the elements of ideal (starting from the first) and
consequently dropping out all  elements that can be obtained from
previous elements after multiplying them by any division ring
element. Since element choice may depend on its position in the
ideal list, the element ordering plays an important role.

\item \textit{Matrix representation of ideal basis vectors in GA}.
Matrix reps of \cl{2}{2}.\index{matrix reps of algebra!\cl{2}{2}}
Once the four-component ideal basis $S=\{S(1),S(2),S(3),S(4)\}$
and single-component field $K=\{K(1)\}$ are known one may find the
matrix representation of ideal basis vectors,
$E_{ij}(\e{k})=\reverse{S}(i)^{\sharp}\e{k}S(j)$, where
$S(i)^{\sharp}$ is the ideal element in reciprocal basis. For
example, $S(1)=\tfrac{1}{4} (1+\e{1}+\e{23}+\e{123})$,
$S(1)^\sharp=\tfrac{1}{4}
(1+\e{}^{1}+\e{}^{23}+\e{}^{123})=\tfrac{1}{4}
(1+\e{1}-\e{23}-\e{123})$, because in \cl{2}{2} the orthonormal
basis is automatically  reciprocal, i.e.,  $\e{}^1=\e{1}$,
$\e{}^2=\e{2}$, $\e{}^3=-\e{3}$, and $\e{}^4=-\e{4}$. In this way,
for each pair of indices $(i,j)$ in the ideal basis we obtain
matrix element $E_{ij}(\e{k})$ for a basis vector $\e{k}$. In our
case $i,j=1,2,3,4$ and $k=1,2,3,4$. The matrices $\hat{e}_1$,
$\hat{e}_2$, $\hat{e}_3$, and $\hat{e}_4$ calculated with
$E_{ij}(\e{k})$ then are, respectively,
\[
\hat{e}_1=
\Biggl[\begin{smallmatrix}
K&0&0&0\\
0&-K&0&0\\
0&0&-K&0\\
0&0&0&K
\end{smallmatrix}\Biggr],
\hat{e}_2=\Biggl[
\begin{smallmatrix}
0&K&0&0\\
K&0&0&0\\
0&0&0&K\\
0&0&K&0
\end{smallmatrix}\Biggr],
\hat{e}_3=\Biggl[
\begin{smallmatrix}
0&-K&0&0\\
K&0&0&0\\
0&0&0&-K\\
0&0&K&0
\end{smallmatrix}\Biggr],
\hat{e}_4=\Biggl[
\begin{smallmatrix}
0&0&-K&0\\
0&0&0&K\\
K&0&0&0\\
0&-K&0&0
\end{smallmatrix}\Biggr],
\]
where $K$ is a shortcut for division ring element $K=\{K(1)\}$,
which in item~4 for \cl{2}{2} is represented by a single entry,
unity in $4\times 4$ matrix $E_{ij}$.

If the division ring $K$
contains more elements, then all such elements will appear in the
calculated $\hat{e}_i$ matrices. In order to get $\bbR$, $\bbC$,
or $\bbH$ matrix representations we need to replace the division
ring element by the corresponding isomorphic element, namely,
$(1)$ or $(1,\ii)$, or $(q_0=1,q_1,q_2,q_3)$, where $q_i$ denotes
the quaternion components $\{\ii,\jj,\kk\}$.

Finally, after calculation of  matrix reps, where $\e{k}$ is
replaced by elements of ideal basis $S_k$, namely
$E_{ij}(S_k)=\reverse{S}^iS_kS_j$, we get the needed matrices:
\[E_{11}=\Biggl[
\begin{smallmatrix}1&0&0&0\\
0&0&0&0\\
0&0&0&0\\
0&0&0&0
\end{smallmatrix}\Biggr],\quad
E_{21}=\Biggl[
\begin{smallmatrix}0&0&0&0\\
1&0&0&0\\
0&0&0&0\\
0&0&0&0
\end{smallmatrix}\Biggr],\quad
E_{31}=\Biggl[
\begin{smallmatrix}0&0&0&0\\
0&0&0&0\\
1&0&0&0\\
0&0&0&0
\end{smallmatrix}\Biggr],\quad
E_{41}=\Biggl[
\begin{smallmatrix}0&0&0&0\\
0&0&0&0\\
0&0&0&0\\
1&0&0&0
\end{smallmatrix}\Biggr],\]
plus/minus combinations of which yield,
\[\hat{\e{}}_1=\Biggl[
\begin{smallmatrix}1&0&0&0\\
0&-1&0&0\\
0&0&-1&0\\
0&0&0&1
\end{smallmatrix}\Biggr],\quad
\hat{\e{}}_2=\Biggl[
\begin{smallmatrix}0&1&0&0\\
1&0&0&0\\
0&0&0&1\\
0&0&1&0
\end{smallmatrix}\Biggr],\quad
\hat{\e{}}_3=\Biggl[
\begin{smallmatrix}0&-1&0&0\\
1&0&0&0\\
0&0&0&-1\\
0&0&3 &0
\end{smallmatrix}\Biggr],\quad
\hat{\e{}}_4=\Biggl[
\begin{smallmatrix}0&0&-1&0\\
0&0&0&1\\
1&0&0&0\\
0&0&-1&0
\end{smallmatrix}\Biggr]. \]
\end{enumerate}

\vspace{3mm}

\begin{flushleft}

\begin{tabular}{l|lLL}
  & Item number and  name & \cl{1}{0}[{}^2\bbR(1)] &\cl{0}{1}[\bbC(1)]\index{algebra!\cl{1}{0}}\index{algebra!\cl{0}{1}}
\\ \hline
1&\textrm{Idempotent, $P_i$}&\index{idempotent!real algebras}
\{P_1=\tfrac12 (1 + \e{1}),P_2=(-)=\gradeinverse{P_1}\}
&
\begin{multlined}[t][0.1\columnwidth]
\{P_1=1 \}
\end{multlined}
\\
2&\textrm{Double-sided ideal}, $K_1$&\index{ideal!double-sided}
\{P_1\}
&
\begin{multlined}[t][0.1\columnwidth]
\{1,\e{1}\}
\end{multlined}
\\
3&\textrm{Ideal basis}, $S_1$&\index{basis!ideal}
\{P_1\}\cup \{\gradeinverse{P_1}\}
&
\begin{multlined}[t][0.1\columnwidth]
\{P_1\}
\end{multlined}
\\
4&\textrm{Vector matrices, $\hat{e}_j$}&\index{vector!matrix
representation}
\{E_{11} - E_{22}\}
&
\begin{multlined}[t][0.1\columnwidth]
\{\ii E_{11}\}
\end{multlined}
\end{tabular}
\vskip 10pt

\begin{tabular}{l|>{$}l<{$}>{$}l<{$}>{$}l<{$}}
  &  \cl{2}{0}[\bbR(2)] & \cl{1}{1}[\bbR(2)] &\cl{0}{2}[\bbH]
\\ \hline
1&
\begin{multlined}[t][0.1\columnwidth]
\{P_1=\tfrac12 (1 + \e{1}),\\[-3ex]P_2=(-)\}
\end{multlined}
&
\begin{multlined}[t][0.1\columnwidth]
\{P_1=\tfrac12 (1 + \e{1}),\\[-3ex]P_2=(-)\}
\end{multlined}
&
\begin{multlined}[t][0.1\columnwidth]
\{P_1=1 \}
\end{multlined}
\\
2&
\{P_1\}
&
\{P_1\}
&
\begin{multlined}[t][0.1\columnwidth]
\{1,\e{1}, \e{2}\}
\end{multlined}
\\
3&
\begin{multlined}[t][0.1\columnwidth]
  \{P_1,\e{2}P_1\}
\end{multlined}
&
\begin{multlined}[t][0.1\columnwidth]
 \{P_1,\e{2}P_1\}
\end{multlined}
&
\begin{multlined}[t][0.1\columnwidth]
\{P_1\}
\end{multlined}
\\
4&
\begin{multlined}[t][0.1\columnwidth]\{E_{11} - E_{22},\\[-3ex] E_{12} + E_{21}\}
\end{multlined}
&
\begin{multlined}[t][0.1\columnwidth]
\{E_{11} - E_{22},\\[-3ex] -E_{12} + E_{21}\}
\end{multlined}
&
\begin{multlined}[t][0.1\columnwidth]
\{q_1 E_{11}, q_2 E_{11}\}
\end{multlined}
\end{tabular}
\vspace{5mm}

\begin{tabular}{l|>{$}l<{$}>{$}l<{$}>{$}l<{$}>{$}l<{$}}
& \cl{3}{0}[\bbC(2)]& \cl{2}{1}[{^2\bbR(2)}]& \cl{1}{2}[\bbC(2)]
 \index{algebra!\cl{3}{0}}\index{algebra!\cl{2}{1}}\index{algebra!\cl{1}{2}}
\\ \hline
1&
\begin{multlined}[t][0.1\columnwidth]
  \{P_1=\tfrac12 (1 + \e{1}),\\[-3ex]P_2=(-)\}
\end{multlined}
&
\begin{multlined}[t][0.1\columnwidth]
\{P_1=\tfrac14 (1 + \e{1})(1 + \e{23}),\\[-3ex]
  P_2=(+,-),
  P_3=(-,+),\\[-1ex]
  P_4=(-,-)
\}
\end{multlined}
&
\begin{multlined}[t][0.1\columnwidth]
  \{P_1=\tfrac12 (1 + \e{1}), \\[-3ex]P_2=(-)\}
\end{multlined}
\\
2&
\begin{multlined}[t][0.1\columnwidth]
\{P_1 1\, P_1, P_1 \e{23} P_1\}
\end{multlined}
&
\{P_1 1\, P_1\}
&
\begin{multlined}[t][0.1\columnwidth]
  \{P_1 1\, P_1, P_1 \e{23} P_1\}
\end{multlined}
\\
3&
\begin{multlined}[t][0.1\columnwidth]
\{P_1,\e{2} P_1\}
\end{multlined}
&
\begin{multlined}[t][0.1\columnwidth]
  \{P_1, \e{2} P_1\}\cup \{\gradeinverse{P_1}, \gradeinverse{\e{2} P_1}\}
\end{multlined}
&
\begin{multlined}[t][0.1\columnwidth]
\{P_1,
 \e{2}P_1\}
\end{multlined}
\\
4&
\begin{multlined}[t][0.1\columnwidth]
  \{E_{11} - E_{22},\\[-3ex] E_{12} + E_{21},\\[-1ex] \ii (-E_{12} + E_{21})
\}
\end{multlined}
&
\begin{multlined}[t][0.1\columnwidth]\{E_{11} - E_{22} - E_{33} + E_{44},\\[-3ex]
  E_{12} + E_{21} - E_{34} - E_{43},\\[-1ex]
-E_{12} + E_{21} + E_{34} - E_{43}\}
\end{multlined}
&
\begin{multlined}[t][0.1\columnwidth]
  \{E_{11} - E_{22},\\[-3ex] -E_{12} + E_{21},\\[-1ex] -\ii E_{12} - \ii E_{21}
\}
\end{multlined}
\end{tabular}
\vskip 10pt

\begin{tabular}{l|>{$}l<{$}>{$}l<{$}>{$}l<{$}}
& \cl{0}{3}[{^2\bbH(1)}]& \cl{4}{0}[\bbH(2)]
 \index{algebra!\cl{0}{3}}\index{algebra!\cl{4}{0}}
\\ \hline
1&
\begin{multlined}[t][0.1\columnwidth]
\{P_1=\tfrac12 (1 + \e{123}),
 P_2=\tfrac12(1 - \e{123})
 \}
\end{multlined}
&
\begin{multlined}[t][0.1\columnwidth]
\{P_1=\tfrac12 (1 + \e{1}),
 P_2=\tfrac12(1 - \e{1})
 \}
\end{multlined}
\\
2&
\begin{multlined}[t][0.1\columnwidth]
\{P_1 1 P_1, P_1 \e{1} P_1, P_1 \e{2} P_1, P_1 \e{3} P_1
 \}
\end{multlined}
&
\begin{multlined}[t][0.1\columnwidth]
  \{P_1 1 P_1, P_1 \e{23} P_1, P_1 \e{24} P_1, P_1 \e{34} P_1 \}
\end{multlined}
\\
3&
\begin{multlined}[t][0.1\columnwidth]
\{\e{3} P_1\}\cup \{\gradeinverse{\e{3} P_1}\}
\end{multlined}
&
\begin{multlined}[t][0.1\columnwidth]
\{P_1, \e{2}P_1\}
\end{multlined}
\\
4&
\begin{multlined}[t][0.1\columnwidth]
\{-q_{1} E_{11} + q_{1}E_{22},
-q_{2}E_{11} + q_{2} E_{22},\\[-3ex]
-q_{3}E_{11} + q_{3} E_{22}
\}
\end{multlined}
&
\begin{multlined}[t][0.1\columnwidth]
\{E_{11} - E_{22},
 E_{12} + E_{21},\\[-3ex]
-q_{1}E_{12} + q_{1}E_{21},
-q_{2}E_{12} + q_{2}E_{21}\}
\end{multlined}
\end{tabular}
\vskip 10pt

\begin{tabular}{l|>{$}l<{$}>{$}l<{$}>{$}l<{$}}
& \cl{3}{1}[\bbR(4)]& \cl{2}{2}[\bbR(4)]
 \index{algebra!\cl{3}{1}}\index{algebra!\cl{2}{3}}
\\ \hline
1&
\begin{multlined}[t][0.1\columnwidth]
\{P_1=\tfrac14 (1 + \e{1})(1 + \e{24}),\\[-3ex]
 P_2=(+,-), P_3=(-,+),\\[-1ex]
  P_4=(-,-)
\}
\end{multlined}
&
\begin{multlined}[t][0.1\columnwidth]
\{P_1=\tfrac14 (1 + \e{1})(1 + \e{23}),\\[-3ex]
 P_2=(+,-),
  P_3=(-,+),\\[-1ex]
  P_4=(-,-)
 \}
\end{multlined}
\\
2&
\begin{multlined}[t][0.1\columnwidth]
  \{P_1 1 P_1\}
\end{multlined}
&
\begin{multlined}[t][0.1\columnwidth]
  \{P_1 1 P_1\}
\end{multlined}
\\
3&
\begin{multlined}[t][0.1\columnwidth]
\{P_1, \e{2}P_1,\e{3}P_1,\e{23}P_1\}
\end{multlined}
&
\begin{multlined}[t][0.1\columnwidth]
\{P_1,\e{2}P_1,\e{4}P_1,\e{24}P_1\}
\end{multlined}
\\
4&
\begin{multlined}[t][0.1\columnwidth]
\{E_{11} - E_{22} - E_{33} + E_{44},\\[-3ex]
E_{12} + E_{21} + E_{34} + E_{43},\\[-1ex]
E_{13} - E_{24} + E_{31} - E_{42},\\[-1ex]
 -E_{12} + E_{21} - E_{34} + E_{43}
\}
\end{multlined}
&
\begin{multlined}[t][0.1\columnwidth]
\{E_{11} - E_{22} - E_{33} + E_{44},\\[-3ex]
E_{12} + E_{21} + E_{34} + E_{43},\\[-1ex]
-E_{12} + E_{21} - E_{34} + E_{43},\\[-1ex]
 -E_{13} + E_{24} + E_{31} - E_{42}
\}
\end{multlined}
\end{tabular}
\vskip 10pt

\begin{tabular}{l|>{$}l<{$}>{$}l<{$}>{$}l<{$}>{$}l<{$}}
&  \cl{1}{3}[\bbH(2)]& \cl{0}{4}[\bbH(2)]
 \index{algebra!\cl{1}{3}}\index{algebra!\cl{0}{4}}
\\ \hline
1&
\begin{multlined}[t][0.1\columnwidth]
\{P_1=\tfrac12 (1 + \e{1}),
 P_2=\tfrac12 (1 - \e{1})
 \}
\end{multlined}
&
\begin{multlined}[t][0.1\columnwidth]
\{P_1=\tfrac12 (1 + \e{123}),
 P_2=\tfrac12 (1 - \e{123})
 \}
\end{multlined}
\\
2&
\begin{multlined}[t][0.1\columnwidth]
\{P_1 1 P_1, P_1 \e{23} P_1, P_1 \e{24} P_1, P_1 \e{34} P_1 \}
\end{multlined}
&
\begin{multlined}[t][0.1\columnwidth]
\{P_1 1 P_1, P_1 \e{1} P_1, P_1 \e{2} P_1, P_1 \e{3} P_1 \}
\end{multlined}
\\
3&
\begin{multlined}[t][0.1\columnwidth]
\{P_1, \e{2} P_1\}
\end{multlined}
&
\begin{multlined}[t][0.1\columnwidth]
  \{\e{3} P_1, \e{34} P_1\}
\end{multlined}
\\
4&
\begin{multlined}[t][0.1\columnwidth]
\{E_{11} - E_{22},
  -E_{12} + E_{21},\\[-3ex]
-q_{1}E_{12} - q_{1}E_{21},
-q_{2}E_{12} - q_{2}E_{21}\}
\end{multlined}
&
\begin{multlined}[t][0.1\columnwidth]
  \{-q_{1}E_{11} + q_{1}E_{22},
-q_{2}E_{11} + q_{2} E_{22},\\[-3ex]
-q_{3}E_{11}+q_{3}E_{22},
 E_{12} - E_{21}
\}
\end{multlined}
\end{tabular}
\vskip 10pt

\begin{tabular}{l|>{$}l<{$}>{$}l<{$}>{$}l<{$}}
& \cl{5}{0}[{^2\bbH(2)}]& \cl{4}{1}[\bbC(4)]
 \index{algebra!\cl{5}{0}}\index{algebra!\cl{4}{1}}
\\ \hline
1&
\begin{multlined}[t][0.1\columnwidth]
\{P_1=\tfrac14 (1 + \e{1})(1 + \e{2345}),\\[-3ex]
 P_2=\tfrac14 (1 + \e{1})(1 - \e{2345}),\\[-1ex]
 P_3=\gradeinverse{P_1}=\tfrac14 (1 - \e{1})(1 + \e{2345}),\\[-1ex]
 P_4=\tfrac14 (1 - \e{1})(1 - \e{2345})\}
\end{multlined}
&
\begin{multlined}[t][0.1\columnwidth]
\{P_1=\tfrac14 (1 + \e{1})(1 + \e{25}),\\[-3ex]
 P_2=\tfrac14 (1 + \e{1})(1 - \e{25}),\\[-1ex]
 P_3=\tfrac14 (1 - \e{1})(1 + \e{25}),\\[-1ex]
 P_4=\tfrac14 (1 - \e{1})(1 - \e{25})\}
\end{multlined}
\\
2&
\begin{multlined}[t][0.1\columnwidth]
  \{P_1 1 P_1,P_1 \e{23} P_1,P_1 \e{24} P_1, P_1 \e{25} P_1\}
\end{multlined}
&
\begin{multlined}[t][0.1\columnwidth]
\{P_1 1 P_1,P_1 \e{34} P_1\}
\end{multlined}
\\
3&
\begin{multlined}[t][0.1\columnwidth]
  \{\e{25}P_1,\e{5}P_1\}\cup \{\gradeinverse{\e{25}P_1},\gradeinverse{\e{5}P_1}\}
\end{multlined}
&
\begin{multlined}[t][0.1\columnwidth]
\{P_1,\e{2}P_1,\e{3}P_1,\e{23}P_1\}
\end{multlined}
\\
4&
\begin{multlined}[t][0.1\columnwidth]
\{E_{11} - E_{22} - E_{33} + E_{44},\\[-3ex]
E_{12} + E_{21} - E_{34} - E_{43},\\[-1ex]
q_{1}(E_{12} - E_{21}  - E_{34}+ E_{43}),\\[-1ex]
q_{2}(E_{12}- E_{21} - E_{34} + E_{43}),\\[-1ex]
q_{3}(-E_{12} + E_{21} + E_{34}  - E_{43})
\}
\end{multlined}
&
\begin{multlined}[t][0.1\columnwidth]
\{E_{11} - E_{22} - E_{33} + E_{44},\\[-3ex]
E_{12} + E_{21} + E_{34} + E_{43},\\[-1ex]
E_{13} - E_{24} + E_{31} - E_{42},\\[-1ex]
\ii(-E_{13} + E_{24} + E_{31} - E_{42}),\\[-1ex]
-E_{12} + E_{21} - E_{34} + E_{43}
\}
\end{multlined}
\end{tabular}
\vskip 10pt

\begin{tabular}{l|>{$}l<{$}}
& \cl{3}{2}[{^2\bbR(4)}] \index{algebra!\cl{3}{2}}
\\ \hline
1&
\begin{multlined}[t][0.1\columnwidth]
\{P_1=\tfrac18 (1 + \e{1})(1 + \e{24})(1 + \e{35}),
P_2=(+,+,-),\\[-3ex] P_3=(+,-,+), P_4=(+,-,-),
P_5=\gradeinverse{P_1}=(-,+,+), \\[-1ex]
P_6=(-,+,-), P_7=(-,-,+), P_8=(-,-,-)
\}
\end{multlined}
\\
2&
\begin{multlined}[t][0.1\columnwidth]
\{P_1 1 P_1\}
\end{multlined}
\\
3&
\begin{multlined}[t][0.1\columnwidth]
  \{P_1, \e{2} P_1,\e{3} P_1,\e{23} P_1\}\cup \{\gradeinverse{P_1}, \gradeinverse{\e{2} P_1},\gradeinverse{\e{3} P_1}, \gradeinverse{\e{23} P_1}
\}
\end{multlined}
\\
4&
\begin{multlined}[t][0.1\columnwidth]
\{E_{11} - E_{22} - E_{33} + E_{44} - E_{55} + E_{66} + E_{77} - E_{88},\\[-3ex]
E_{12} + E_{21} + E_{34} + E_{43} - E_{56} - E_{65} - E_{78} - E_{87},\\[-1ex]
E_{13} - E_{24} + E_{31} - E_{42} - E_{57} + E_{68} - E_{75} + E_{86},\\[-1ex]
-E_{12} + E_{21} - E_{34} + E_{43} + E_{56} - E_{65} + E_{78} - E_{87},\\[-1ex]
-E_{13} + E_{24} + E_{31} - E_{42} + E_{57} - E_{68} - E_{75} + E_{86}
\}
\end{multlined}
\end{tabular}
\vskip 10pt

\begin{tabular}{l|>{$}l<{$}}
& \cl{2}{3}[\bbC(4)]
\\ \hline
1&
\begin{multlined}[t][0.1\columnwidth]
\{P_1=\tfrac14 (1 + \e{1})(1 + \e{23}),
P_2=(+,-), P_3=(-,+), P_4=(-,-)\}
\end{multlined}
\\
2&
\begin{multlined}[t][0.1\columnwidth]
\{P_1 1 P_1,P_1 \e{45} P_1\}
\end{multlined}
\\
3&
\begin{multlined}[t][0.1\columnwidth]
\{P_1, \e{2} P_1,\e{4} P_1,\e{24} P_1\}
\end{multlined}
\\
4&
\begin{multlined}[t][0.1\columnwidth]
\{E_{11} - E_{22} - E_{33} + E_{44},
E_{12} + E_{21} + E_{34} + E_{43},
-E_{12} + E_{21} - E_{34} + E_{43},\\[-3ex]
-E_{13} + E_{24} + E_{31} - E_{42},
\ii(-E_{13} + E_{24} - E_{31} + E_{42})
\}
\end{multlined}
\end{tabular}
\vskip 10pt

\begin{tabular}{l|>{$}l<{$}}
& \cl{1}{4}[{^2\bbH(2)}]\index{algebra!\cl{1}{4}}
\\ \hline
1&
\begin{multlined}[t][0.1\columnwidth]
\{P_1=\tfrac14 (1 + \e{1})(1 + \e{2345}),
P_2=(+,-), P_3=\gradeinverse{P_1}=(-,+), P_4=(-,-)\}
\end{multlined}
\\
2&
\begin{multlined}[t][0.1\columnwidth]
\{P_1 1 P_1, P_1 \e{23} P_1, P_1\e{24} P_1, P_1 \e{25} P_1\}
\end{multlined}
\\
3&
\begin{multlined}[t][0.1\columnwidth]
\{\e{25}P_1,\e{5}P_1\}\cup \{\gradeinverse{\e{25}P_1},\gradeinverse{\e{5}P_1}\}
\end{multlined}
\\
4&
\begin{multlined}[t][0.1\columnwidth]
\{E_{11} - E_{22} - E_{33} + E_{44},
E_{12} - E_{21} - E_{34} + E_{43},
q_{1}(-E_{12} - E_{21} + E_{34} + E_{43}),\\[-3ex]
q_{2}(-E_{12} - E_{21} + E_{34} + E_{43}),
q_{3}(-E_{12} - E_{21} + E_{34} + E_{43})
\}
\end{multlined}
\end{tabular}
\vskip 10pt

\begin{tabular}{l|>{$}l<{$}}
& \cl{0}{5}[\bbC(4)]]\index{algebra!\cl{0}{5}}
\\ \hline
1&
\begin{multlined}[t][0.1\columnwidth]
\{P_1=\tfrac14 (1 + \e{123})(1 + \e{145}),
P_2=(+,-), P_3=(-,+), P_4=(-,-)\}
\end{multlined}
\\
2&
\begin{multlined}[t][0.1\columnwidth]
  \{P_1 1 P_1, P_1 \e{1} P_1\}
\end{multlined}
\\
3&
\begin{multlined}[t][0.1\columnwidth]
\{P_1,\e{2} P_1, \e{4} P_1, \e{24} P_1\}
\end{multlined}
\\
4&
\begin{multlined}[t][0.1\columnwidth]
\{\ii(E_{11} - E_{22} - E_{33} + E_{44}),
-E_{12} + E_{21} - E_{34} + E_{43},\\[-3ex]
\ii(E_{12} + E_{21} + E_{34} + E_{43}),
-E_{13} + E_{24} + E_{31} - E_{42},\\[-1ex]
\ii(E_{13} - E_{24} + E_{31} - E_{42})
\}
\end{multlined}
\end{tabular}
\vskip 10pt

\begin{tabular}{l|>{$}l<{$}>{$}l<{$}}
& \cl{6}{0}[{\bbH(4)}]\index{algebra!\cl{6}{0}}
\\ \hline
1&
\begin{multlined}[t][0.1\columnwidth]
\{P_1=\tfrac14 (1 + \e{1})(1 + \e{2345}),
P_2=(+,-), P_3=\gradeinverse{P_1}=(-,+), P_4=(-,-)\}
\end{multlined}
\\
2&
\begin{multlined}[t][0.1\columnwidth]
  \{P_1 1 P_1, P_1 \e{23} P_1, P_1 \e{24} P_1, P_1 \e{25} P_1\}
\end{multlined}
\\
3&
\begin{multlined}[t][0.1\columnwidth]
  \{\e{25} P_1,\e{5} P_1,\e{256} P_1,\e{56} P_1\}
\end{multlined}
\\
4&
\begin{multlined}[t][0.1\columnwidth]
\{E_{11} - E_{22} - E_{33} + E_{44},
E_{12} + E_{21} + E_{34} + E_{43},
q_{1}(E_{12} - E_{21}+ E_{34} - E_{43}),\\[-3ex]
q_{2}(E_{12} - E_{21}  + E_{34} - E_{43}),
q_{3}(-E_{12}+ E_{21} - E_{34}+ E_{43}),\\[-1ex]
E_{13} - E_{24} + E_{31} - E_{42}
\}
\end{multlined}
\end{tabular}
\vskip 10pt

\begin{tabular}{l|>{$}l<{$}}
 & \cl{5}{1}[\bbH(4)]\index{algebra!\cl{5}{1}}
\\ \hline
1&
\begin{multlined}[t][0.1\columnwidth]
\{P_1=\tfrac14 (1 + \e{1})(1 + \e{26}),
P_2=(+,-), P_3=(-,+), P_4=(-,-)\}
\end{multlined}
\\
2&
\begin{multlined}[t][0.1\columnwidth]
  \{P_1 1 P_1, P_1 \e{34} P_1, P_1 \e{35} P_1, P_1 \e{45} P_1\}
\end{multlined}
\\
3&
\begin{multlined}[t][0.1\columnwidth]
\{P_1,\e{2}P_1,\e{3}P_1,\e{23}P_1\}
\end{multlined}
\\
4&
\begin{multlined}[t][0.1\columnwidth]
\{E_{11} - E_{22} - E_{33} + E_{44},
 E_{12} + E_{21} + E_{34} + E_{43},\\[-3ex]
E_{13} - E_{24} + E_{31} - E_{42},
q_{1}(-E_{13} + E_{24} + E_{31}- E_{42}),\\[-1ex]
q_{2}(-E_{13} + E_{24} + E_{31}- E_{42}),
-E_{12} + E_{21} - E_{34} + E_{43}
\}
\end{multlined}
\end{tabular}
\vskip 10pt

\begin{tabular}{l|>{$}l<{$}}
 &  \cl{4}{2}[{\bbR(8)}]\index{algebra!\cl{4}{2}}
\\ \hline
1&
\begin{multlined}[t][0.1\columnwidth]
\{P_1=\tfrac18 (1 + \e{1})(1 + \e{25})(1 + \e{36}),
P_2=(+,+,-),\\[-3ex] P_3=(+,-,+), P_4=(+,-,-),
P_5=(-,+,+), \\[-1ex]
P_6=(-,+,-), P_7=(-,-,+), P_8=(-,-,-)
\}
\end{multlined}
\\
2&
\begin{multlined}[t][0.1\columnwidth]
  \{P_1 1 P_1\}
\end{multlined}
\\
3&
\begin{multlined}[t][0.1\columnwidth]
  \{P_1,\e{2}P_1,\e{3}P_1,\e{23}P_1,\e{4}P_1,\e{24}P_1,\e{34}P_1,\e{234}P_1\}
\end{multlined}
\\
4&
\begin{multlined}[t][0.1\columnwidth]
\{E_{11} - E_{22} - E_{33} + E_{44} - E_{55} + E_{66} + E_{77} - E_{88},\\[-3ex]
E_{12} + E_{21} + E_{34} + E_{43} + E_{56} + E_{65} + E_{78} + E_{87},\\[-1ex]
E_{13} - E_{24} + E_{31} - E_{42} + E_{57} - E_{68} + E_{75} - E_{86},\\[-1ex]
E_{15} - E_{26} - E_{37} + E_{48} + E_{51} - E_{62} - E_{73} + E_{84},\\[-1ex]
-E_{12} + E_{21} - E_{34} + E_{43} - E_{56} + E_{65} - E_{78} + E_{87},\\[-1ex]
-E_{13} + E_{24} + E_{31} - E_{42} - E_{57} + E_{68} + E_{75} - E_{86}
\}
\end{multlined}
\end{tabular}
\vskip 10pt

\begin{tabular}{l|>{$}l<{$}}
 & \cl{3}{3}[\bbR(8)]\index{algebra!\cl{3}{3}}
\\ \hline
1&
\begin{multlined}[t][0.1\columnwidth]
\{P_1=\tfrac18 (1 + \e{1})(1 + \e{24})(1 + \e{35}),
P_2=(+,+,-),\\[-3ex] P_3=(+,-,+), P_4=(+,-,-),
P_5=(-,+,+), \\[-1ex]
P_6=(-,+,-), P_7=(-,-,+), P_8=(-,-,-)
\}
\end{multlined}
\\
2&
\begin{multlined}[t][0.1\columnwidth]
\{P_1 1 P_1 \}
\end{multlined}
\\
3&
\begin{multlined}[t][0.1\columnwidth]
\{P_1,\e{2}P_1,\e{3}P_1,\e{23}P_1,\e{6}P_1,\e{26}P_1,\e{36}P_1,\e{236}P_1\}
\end{multlined}
\\
4&
\begin{multlined}[t][0.1\columnwidth]
\{E_{11} - E_{22} - E_{33} + E_{44} - E_{55} + E_{66} + E_{77} - E_{88},\\[-3ex]
E_{12} + E_{21} + E_{34} + E_{43} + E_{56} + E_{65} + E_{78} + E_{87},\\[-1ex]
E_{13} - E_{24} + E_{31} - E_{42} + E_{57} - E_{68} + E_{75} - E_{86},\\[-1ex]
-E_{12} + E_{21} - E_{34} + E_{43} - E_{56} + E_{65} - E_{78} + E_{87},\\[-1ex]
-E_{13} + E_{24} + E_{31} - E_{42} - E_{57} + E_{68} + E_{75} - E_{86},\\[-1ex]
-E_{15} + E_{26} + E_{37} - E_{48} + E_{51} - E_{62} - E_{73} + E_{84}
\}
\end{multlined}
\end{tabular}
\vskip 10pt

\begin{tabular}{l|>{$}l<{$}}
& \cl{2}{4}[\bbH(4)]\index{algebra!\cl{2}{4}}
\\ \hline
1&
\begin{multlined}[t][0.1\columnwidth]
\{P_1=\tfrac14 (1 + \e{1})(1 + \e{23}),
P_2=(+,-), P_3=(-,+), P_4=(-,-)\}
\end{multlined}
\\
2&
\begin{multlined}[t][0.1\columnwidth]
 \{P_1 1 P_1, P_1 \e{45} P_1, P_1 \e{46} P_1, P_1 \e{56} P_1\}
\end{multlined}
\\
3&
\begin{multlined}[t][0.1\columnwidth]
\{P_1,\e{2}P_1,\e{4}P_1,\e{24}P_1\}
\end{multlined}
\\
4&
\begin{multlined}[t][0.1\columnwidth]
\{E_{11} - E_{22} - E_{33} + E_{44},
E_{12} + E_{21} + E_{34} + E_{43},
-E_{12} + E_{21} - E_{34} + E_{43},\\[-3ex]
-E_{13} + E_{24} + E_{31} - E_{42},
q_{1}(-E_{13}+ E_{24}- E_{31}+ E_{42}),\\[-1ex]
q_{2}(-E_{13} + E_{24}- E_{31} + E_{42})
\}
\end{multlined}
\end{tabular}
\vskip 10pt

\begin{tabular}{l|>{$}l<{$}}
& \cl{1}{5}[\bbH(4)]\index{algebra!\cl{1}{5}}
\\ \hline
1&
\begin{multlined}[t][0.1\columnwidth]
\{P_1=\tfrac14 (1 + \e{1})(1 + \e{2345}),
P_2=(+,-), P_3=(-,+), P_4=(-,-)\}
\end{multlined}

\\
2&
\begin{multlined}[t][0.1\columnwidth]
 \{P_1 1 P_1, P_1 \e{23} P_1, P_1 \e{24} P_1, P_1 \e{25} P_1\}
\end{multlined}
\\
3&
\begin{multlined}[t][0.1\columnwidth]
\{\e{25}P_1,\e{5}P_1,\e{256}P_1,\e{56}P_1\}
\end{multlined}
\\
4&
\begin{multlined}[t][0.1\columnwidth]
\{E_{11} - E_{22} - E_{33} + E_{44},
E_{12} - E_{21} + E_{34} - E_{43},\\[-3ex]
q_{1}(-E_{12} - E_{21} - E_{34}- E_{43}),
q_{2}(-E_{12} - E_{21} - E_{34} - E_{43}),\\[-1ex]
q_{3}(-E_{12} - E_{21} - E_{34}- E_{43}),
-E_{13} + E_{24} + E_{31} - E_{42}
\}
\end{multlined}
\end{tabular}
\vskip 10pt

\begin{tabular}{l|>{$}l<{$}}
& \cl{0}{6}[\bbR(8)]\index{algebra!\cl{0}{6}}
\\ \hline
1&
\begin{multlined}[t][0.1\columnwidth]
\{P_1=\tfrac18 (1 + \e{123})(1 + \e{145})(1 + \e{246}),
P_2=(+,+,-),\\[-3ex] P_3=(+,-,+), P_4=(+,-,-),
P_5=(-,+,+), \\[-1ex]
P_6=(-,+,-), P_7=(-,-,+), P_8=(-,-,-)
\}
\end{multlined}
\\
2&
\begin{multlined}[t][0.1\columnwidth]
  \{P_1 1 P_1 \}
\end{multlined}
\\
3&
\begin{multlined}[t][0.1\columnwidth]
\{P_1,\e{1}P_1,\e{2}P_1,\e{3}P_1,\e{4}P_1,\e{5}P_1,\e{6}P_1,\e{16}P_1\}
\end{multlined}
\\
4&
\begin{multlined}[t][0.1\columnwidth]
\{-E_{12} + E_{21} + E_{34} - E_{43} + E_{56} - E_{65} - E_{78} + E_{87},\\[-3ex]
-E_{13} - E_{24} + E_{31} + E_{42} + E_{57} + E_{68} - E_{75} - E_{86},\\[-1ex]
 -E_{14} + E_{23} - E_{32} + E_{41} + E_{58} - E_{67} + E_{76} - E_{85},\\[-1ex]
-E_{15} - E_{26} - E_{37} - E_{48} + E_{51} + E_{62} + E_{73} + E_{84},\\[-1ex]
-E_{16} + E_{25} - E_{38} + E_{47} - E_{52} + E_{61} - E_{74} + E_{83},\\[-1ex]
-E_{17} + E_{28} + E_{35} - E_{46} - E_{53} + E_{64} + E_{71} - E_{82}
\}
\end{multlined}
\\
\end{tabular}
\vskip 10pt
\end{flushleft}

\bibliographystyle{REPORT}

\bibliography{SqrtRootMV}

\end{document}